\documentclass[11pt]{article}
\usepackage{amssymb}
\usepackage{multirow}
\usepackage{amsmath}
\usepackage{amsfonts}
 \usepackage{pifont}
\usepackage{amsmath,amsfonts,amssymb,theorem,graphicx,listings,txfonts}
\usepackage{graphics}
 \usepackage{caption2}
\usepackage{flafter}
 \usepackage{indentfirst}
\usepackage{epsfig}
\usepackage{epstopdf}
  \usepackage{mathrsfs}
  \usepackage{float}
\usepackage{subfigure}
 \usepackage{cite}
\newtheorem{theorem}{Theorem}[section]

\newtheorem{definition}[theorem]{Definition}
\newtheorem{algorithm}[theorem]{Algorithm}

\theoremstyle{example}
\newtheorem{example}[theorem]{Example}
\theoremstyle{programme}

\theoremstyle{property}

\theoremstyle{problem}

\title{Incremental approaches to updating attribute reducts
when refining and coarsening coverings}

\author
{Mingjie Cai$^{1}$ \hspace{0.3cm} Guangming Lang$^{2}$
\thanks{Corresponding author:\quad langguangming1984@126.com
\newline\mbox{}\hspace{0.55cm}
E-mail address: cmjlong@163.com}\hspace{0.3cm}\\
\small {$^{1}$ College of Mathematics and Econometrics, Hunan University}\\
\small {Changsha, Hunan 410004, P.R. China}\\
\small {$^{2}$ School of Mathematics and Statistics, Changsha University of Science and Technology}\\
\small {Changsha, Hunan 410114, P.R. China}\\
}
\date{}

\begin{document}
\maketitle \baselineskip=17pt
\begin{center}
\begin{quote}
{{\bf Abstract:}
In a dynamic environment, knowledge reduction of information systems with variations of object sets, attribute sets and attribute values is an important topic of rough set theory, and related family-based attribute reduction of dynamic covering information systems when refining and coarsening coverings has attracted little attention. In this paper, firstly, we
introduce the concepts of the refinement and coarsening of a covering and
provide the mechanisms of updating related families of dynamic covering decision information systems with refining and coarsening coverings. Meanwhile, we investigate how to construct attribute reducts with the updated related families and propose the incremental algorithms for computing
attribute reducts of dynamic covering decision information systems. Finally, the experimental results verify that the proposed algorithms are more effective than the non-incremental algorithms for attribute reduction of dynamic covering decision information systems in terms of stability and computational time.

\\}
\end{quote}
\end{center}
\renewcommand{\thesection}{\arabic{section}}

\section{Introduction}

Covering rough set theory\cite{Zakowski}, introduced by Zakowski as an extension of Pawlak rough set theory\cite{Pawlak1}, is an important tool for knowledge discovery of information systems with incomplete, inconsistent and insufficient information. With more than 30 years of development, covering rough set theory has been combined with topology theory, fuzzy set theory, matrix theory, lattice theory, graph theory and so on. Especially, it has been successfully applied to many fields such as machine learning, feature selection, pattern recognition and image processing.

Many researchers\cite{Chen4,Chen5,Deer1,Deer2,Deer3,Deer4,Ge,Huang1,Li5,Lang1,Lang2,Lang3,Lang4,
Liu1,Restrepo1,Restrepo2,Shakiba,Tan2,Wang4,Wang5,Wu2,Wu3,Yang1,
Yang2,Yang4,Yao2,Yao3,Zhang3,Zhu1} have designed many types of covering rough set models with different criteria and derived significant results of knowledge reduction of covering information systems. For example, Chen et al.\cite{Chen4} translated the problem of attribute reduction of covering decision information systems into a graph model and proved that attribute reduction of a covering decision information system is equivalent to finding the minimal vertex cover of a derivative hypergraph. Lang et al.\cite{Lang1} provided incremental algorithms to compute the second and sixth lower and upper approximations of sets in dynamic covering approximation spaces and performd knowledge reduction of dynamic covering decision information systems with the incremental approaches.
Shakiba et al.\cite{Shakiba} investigated whether consistent mappings can be used as homomorphism mappings between a covering approximation space and its image with respect to twenty-two pairs of covering upper and lower approximation operators.
Tan et al.\cite{Tan2} proposed fast approaches to knowledge acquisition in covering information systems by employing novel matrix operations and employed the experimental results to illustrate that the new algorithms can dramatically reduce the time consumptions for computing reducts of a covering information systems. Wang et al.\cite{Wang4} provided a new method for constructing  simpler discernibility matrix with covering based rough sets and studied attribute reduction of decision information systems based on a different strategy of identifying objects.

Knowledge reduction of dynamic information systems\cite{Lang1,Lang2,Lang3,Lang4,Li2,Yang5,Li1,Li4,Liu2,Cai,Jing1,Jing2,Jing3,Jing4,
Li3,Luo1,Qian1,Sang,Yang6,Zhang2,Huang2,Huang3,Das,Hu1,Hu2,Hu3,Hu4,Li6,Liang1,Luo2,Luo3,
Wang1,Xie1,
Xu1,Yang3,Yang7,Yu,Wei} has attracted more attention. Especially, researchers have focused on attribute reduction of dynamic information systems with variation of attribute values. For example,
Cai et al.\cite{Cai} introduced the incremental approaches to computing the type-1 and type-2 characteristic matrices for constructing the second and sixth lower and upper approximations of sets in dynamic covering approximation spaces caused by revising attribute attributes. Hu et al.\cite{Hu4} presented the dynamic mechanisms for updating approximations in multigranulation rough sets while refining or coarsening attribute values and designed the corresponding dynamic algorithms for updating multigranulation approximations.
Jing et al.\cite{Jing4} developed a group incremental reduction algorithm with varying data values and employed the experimental results to validate that the proposed incremental algorithms are effective to update the reduction with the variation of attribute values. Li et al.\cite{Li4} proposed the incremental approach to maintaining approximations of dominance-based rough sets approach when attribute values vary over time. Luo et al.\cite{Luo1} presented the updating properties for dynamic maintenance of approximations when the criteria values in the set-valued decision system evolve with time and provided two incremental algorithms for computing rough approximations corresponding to the addition and removal of criteria values. Qian et al.\cite{Qian1} addressed the attribute reduction problem for sequential three-way decisions under dynamic granulation and discussed the relationships of the different attribute reducts, the probabilistic positive regions and the probabilistic positive rules for decision-theoretic rough set models under global view, local view and sequential three-way decisions. Wei et al.\cite{Wei}
proposed an incremental attribute reduction algorithm based on the discernibility matrix of a compact decision table, and the theoretical analyses and experimental results indicate that the proposed algorithm requires much less time to find reducts.
Xie et al.\cite{Xie1} proved that attribute reduction based on the inconsistency degree is equivalent to that based on the positive region and provided three update strategies of inconsistency degree for dynamic incomplete decision systems.

Actually, many scholars\cite{Chen6,Skowron,Feng,Lang4,Konecny,Yao1,Yang5} have focused on discernibility matrix methods for knowledge reduction of information systems. For example, Chen et al.\cite{Chen6}
introduced two Boolean row vectors to characterize the discernibility matrix and reduct in variable precision rough sets and employed an incremental manner to update minimal elements in the discernibility matrix at the arrival of an incremental sample.
Feng et al.\cite{Feng} provided the notion of soft discernibility matrix in soft sets and proposed a novel algorithm based on soft discernibility matrix to solve the problems of decision making.
Yao et al.\cite{Yao1} put forward the elementary matrix simplification
operations and transformed a
discernibility matrix into one of its minimum forms for attribute reduction of information systems.
Although discernibility matrices based methods are very effective and efficient for computing attribute reducts of information systems, we observe that they are not applicable for knowledge reduction of covering decision information systems with respect to the third type of covering-based approximation operators. To tackle this problem, Yang et al.\cite{Yang4} provided the related families based attribute reduction approaches for covering decision information systems, and avoided the above disadvantages of discernibility matrices to some degree.
In a dynamic environment, there are many dynamic covering decision information systems with refining and coarsening coverings, which makes the non-incremental approaches extremely inefficient for knowledge reduction of dynamic covering decision information systems. Specially, we have not seen investigations on the related families based attribute reduction of dynamic covering decision information systems when refining and coarsening coverings, and the non-incremental algorithms are very time-consuming for attribute reduction of this type of information systems, it motivates us to develop more effective approaches for feature selection of dynamic covering decision information systems.

The purpose of this work is to investigate knowledge reduction of dynamic covering decision information systems. First,
we provide the related families based incremental learning methods for attribute reduction of dynamic covering decision information systems when refining coverings. Concretely, we
introduce the concepts of the refinement and coarsening of a covering and study the relationship between related sets of covering decision information systems and those of dynamic covering decision information systems with refining coverings. We show how to compute attribute reducts of dynamic covering decision information systems with the updated related families. Second, we propose the related families based incremental approaches for attribute reduction of dynamic covering decision information systems while coarsening coverings. Concretely, we study the relationship between related sets of covering decision information systems and those of dynamic covering decision information systems. We propose the incremental algorithms for attribute reduction of dynamic covering decision information systems with coarsening coverings.
Finally, we employ the experimental results on data sets\cite{Frank1} downloaded from UCI Machine Learning Repository to illustrate that the proposed algorithms are effective for knowledge reduction of dynamic covering decision information systems with refining and coarsening coverings.

This paper is structured as follows: In Section 2, we briefly
review some concepts of covering-based rough set theory. In
Section 3, we provide the related families based incremental methods for attribute reduction of dynamic covering decision information systems when refining coverings. In Section 4, we develop the related families based incremental approaches for attribute reduction of dynamic covering decision information systems when coarsening coverings. In Section 5, the experimental results illustrate that the proposed algorithms have better performance than non-incremental algorithms for dynamic covering decision information systems.
All conclusions and further research are drawn in Section 6.

\section{Preliminaries}

In this section, we briefly review some concepts of covering information system.

\begin{definition}\cite{Pawlak1}
Let $S=(U,A,V,f)$ be an information system, where $U = \{x_{1},x_{2},...,x_{n}\}$ is a non-empty universe, $A = \{a_{1},a_{2},...,a_{m}\}$ is a non-empty attribute set, $V = \bigcup_{a \in A}V_{a}$ is the set of attribute values, where $V_{a}$ is the domain of attribute $a$, $f:\ U\times A \rightarrow V$ is an information function such that $f(x,a) \in V_{a}$ for any $ a\in A$ and $x\in U.$
\end{definition}

An information system, where objects are measured by using a finite
number of attributes, represents all available information and
knowledge. Additionally, if the function $f$ is total, then the
information system is called complete. Otherwise, the system is
incomplete. Especially, we denote  $[x]_{A}= \{y \in U \mid f(x,a) = f(y,a),\forall a\in A\}$ as
the equivalence class of $x$ with respect to $A$.

\begin{definition}\cite{Hu4}
Let $S=(U,A,V,f)$ be an information system, and $[x]_{a} = \{y \in U \mid f(x,a) = f(y,a)\}$, where $x,y\in U$, and $a\in A$. If we have $f(y,a)=v\notin V_{a}$ for some $y\in [x]_{a}$, then we say that $f(y,a)$ is refined to $v$.
\end{definition}

For simplicity, $[x]_{a}$ denotes the equivalence class of $x$ with respect to $a\in A$, and $[x]^{+}_{a}$ means the equivalence class of $x$ with respect to $a$ after refining attribute values. Furthermore, $[x]_{A}$ denotes the equivalence class of $x$ with respect to $A$, and $[x]^{+}_{A}$ means the equivalence class of $x$ with respect to $A$ after refining attribute values. Especially, we denote $U/A=\{[x]_{A}|x\in U\}$ as a partition of the universe $U$ with respect to $A$.

\begin{definition}\cite{Hu4}
Let $S=(U,A,V,f)$ be an information system, and $f(x,a) \neq f(y,a)$, where $x,y\in U$ and $ a\in A$, $[x]_{a} = \{z \in U \mid f(z,a) = f(x,a)\}$. If we have $f(z,a) = f(y,a)$ for any $z\in [x]_{a},$ then $f(x,a)$ is coarsened to $f(y,a)$.
\end{definition}

For simplicity, $[x]_{a}$ is the equivalence class of $x$ with respect to $a\in A$, and $[x]^{-}_{a}$ denotes the equivalence class of $x$  with respect to $a$ after coarsening attribute values. Furthermore, $[x]_{A}$ is the equivalence class of $x$ with respect to $A$, and $[x]^{-}_{A}$ means the equivalence class of $x$ with respect to $A$ after coarsening attribute values.

The condition of Pawlak rough set model is so strict that limits its applications in practical situations, and the concept of partition of the universe is generalized to the concept of covering as follows: if $\mathscr{C}$ is a family of non-empty subsets of $U$ and $\bigcup\{C\mid C\in \mathscr{C}\}=U$, then $\mathscr{C}$ is called a covering of the universe $U$.

\begin{definition}\cite{Zhu1}
Let $(U,\mathscr{C})$ be a covering approximation space, where $U$ is a non-empty finite universe of discourse, $\mathscr{C}$ is a covering of $U$, and $Md_{\mathscr{C}}(x)=\{K\in \mathscr{C}\mid x\in K\wedge (\forall S\in \mathscr{C}\wedge x\in S \wedge S\subseteq K\Rightarrow K=S)\}$ the minimal description of $x\in U$. Then the third lower and upper approximations of $X\subseteq U$ with respect to $\mathscr{C}$ are defined as follows:
\begin{eqnarray*}
CL_{\mathscr{C}}(X)&=&\cup\{K\in \mathscr{C}\mid K\subseteq X\};\\
CH_{\mathscr{C}}(X)&=&\cup\{K\in Md_{\mathscr{C}}(x)\mid x\in X\}.
\end{eqnarray*}
\end{definition}

If $U$ is a non-empty finite universe of discourse, and $\Delta = \{\mathscr{C}_{1},\mathscr{C}_{2},...,\mathscr{C}_{m}\}$ is a family of coverings of $U$, then $(U,\Delta)$ is called a covering information system. Especially, if $\mathscr{D}=\{D_{1},D_{2},...,D_{k}\}$ is a partition based on decision attributes, then $(U,\Delta,\mathscr{D})$ is called a covering decision information system. For convenience, we denote
$POS_{\cup\Delta}(X)=CL_{\cup\Delta}(X), BND_{\cup\Delta}(X)=CH_{\cup\Delta}(X)\backslash CL_{\cup\Delta}(X)$ and $ NEG_{\cup\Delta}(X)=U\backslash CH_{\cup\Delta}(X).$

\begin{definition}\cite{Yang4}
Let $(U,\Delta, \mathscr{D})$ be a covering decision information system, where $U = \{x_{1},x_{2},...,x_{n}\}$, $\Delta = \{\mathscr{C}_{1},\mathscr{C}_{2},...,\mathscr{C}_{m}\}$, and $\mathscr{D}=\{D_{1},D_{2},...,D_{k}\}$.

$(1)$ If there exists $K\in Md_{\cup\Delta}(y)$ and $D_{j}\in \mathscr{D}$ such that $x\in K\subseteq D_{j}$ for any $x\in U$, then $(U,\Delta,\mathscr{D})$ is called a consistent covering decision information system.

$(2)$ If there exists $x\in U$ but $\overline{\exists} K\in \cup\Delta \text{ and } D_{j}\in \mathscr{D}$ such that $x\in K\subseteq D_{j}$, then $(U,\Delta,\mathscr{D})$ is called an inconsistent covering decision information system.
\end{definition}

According to Definition 2.5, all covering decision information systems are classified into two categories as follows: consistent covering decision information systems and inconsistent covering decision information systems.

\begin{example}
$(1)$ Let $(U,\Delta,\mathscr{D})$ be a consistent covering decision information system, where $U=\{x_{1},x_{2},$ $...,x_{8}\}$, $\Delta=\{\mathscr{C}_{1},
\mathscr{C}_{2},\mathscr{C}_{3},\mathscr{C}_{4},\mathscr{C}_{5}\}$, $\mathscr{D}=\{\{x_{1},x_{2}\},\{x_{3},x_{4},x_{5}\},\{x_{6},x_{7},x_{8}\}\}$, where
$\mathscr{C}_{1}=\{\{x_{1}\},\{x_{1},x_{2}\},\{x_{3},x_{5}\},$ $\{x_{3},x_{4},x_{5}\},\{x_{6}\},\{x_{5},x_{7},x_{8}\}\},
\mathscr{C}_{2}=\{\{x_{1},x_{2},x_{3}\},\{x_{3},x_{4}\},\{x_{4},x_{5},x_{6}\},\{x_{5}\},$ $\{x_{6},x_{7},x_{8}\}\},
\mathscr{C}_{3}=\{\{x_{1},x_{2},x_{4}\},$ $\{x_{3},x_{4},x_{5}\},\{x_{3}\},\{x_{4},x_{5},x_{6}\},\{x_{6},x_{7},x_{8}\}\},
\mathscr{C}_{4}=\{\{x_{1},x_{2}\},\{x_{2},x_{3},x_{4}\},\{x_{3}\},\{x_{4},$ $x_{5}\},\{x_{6}\},\{x_{5},x_{7},x_{8}\}\},$ and $
\mathscr{C}_{5}$ $=\{\{x_{1},x_{2}\},\{x_{2},x_{3},x_{5}\},\{x_{4},x_{5},x_{7}\},\{x_{6}\},\{x_{3},x_{7},x_{8}\}\}.
$

$(2)$ Let $(U,\Delta,\mathscr{D})$ be an inconsistent covering decision information system, where
$U=\{x_{1},x_{2},$ $...,x_{8}\}$, $\Delta=\{\mathscr{C}_{1},
\mathscr{C}_{2},\mathscr{C}_{3},\mathscr{C}_{4},\mathscr{C}_{5}\}$, $\mathscr{D}=\{\{x_{1},x_{2}\},\{x_{3},x_{4},x_{5}\},\{x_{6},x_{7},x_{8}\}\}$, where
$\mathscr{C}_{1}=\{\{x_{1},x_{3}\},\{x_{2},x_{4},x_{5}\},\{x_{4},x_{5},x_{6}\},$ $\{x_{6},x_{7},x_{8}\}\},
\mathscr{C}_{2}=\{\{x_{1}\},\{x_{2},x_{3},x_{4},x_{6}\},\{x_{4},x_{5}\},\{x_{5},x_{6},x_{7},x_{8}\}\},
\mathscr{C}_{3}$ $=\{\{x_{1},x_{2},x_{3},x_{7}\},\{x_{4},x_{5}\},\{x_{4},x_{5},x_{6},x_{7}\},$ $\{x_{7},x_{8}\}\},
\mathscr{C}_{4}=\{\{x_{1}\},\{x_{2},x_{3},x_{5},x_{6}\},\{x_{4},x_{5}\},\{x_{4},x_{6},x_{7},x_{8}\}\}, $ and $
\mathscr{C}_{5}$ $= \{\{x_{1},x_{2},x_{3},x_{5}\},\{x_{2},x_{4},x_{5},x_{6}\},\{x_{6}\},$ $\{x_{7},x_{8}\}\}.
$
\end{example}

\begin{definition}\cite{Yang4}
Let $(U,\Delta, \mathscr{D})$ be a covering decision information system, where $U = \{x_{1},x_{2},...,x_{n}\}$, $\Delta = \{\mathscr{C}_{1},\mathscr{C}_{2},...,\mathscr{C}_{m}\}$, and $\mathscr{D}=\{D_{1},D_{2},...,D_{k}\}$.

$(1)$ If $POS_{\cup\Delta}(\mathscr{D})=POS_{\cup\Delta-\{\mathscr{C}_{i}\}}(\mathscr{D})$ for any $\mathscr{C}_{i}\in \Delta$, where $POS_{\cup\Delta}(\mathscr{D})=\bigcup \{POS_{\cup\Delta}(D_{i})$ $\mid D_{i}\in \mathscr{D}\}$, then $\mathscr{C}_{i}$ is called superfluous relative to $\mathscr{D}$; otherwise, $\mathscr{C}_{i}$ is called indispensable relative to $\mathscr{D}$;

$(2)$ If every element of $P\subseteq \Delta$ is indispensable relative to $\mathscr{D}$ and $POS_{\cup P}(\mathscr{D}) = POS_{\cup\Delta}(\mathscr{D})$, then $P$ is called a reduct of $\Delta$ relative to $\mathscr{D}$.
\end{definition}

Suppose $(U,\Delta, \mathscr{D})$ is a covering decision information system, where $U=\{x_{1},x_{2},...,x_{n}\}$, $\Delta=\{\mathscr{C}_{1},\mathscr{C}_{2},...,$ $\mathscr{C}_{m}\}$,
$\mathscr{A}_{\Delta}=\{C_{k}\in \cup \Delta\mid \exists D_{j}\in \mathscr{D}, $ s.t. $ C_{k}\subseteq D_{j}\}$, $r(x)=\{\mathscr{C}\in \Delta\mid \exists C_{k}\in \mathscr{A}_{\Delta}, $ s.t. $  x\in C_{k}\in \mathscr{C}\}$, and the related family $R(U,\Delta,\mathscr{D})=\{r(x)\mid x\in POS_{\cup\Delta}(\mathscr{D})\}$.

\begin{definition}\cite{Yang4}
Let $(U,\Delta, \mathscr{D})$ be a covering decision information system, where $U=\{x_{1},x_{2},...,x_{n}\}$, $\Delta=\{\mathscr{C}_{1},\mathscr{C}_{2},...,$ $\mathscr{C}_{m}\}$, $\mathscr{D}=\{D_{1},D_{2},...,D_{k}\}$, and
$R(U,\Delta,\mathscr{D})$ the related family of $(U,\Delta, \mathscr{D})$. Then

$(1)$ $f(U,\Delta,\mathscr{D})=\bigwedge\{\bigvee r(x)\mid r(x)\in R(U,\Delta,\mathscr{D})\}$ is the related function, where $\bigvee r(x)$ is the disjunction of all elements in $r(x)$;

$(2)$ $g(U,\Delta,\mathscr{D})=\bigvee^{l}_{i=1}\{\bigwedge \Delta_{i}\mid \Delta_{i}\subseteq\Delta\}$ is the reduced disjunctive form of $f(U,\Delta,\mathscr{D})$ with the multiplication and absorption laws.
\end{definition}

According to Definition 2.8, we have the attribute reduct set $\mathscr{R}(\Delta,U,D)=\{\Delta_{1},\Delta_{1},...,\Delta_{l}\}$ for the covering decision information system $(U,\Delta, \mathscr{D})$, which is similar to construct attribute reducts of information systems using discernibility matrices.

\begin{algorithm}\cite{Yang4}
Let $(U,\Delta, \mathscr{D})$ be a covering decision information system, where $U=\{x_{1},x_{2},...,x_{n}\}$, $\Delta=\{\mathscr{C}_{1},\mathscr{C}_{2},...,\mathscr{C}_{m}\}$, and $\mathscr{D}=\{D_{1},D_{2},...,D_{k}\}$. Then

Step 1: Input $(U,\Delta, \mathscr{D})$;

Step 2: Construct $POS_{\cup\Delta}(\mathscr{D})=\bigcup \{POS_{\cup\Delta}(D_{i})$ $\mid D_{i}\in \mathscr{D}\}$;

Step 3: Compute $R(U,\Delta,\mathscr{D})=\{r(x)\mid x\in POS_{\cup\Delta}(\mathscr{D})\}$;

Step 4: Construct $f(U,\Delta,\mathscr{D})=\bigwedge\{\bigvee r(x)\mid r(x)\in R(U,\Delta,\mathscr{D})\}=\bigvee^{l}_{i=1}\{\bigwedge \Delta_{i}\mid \Delta_{i}\subseteq\Delta\}$;

Step 5: Output $\mathscr{R}(U,\Delta,\mathscr{D})$.
\end{algorithm}


We employ the following example to illustrate how to compute attribute reducts of consistent covering decision information systems and inconsistent covering decision information systems.

\begin{example}(Continuation from Example 2.6)
$(1)$ Firstly, by Definition 2.8, we have
$r(x_{1})=\{\mathscr{C}_{1},\mathscr{C}_{4},\mathscr{C}_{5}\},$ $
r(x_{2})$ $=\{\mathscr{C}_{1},\mathscr{C}_{4},\mathscr{C}_{5}\},
r(x_{3})=\{\mathscr{C}_{1},\mathscr{C}_{2},\mathscr{C}_{3},\mathscr{C}_{4}\},$ $
r(x_{4})=\{\mathscr{C}_{1},\mathscr{C}_{2},\mathscr{C}_{3},\mathscr{C}_{4}\},
r(x_{5})=\{\mathscr{C}_{1},\mathscr{C}_{2},\mathscr{C}_{3},\mathscr{C}_{4}\},
r(x_{6})$ $=\{\mathscr{C}_{1},\mathscr{C}_{2},\mathscr{C}_{3},\mathscr{C}_{4},\mathscr{C}_{5}\},
r(x_{7})=\{\mathscr{C}_{2},\mathscr{C}_{3}\},$ and $
r(x_{8})=\{\mathscr{C}_{2},\mathscr{C}_{3}\}$.
It follows that $R(U,\Delta,\mathscr{D})=\{\{\mathscr{C}_{1},\mathscr{C}_{4},\mathscr{C}_{5}\},$ $
\{\mathscr{C}_{1},$ $\mathscr{C}_{2},\mathscr{C}_{3},\mathscr{C}_{4}\},
\{\mathscr{C}_{1},\mathscr{C}_{2},\mathscr{C}_{3},\mathscr{C}_{4},$ $\mathscr{C}_{5}\},
\{\mathscr{C}_{2},\mathscr{C}_{3}\}\}.$
Secondly, we get
\begin{eqnarray*}
f(U,\Delta,\mathscr{D})
&=&(\mathscr{C}_{1}\vee\mathscr{C}_{4}\vee\mathscr{C}_{5})\wedge(\mathscr{C}_{1}\vee\mathscr{C}_{2}\vee\mathscr{C}_{3}\vee\mathscr{C}_{4})\wedge
(\mathscr{C}_{1}\vee\mathscr{C}_{2}\vee\mathscr{C}_{3}\vee\mathscr{C}_{4}\vee\mathscr{C}_{5})
\wedge(\mathscr{C}_{2}\vee\mathscr{C}_{3})\\
&=&(\mathscr{C}_{1}\vee\mathscr{C}_{4}\vee\mathscr{C}_{5})\wedge (\mathscr{C}_{2}\vee\mathscr{C}_{3})\\
&=&(\mathscr{C}_{1}\wedge\mathscr{C}_{2})\vee (\mathscr{C}_{1}\wedge\mathscr{C}_{3})\vee (\mathscr{C}_{2}\wedge\mathscr{C}_{4})\vee(\mathscr{C}_{2}\wedge\mathscr{C}_{5})\vee
(\mathscr{C}_{3}\wedge\mathscr{C}_{4})\vee(\mathscr{C}_{3}\wedge\mathscr{C}_{5}).
\end{eqnarray*}

Therefore, we have $\mathscr{R}(U,\Delta,\mathscr{D})=\{\{\mathscr{C}_{1},\mathscr{C}_{2}\}, \{\mathscr{C}_{1},\mathscr{C}_{3}\}, \{\mathscr{C}_{2},\mathscr{C}_{4}\},\{\mathscr{C}_{2},\mathscr{C}_{5}\},
\{\mathscr{C}_{3},\mathscr{C}_{4}\},\{\mathscr{C}_{3},\mathscr{C}_{5}\}\}.$

$(2)$ Firstly, by Definition 2.8, we get
$r(x_{1})=\{\mathscr{C}_{2},\mathscr{C}_{4}\},
r(x_{2})$ $=\emptyset,
r(x_{3})=\emptyset,
r(x_{4})=\{\mathscr{C}_{2},\mathscr{C}_{3},\mathscr{C}_{4}\},
r(x_{5})=\{\mathscr{C}_{2},\mathscr{C}_{3},\mathscr{C}_{4}\},
r(x_{6})=\{\mathscr{C}_{1},\mathscr{C}_{5}\},
r(x_{7})=\{\mathscr{C}_{1},\mathscr{C}_{3},\mathscr{C}_{5}\},$ and $
r(x_{8})=\{\mathscr{C}_{1},\mathscr{C}_{3},\mathscr{C}_{5}\}.$
It implies that $R(U,\Delta,\mathscr{D})=\{\{\mathscr{C}_{2},\mathscr{C}_{4}\},\{\mathscr{C}_{2},\mathscr{C}_{3},\mathscr{C}_{4}\},
\{\mathscr{C}_{1},\mathscr{C}_{5}\},$ $\{\mathscr{C}_{1},\mathscr{C}_{3},$ $\mathscr{C}_{5}\}\}.$
Secondly, according to Definition 2.8, we obtain
\begin{eqnarray*}
f(U,\Delta,\mathscr{D})
&=&(\mathscr{C}_{2}\vee\mathscr{C}_{4})\wedge(\mathscr{C}_{2}\vee\mathscr{C}_{3}\vee\mathscr{C}_{4})\wedge
(\mathscr{C}_{1}\vee\mathscr{C}_{5})\wedge(\mathscr{C}_{1}\vee\mathscr{C}_{3}\vee\mathscr{C}_{5})\\
&=&(\mathscr{C}_{2}\vee\mathscr{C}_{4})\wedge (\mathscr{C}_{1}\vee\mathscr{C}_{5})\\
&=&(\mathscr{C}_{1}\wedge\mathscr{C}_{2})\vee (\mathscr{C}_{1}\wedge\mathscr{C}_{4})\vee
(\mathscr{C}_{2}\wedge\mathscr{C}_{5})\vee (\mathscr{C}_{4}\wedge\mathscr{C}_{5}).
\end{eqnarray*}

Therefore, we have $\mathscr{R}(U,\Delta,\mathscr{D})=\{\{\mathscr{C}_{1},\mathscr{C}_{2}\}, \{\mathscr{C}_{1},\mathscr{C}_{4}\}, \{\mathscr{C}_{2},\mathscr{C}_{5}\}, \{\mathscr{C}_{4},\mathscr{C}_{5}\}\}.$
\end{example}

Suppose $(U,\Delta, \mathscr{D})$ is a covering decision information system, we denote $SR(U,\Delta,\mathscr{D})=\{r(x)\in R(U,\Delta,\mathscr{D})\mid x\in POS_{\cup\Delta}(\mathscr{D}), (\forall y\in POS_{\cup\Delta}(\mathscr{D}), r(y)\nsubseteq r(x)\text{ and } r(y)\in R(U,\Delta,\mathscr{D}))\}$, and $\|\mathscr{C}\|$ denotes the number of times for a covering $\mathscr{C}$ appeared in $SR(U,\Delta,\mathscr{D})$.

\begin{algorithm}\cite{Yang4}(Heuristic Algorithm of Computing a Reduct of $(U,\Delta,\mathscr{D})$)(NIHV).

Step 1: Input $(U,\Delta, \mathscr{D})$;

Step 2: Construct $POS_{\cup\Delta}(\mathscr{D})=\bigcup \{POS_{\cup\Delta}(D_{i})$ $\mid D_{i}\in \mathscr{D}\}$;

Step 3: Compute $R(U,\Delta,\mathscr{D})=\{r(x)\mid x\in POS_{\cup\Delta}(\mathscr{D})\}$;

Step 4: Construct a reduct $\bigtriangleup^{\ast}=\{\mathscr{C}_{i_{1}},\mathscr{C}_{i_{2}},...,\mathscr{C}_{i_{j}}\}$, where
$SR_{1}(U,\Delta,\mathscr{D})=SR(U,\Delta,\mathscr{D})$, $\|\mathscr{C}_{i_{1}}\|=max\{\|\mathscr{C}_{i}\|\mid \mathscr{C}_{i}\in r(x)\in SR_{1}(U,\Delta,\mathscr{D})\}$;
$SR_{2}(U,\Delta,\mathscr{D})=\{r(x)\in SR(U,\Delta,\mathscr{D})\mid \mathscr{C}_{i_{1}}\notin r(x)\}$, $\|\mathscr{C}_{i_{2}}\|=max\{\|\mathscr{C}_{i}\|\mid \mathscr{C}_{i}\in r(x)\in SR_{2}(U,\Delta,\mathscr{D})\}$;
$SR_{3}(U,\Delta,\mathscr{D})=\{r(x)\in SR(U,\Delta,\mathscr{D})\mid \mathscr{C}_{i_{1}}\notin r(x)\text{ or } \mathscr{C}_{i_{2}}\notin r(x)\}$, $\|\mathscr{C}_{i_{3}}\|=max\{\|\mathscr{C}_{i}\|\mid \mathscr{C}_{i}\in r(x)\in SR_{3}(U,\Delta,\mathscr{D})\}$;
...;
$SR_{j}(U,\Delta,\mathscr{D})=\{r(x)\in SR(U,\Delta,\mathscr{D})\mid \mathscr{C}_{i_{1}}\notin r(x)\text{ or } \mathscr{C}_{i_{2}}\notin r(x)\text{ or }... \text{ or }\mathscr{C}_{i_{j-1}}\notin r(x)\}$, $\|\mathscr{C}_{i_{j}}\|=max\{\|\mathscr{C}_{i}\|\mid \mathscr{C}_{i}\in r(x)\in SR_{j}(U,\Delta,\mathscr{D})\}$, and $SR(U,\Delta,\mathscr{D})=\{r(x)\mid \exists \mathscr{C}_{i_{k}}\in r(x),1\leq k\leq j\}$;

Step 5: Output the reduct $\bigtriangleup^{\ast}$.
\end{algorithm}

We observe that constructing all attribute reducts of covering decision information systems by Algorithm 2.9 is a NP hard problem, and it is enough to compute a reduct for covering decision information systems by Algorithm 2.11. Furthermore, if there exist two coverings $\mathscr{C}_{i}$  and $\mathscr{C}_{j}$ such that $\|\mathscr{C}_{i}\|=\|\mathscr{C}_{j}\|=max\{\|\mathscr{C}_{i}\|\mid \mathscr{C}_{i}\in r(x)\in SR_{k}(U,\Delta,\mathscr{D})\}$, then we select $\|\mathscr{C}_{i}\|=max\{\|\mathscr{C}_{i}\|\mid \mathscr{C}_{i}\in r(x)\in SR_{k}(U,\Delta,\mathscr{D})\}$ or $\|\mathscr{C}_{j}\|=max\{\|\mathscr{C}_{i}\|\mid \mathscr{C}_{i}\in r(x)\in SR_{k}(U,\Delta,\mathscr{D})\}$.

\begin{example}(Continuation from Example 2.10)
$(1)$ In Example 2.10(1), we derive $SR(U,\Delta,\mathscr{D})=\{\{\mathscr{C}_{1},\mathscr{C}_{4},$ $\mathscr{C}_{5}\},
$ $\{\mathscr{C}_{2},\mathscr{C}_{3}\}\}.$
By Algorithm 2.11, firstly, we obtain $SR_{1}(U,\Delta,\mathscr{D})=\{\{\mathscr{C}_{1},\mathscr{C}_{4},\mathscr{C}_{5}\},
\{\mathscr{C}_{2},\mathscr{C}_{3}\}\}$ and $\|\mathscr{C}_{1}\|=max\{\|\mathscr{C}_{i}\|\mid \mathscr{C}_{i}\in r(x)\in SR_{1}(U,\Delta,\mathscr{D})\}$. Secondly, we get $SR_{2}(U,\Delta,\mathscr{D})=\{\{\mathscr{C}_{2},\mathscr{C}_{3}\}\},$ and
$\|\mathscr{C}_{2}\|=max\{\|\mathscr{C}_{i}\|\mid \mathscr{C}_{i}\in r(x)\in SR_{2}(U,\Delta,\mathscr{D})\}$. Finally, we have a reduct $\bigtriangleup^{\ast}=\{\mathscr{C}_{1},\mathscr{C}_{2}\}$ of  $(U,\Delta,\mathscr{D})$.

$(2)$ In Example 2.10(2), we derive $SR(U,\Delta,\mathscr{D})=\{
\{\mathscr{C}_{1},\mathscr{C}_{5}\},\{\mathscr{C}_{2},\mathscr{C}_{4}\}\}.$
By Algorithm 2.11, firstly, we obtain $SR_{1}(U,\Delta,\mathscr{D})=\{
\{\mathscr{C}_{1},\mathscr{C}_{5}\},\{\mathscr{C}_{2},\mathscr{C}_{4}\}\}$ and $\|\mathscr{C}_{1}\|=max\{\|\mathscr{C}_{i}\|\mid \mathscr{C}_{i}\in r(x)\in SR_{1}(U,\Delta,\mathscr{D})\}$. Secondly, we get $SR_{2}(U,\Delta,\mathscr{D})=\{\{\mathscr{C}_{2},\mathscr{C}_{4}\}\},$ and
$\|\mathscr{C}_{2}\|=max\{\|\mathscr{C}_{i}\|\mid \mathscr{C}_{i}\in r(x)\in SR_{2}(U,\Delta,\mathscr{D})\}$. Finally, we have a reduct $\bigtriangleup^{\ast}=\{\mathscr{C}_{1},\mathscr{C}_{2}\}$ of $(U,\Delta,\mathscr{D})$.
\end{example}

\section{Updating attribute reducts of dynamic covering decision information systems with refining coverings}

In this section, we investigate how to update attribute reducts of dynamic covering decision information systems with refining coverings.

\begin{definition}
Let $\mathscr{C}_{1}$ and $\mathscr{C}_{2}$ be coverings of the universe $U$,
where $U=\{x_{1},x_{2},...,x_{n}\}$, $\mathscr{C}_{1}=\{C_{11},C_{12},...,$ $C_{1m_{1}}\}$, and $\mathscr{C}_{2}=\{C_{21},C_{22},...,C_{2m_{2}}\}$. For any $C_{2i}\in \mathscr{C}_{2}$,
if there exists $C_{1j}\in \mathscr{C}_{1}$ such that $C_{2i}\subseteq C_{1j}$, then $\mathscr{C}_{2}$ is called a refinement of $\mathscr{C}_{1}$. Otherwise, $\mathscr{C}_{1}$ is called a coarsening of $\mathscr{C}_{2}$.
\end{definition}

For convenience, we refer $\mathscr{C}^{+}$ and $\mathscr{C}^{-}$ to as the refinement and coarsening of $\mathscr{C}$, respectively. Especially, the refinement and coarsening of the covering given by Definition 3.1 are generalizations of concepts given by Definitions 2.2 and 2.3.

\begin{definition}
Let $(U,\Delta, \mathscr{D})$ and $(U,\Delta^{+}, \mathscr{D})$ be covering decision information systems,
where $U=\{x_{1},x_{2},$ $...,x_{n}\}$, $\Delta=\{\mathscr{C}_{1},\mathscr{C}_{2},...,\mathscr{C}_{m-1},\mathscr{C}_{m}\}$,  $\Delta^{+}=\{\mathscr{C}_{1},\mathscr{C}_{2},...,\mathscr{C}_{m-1},\mathscr{C}^{+}_{m}\}$, and
$\mathscr{D}=\{D_{1},D_{2},...,D_{k}\}$. Then $(U,\Delta^{+}, \mathscr{D})$ is called a dynamic covering decision information system of $(U,\Delta, \mathscr{D})$.
\end{definition}

In practical situations, there are many types of dynamic covering decision information systems with refining coverings. For simplicity, we only consider the dynamic covering decision information system with a refining covering in this section.

\begin{example}(Continuation from Example 2.6)
$(1)$ Let $(U,\Delta,\mathscr{D})$ and $(U,\Delta^{+},\mathscr{D})$ be covering decision information systems, where $U=\{x_{1},x_{2},$ $...,x_{8}\}$, $\Delta=\{\mathscr{C}_{1},\mathscr{C}_{2},\mathscr{C}_{3},\mathscr{C}_{4},\mathscr{C}_{5}\}$, $\Delta^{+}=\{\mathscr{C}_{1},
\mathscr{C}_{2},\mathscr{C}_{3},\mathscr{C}_{4},\mathscr{C}^{+}_{5}\}$, $\mathscr{D}=\{\{x_{1},x_{2}\},\{x_{3},x_{4},x_{5}\},\{x_{6},x_{7},x_{8}\}\}$, $\mathscr{C}_{5}=\{\{x_{1},x_{2}\},\{x_{2},x_{3},x_{5}\},\{x_{4},x_{5},x_{7}\},\{x_{6}\},\{x_{3},x_{7},x_{8}\}\},$
and $\mathscr{C}^{+}_{5}=\{\{x_{1},x_{2}\},\{x_{2},x_{3},x_{5}\},\{x_{4},x_{5},x_{7}\},\{x_{6}\},\{x_{3},x_{7}\},\{x_{8}\}\}$.
Therefore, we see that $(U,\Delta^{+}, \mathscr{D})$ is a dynamic covering decision information system of $(U,\Delta, \mathscr{D})$. Especially, $(U,\Delta^{+}, \mathscr{D})$ is a consistent covering decision information system.

$(2)$ Let $(U,\Delta,\mathscr{D})$ and $(U,\Delta^{+},\mathscr{D})$ be covering decision information systems, where $U=\{x_{1},x_{2},$ $...,x_{8}\}$,
$\Delta = \{\mathscr{C}_{1},\mathscr{C}_{2},\mathscr{C}_{3},\mathscr{C}_{4},\mathscr{C}_{5}\}$,
$\Delta^{+} = \{\mathscr{C}_{1},\mathscr{C}_{2},\mathscr{C}_{3},\mathscr{C}_{4},\mathscr{C}^{+}_{5}\}$,
and $\mathscr{D}=\{\{x_{1},x_{2}\},\{x_{3},x_{4},x_{5}\},\{x_{6},x_{7},x_{8}\}\}$,
$\mathscr{C}_{5}=\{\{x_{1},x_{2},x_{3},x_{5}\},\{x_{2},x_{4},x_{5},x_{6}\},\{x_{6}\},\{x_{7},x_{8}\}\},$
and $\mathscr{C}^{+}_{5}=\{\{x_{1}\},\{x_{2},x_{3},x_{5}\},\{x_{2},x_{4}\},\{x_{5},x_{6}\},\{x_{6}\},\{x_{7},x_{8}\}\}.$
Therefore, we notice that $(U,\Delta^{+}, \mathscr{D})$ is a dynamic covering decision information system of $(U,\Delta, \mathscr{D})$. Especially, $(U,\Delta^{+}, \mathscr{D})$ is an inconsistent covering decision information system.
\end{example}

Suppose $(U,\Delta, \mathscr{D})$ and $(U,\Delta^{+}, \mathscr{D})$ are covering decision information systems, where $U=\{x_{1},x_{2},...,x_{n}\}$, $\Delta=\{\mathscr{C}_{1},\mathscr{C}_{2},...,\mathscr{C}_{m-1},\mathscr{C}_{m}\}$, and $\Delta^{+}=\{\mathscr{C}_{1},\mathscr{C}_{2},...,\mathscr{C}_{m-1},\mathscr{C}^{+}_{m}\}$,
$\mathscr{A}_{\Delta}=\{C\in \cup \Delta\mid \exists D_{j}\in \mathscr{D}, \text{ s.t. } C\subseteq D_{j}\}$, $\mathscr{A}_{\Delta^{+}}=\{C\in \cup \Delta^{+}\mid \exists D_{j}\in \mathscr{D}, \text{ s.t. } C\subseteq D_{j}\}$, $\mathscr{A}_{\mathscr{C}_{m}}=\{C\in \mathscr{C}_{m}\mid \exists D_{j}\in \mathscr{D}, \text{ s.t. } C\subseteq D_{j}\}$,
$\mathscr{A}_{\mathscr{C}^{+}_{m}}=\{C\in \mathscr{C}^{+}_{m}\mid \exists D_{j}\in \mathscr{D}, \text{ s.t. } C\subseteq D_{j}\}$, $r(x)=\{\mathscr{C}\in \Delta\mid \exists C\in \mathscr{A}_{\Delta}, \text{ s.t. } x\in C\in \mathscr{C}\},$ and  $r^{+}(x)=\{\mathscr{C}\in \Delta^{+}\mid \exists C\in \mathscr{A}_{\Delta^{+}}, \text{ s.t. } x\in C\in \mathscr{C}\}.$

\begin{theorem}
Let $(U,\Delta, \mathscr{D})$ and $(U,\Delta^{+}, \mathscr{D})$ be covering decision information systems, where $U=\{x_{1},x_{2},$ $...,x_{n}\}$, $\Delta=\{\mathscr{C}_{1},\mathscr{C}_{2},...,\mathscr{C}_{m-1},\mathscr{C}_{m}\}$, $\Delta^{+}=\{\mathscr{C}_{1},\mathscr{C}_{2},...,\mathscr{C}_{m-1},\mathscr{C}^{+}_{m}\}$,
and $\mathscr{D}=\{D_{1},D_{2},...,D_{k}\}$. Then we have
\makeatother $$r^{+}(x)=\left\{
\begin{array}{ccc}
(r(x)\backslash\{\mathscr{C}_{m}\})\cup \{\mathscr{C}^{+}_{m}\},&{\rm if}& x\in \cup\mathscr{A}_{\mathscr{C}^{+}_{m}};\\\\
r(x),&{\rm }& otherwise.
\end{array}
\right. $$
\end{theorem}

\noindent\textbf{Proof:} According to Definition 2.8, we have $r(x)=\{\mathscr{C}\in \Delta\mid \exists C\in \mathscr{A}_{\Delta}, \text{ s.t. } x\in C\in \mathscr{C}\}$ and $r^{+}(x)=\{\mathscr{C}\in \Delta^{+}\mid \exists C\in \mathscr{A}_{\Delta^{+}}, \text{ s.t. } x\in C\in \mathscr{C}\}$.
Since $\Delta =\{\mathscr{C}_{1},\mathscr{C}_{2},...,\mathscr{C}_{m-1},\mathscr{C}_{m}\}$ and $\Delta^{+}=\{\mathscr{C}_{1},\mathscr{C}_{2},...,\mathscr{C}_{m-1},\mathscr{C}^{+}_{m}\}$, then $r^{+}(x)=(\{\mathscr{C}\in \Delta\mid \exists C\in \mathscr{A}_{\Delta}, \text{ s.t. } x\in C\in \mathscr{C}\}\backslash\{\mathscr{C}_{m}\})\cup \{\mathscr{C}^{+}_{m}\mid \exists C\in \mathscr{A}_{\mathscr{C}^{+}_{m}}, \text{ s.t. } x\in C\in \mathscr{C}^{+}_{m}\}$ for $x\in U$. Especially, we have $\cup\mathscr{A}_{\mathscr{C}_{m}}$ $\subseteq$ $\cup\mathscr{A}_{\mathscr{C}^{+}_{m}}$. It follows that  $r^{+}(x)=(r(x)\backslash\{\mathscr{C}_{m}\})\cup \{\mathscr{C}^{+}_{m}\}$ and $r^{+}(y)=r(y)$ for $x\in \cup\mathscr{A}_{\mathscr{C}^{+}_{m}}$ and $y\notin \cup\mathscr{A}^{+}_{\mathscr{C}_{m}}$, respectively. Therefore, we obtain
\makeatother $$r^{+}(x)=\left\{
\begin{array}{ccc}
(r(x)\backslash\{\mathscr{C}_{m}\})\cup \{\mathscr{C}^{+}_{m}\},&{\rm if}& x\in \cup\mathscr{A}_{\mathscr{C}^{+}_{m}};\\\\
r(x),&{\rm }& otherwise.
\end{array}
\right. \Box$$

Theorem 3.4 provides an approach to updating $r^{+}(x)$ of $(U,\Delta^{+},\mathscr{D})$ based on $r(x)$ of $(U,\Delta,\mathscr{D})$ with refining coverings. Furthermore, there are two special cases as follows: (1) if $\cup\mathscr{A}_{\mathscr{C}^{+}_{m}}=\emptyset$, then we have $r^{+}(x)=r(x)$ for any $x\in U$; (2) if $\cup\mathscr{A}_{\mathscr{C}^{+}_{m}}=U$, then we have $r^{+}(x)=(r(x)\backslash\{\mathscr{C}_{m}\})\cup \{\mathscr{C}^{+}_{m}\}$ for any $x\in U$.

\begin{theorem}
Let $(U,\Delta, \mathscr{D})$ and $(U,\Delta^{+}, \mathscr{D})$ be covering decision information systems, where $U=\{x_{1},x_{2},$ $...,x_{n}\}$, $\Delta=\{\mathscr{C}_{1},\mathscr{C}_{2},...,\mathscr{C}_{m-1},\mathscr{C}_{m}\}$, $\Delta^{+}=\{\mathscr{C}_{1},\mathscr{C}_{2},...,\mathscr{C}_{m-1},\mathscr{C}^{+}_{m}\}$, $\mathscr{D}=\{D_{1},D_{2},...,D_{k}\}$,
$\lozenge{\mathscr{R}}(U,\Delta,\mathscr{D})=\{\Delta_{i}\mid \mathscr{C}_{m}\notin\Delta_{i}\in \mathscr{R}(U,\Delta,\mathscr{D})\},$
$\blacktriangle f(U,\Delta,\mathscr{D})=\{\mathscr{C}^{+}_{m}\}\bigwedge (\bigwedge_{x\in POS_{\cup\Delta}(\mathscr{D})\wedge x\notin \cup\mathscr{A}_{\mathscr{C}^{+}_{m}}}$ $\bigvee r(x))=\bigvee^{l}_{i=1}\{\bigwedge \Delta'_{i}\mid \Delta'_{i}\subseteq\Delta\}$, and $\blacktriangle\mathscr{R}(U,\Delta,\mathscr{D})=\{\Delta'_{j}\mid \overline{\exists}\Delta_{i}\in \mathscr{R}(U,\Delta,\mathscr{D}), \text{ s.t. } \Delta_{i}\subset\Delta'_{j},1\leq j\leq l\}$. If $POS_{\cup\Delta^{+}}(\mathscr{D})=POS_{\cup\Delta}(\mathscr{D})$, then  $\mathscr{R}(U,\Delta^{+},\mathscr{D})=\lozenge{\mathscr{R}}(U,\Delta,\mathscr{D})\cup (\blacktriangle\mathscr{R}(U,\Delta,\mathscr{D}))$.
\end{theorem}

\noindent\textbf{Proof:} According to Definition 3.1, we have $\lozenge{\mathscr{R}}(U,\Delta,\mathscr{D})\subseteq\mathscr{R}(U,\Delta^{+},\mathscr{D})$. Furthermore, taking $\Delta'_{j}\in \blacktriangle\mathscr{R}(U,\Delta,\mathscr{D})$, we have $POS_{\cup\Delta}(\mathscr{D})= POS_{\cup\Delta'_{j}}(\mathscr{D})$ and $POS_{\cup\Delta'_{j}}(\mathscr{D})\neq POS_{\cup\Delta'_{j}-\{\mathscr{C}_{i}\}}(\mathscr{D})$ for any $\mathscr{C}_{i}\in \Delta'_{j}$. According to Definition 2.8, we obtain $\Delta'_{j}\in \mathscr{R}(U,\Delta^{+},\mathscr{D})$. So
$\lozenge{\mathscr{R}}(U,\Delta,\mathscr{D})\cup (\blacktriangle\mathscr{R}(U,\Delta,\mathscr{D}))\subseteq\mathscr{R}(U,\Delta^{+},\mathscr{D})$.
Subsequently, let
$\mathscr{R}(U,\Delta^{+},\mathscr{D})=\mathscr{R}_{1}(U,\Delta^{+},\mathscr{D})\cup \mathscr{R}_{2}(U,\Delta^{+},\mathscr{D})$, where $\mathscr{R}_{1}(U,\Delta^{+},\mathscr{D})=\{\Delta_{i}\mid \mathscr{C}^{+}_{m}\notin \Delta_{i},\Delta_{i}\in \mathscr{R}(U,\Delta^{+},\mathscr{D})\}$ and $\mathscr{R}_{2}(U ,\Delta^{+},\mathscr{D})=\{\Delta_{i}\mid \mathscr{C}^{+}_{m}\in \Delta_{i}\in \mathscr{R}(U,\Delta^{+},\mathscr{D})\}$. Obviously, we have $\mathscr{R}_{1}(U,\Delta^{+},\mathscr{D})= \lozenge{\mathscr{R}}(U,\Delta,\mathscr{D})$ and $\blacktriangle\mathscr{R}(U,\Delta,\mathscr{D})\subseteq \mathscr{R}_{2}(U,\Delta^{+},\mathscr{D})$. Now prove  $\mathscr{R}_{2}(U,\Delta^{+},\mathscr{D})\subseteq \blacktriangle\mathscr{R}(U,\Delta,\mathscr{D})$. In other words, $\mathscr{R}_{2}(U,\Delta^{+},\mathscr{D})\backslash (\blacktriangle\mathscr{R}(U,\Delta,\mathscr{D}))=\emptyset$.
Suppose we have $ \Delta'=\{\mathscr{C}_{1'},\mathscr{C}_{2'},...,\mathscr{C}_{k'},\\\mathscr{C}^{+}_{m}\}\in \mathscr{R}_{2}(U,\Delta^{+},\mathscr{D})\backslash \blacktriangle\mathscr{R}(U,\Delta,\mathscr{D}),$ there exists $x\in U$ such that $\mathscr{C}_{i'}\in r^{+}(x)(1'\leq i'\leq k').$ If $\mathscr{C}^{+}_{m}\in r^{+}(x)$, then $\mathscr{C}_{i'}$ is superfluous relative to $\mathscr{D}$. It implies that $\mathscr{C}^{+}_{m}\notin r^{+}(x)$, so $\Delta'\in \blacktriangle\mathscr{R}(U,\Delta,\mathscr{D})$, which is contradicted. It follows that $\mathscr{R}_{2}(U,\Delta^{+},\mathscr{D})\backslash (\blacktriangle\mathscr{R}(U,\Delta,\mathscr{D}))=\emptyset$. Thus $\blacktriangle\mathscr{R}(U,\Delta,\mathscr{D})=\mathscr{R}_{2}(U,\Delta^{+},\mathscr{D}).$
 So we obtain $\mathscr{R}(U,\Delta^{+},\mathscr{D})\subseteq \lozenge{\mathscr{R}}(U,\Delta,\mathscr{D})\cup (\blacktriangle\mathscr{R}(U,\Delta,\mathscr{D}))$. Therefore, $\mathscr{R}(U,\Delta^{+},\mathscr{D})=\lozenge{\mathscr{R}}(U,\Delta,\mathscr{D})\cup (\blacktriangle\mathscr{R}(U,\Delta,\mathscr{D}))$.
$\Box$

Theorem 3.5 illustrates how to construct $\mathscr{R}(U,\Delta^{+},\mathscr{D})$ of $(U,\Delta^{+},\mathscr{D})$ based on $\mathscr{R}(U,\Delta,\mathscr{D})$ of $(U,\Delta,\mathscr{D})$ when $POS_{\cup\Delta^{+}}(\mathscr{D})=POS_{\cup\Delta}(\mathscr{D})$. Especially, each reduct of  $\mathscr{R}(U,\Delta,\mathscr{D})$ that does not contain $\mathscr{C}_{m}$ is a reduct of $(U,\Delta^{+},\mathscr{D})$, and we can get all reducts of $(U,\Delta^{+},\mathscr{D})$ by Theorem 3.5.

\begin{theorem}
Let $(U,\Delta, \mathscr{D})$ and $(U,\Delta^{+}, \mathscr{D})$ be covering decision information systems, where $U=\{x_{1},x_{2},$ $...,x_{n}\}$, $\Delta=\{\mathscr{C}_{1},\mathscr{C}_{2},...,\mathscr{C}_{m-1},\mathscr{C}_{m}\}$, $\Delta^{+}=\{\mathscr{C}_{1},\mathscr{C}_{2},...,\mathscr{C}_{m-1},\mathscr{C}^{+}_{m}\}$, $\mathscr{D}=\{D_{1},D_{2},...,D_{k}\}$, $\blacktriangle f(U,\Delta,\mathscr{D})=(\{\mathscr{C}^{+}_{m}\})\bigwedge (\bigwedge_{x\in POS_{\cup\Delta}(\mathscr{D})\wedge x\notin \cup\mathscr{A}_{\mathscr{C}^{+}_{m}}}$ $\bigvee r(x))=\bigvee^{l}_{i=1}\{\bigwedge \Delta'_{i}\mid \Delta'_{i}\subseteq\Delta\}$, and $\blacktriangle\mathscr{R}(U,\Delta,\mathscr{D})=\{\Delta'_{j}\mid 1\leq j\leq l\}$. If $POS_{\cup\Delta^{+}}(\mathscr{D})\neq POS_{\cup\Delta}(\mathscr{D})$, then $\mathscr{R}(U,\Delta^{+},\mathscr{D})=\blacktriangle\mathscr{R}(U,\Delta,\mathscr{D})$.
\end{theorem}

\noindent\textbf{Proof:} Suppose
$\mathscr{R}(U,\Delta^{+},\mathscr{D})=\mathscr{R}_{1}(U,\Delta^{+},\mathscr{D})\cup \mathscr{R}_{2}(U,\Delta^{+},\mathscr{D})$, where $\mathscr{R}_{1}(U,\Delta^{+},\mathscr{D})=\{\Delta_{i}\mid \mathscr{C}_{m}\notin \Delta_{i}\in \mathscr{R}(U,\Delta^{+},\mathscr{D})\}$ and $\mathscr{R}_{2}(U ,\Delta^{+},\mathscr{D})=\{\Delta_{i}\mid \mathscr{C}^{+}_{m}\in \Delta_{i},\Delta_{i}\in \mathscr{R}(U,\Delta^{+},\mathscr{D})\}$. Because $POS_{\cup\Delta^{+}}(\mathscr{D})\neq POS_{\cup\Delta}(\mathscr{D})$, thus $POS_{\cup\Delta}(\mathscr{D})\subseteq POS_{\cup\Delta^{+}}(\mathscr{D})$. It follows that $r(x)^{+} = \{\mathscr{C}^{+}_{m}\}$ for any $x \in POS_{\cup\Delta^{+}}(\mathscr{D}) \backslash POS_{\cup\Delta}(\mathscr{D})$. Thus, we have $\mathscr{R}_{1}(U,\Delta^{+},\mathscr{D}) = \emptyset$. Obviously, $\blacktriangle\mathscr{R}(U,\Delta,\mathscr{D})\subseteq \mathscr{R}_{2}(U,\Delta^{+},\mathscr{D})$. Now prove  $\mathscr{R}_{2}(U,\Delta^{+},\mathscr{D})\subseteq \blacktriangle\mathscr{R}(U,\Delta,\mathscr{D})$. In other words, $\mathscr{R}_{2}(U,\Delta^{+},\mathscr{D})\backslash (\blacktriangle\mathscr{R}(U,\Delta,\mathscr{D}))=\emptyset$.
Suppose $\mathscr{R}_{2}(U,\Delta^{+},\mathscr{D})\backslash \blacktriangle\mathscr{R}(U,\Delta,\mathscr{D}) = \Delta'=\{\mathscr{C}_{1'},\mathscr{C}_{2'},...,\mathscr{C}_{k'},\mathscr{C}^{+}_{m}\},$ there exists $x\in U$ such that $\mathscr{C}_{i'}\in r^{+}(x)(1'\leq i'\leq k').$ If $\mathscr{C}^{+}_{m}\in r^{+}(x)$, then $\mathscr{C}_{i'}$ is superfluous relative to $\mathscr{D}$. It implies that $\mathscr{C}^{+}_{m}\notin r^{+}(x)$, so $\Delta'\in \blacktriangle\mathscr{R}(U,\Delta,\mathscr{D})$, which is contradicted. So $\mathscr{R}_{2}(U,\Delta^{+},\mathscr{D})\backslash (\blacktriangle\mathscr{R}(U,\Delta,\mathscr{D}))=\emptyset$. It follows that $\blacktriangle\mathscr{R}(U,\Delta,\mathscr{D})=\mathscr{R}_{2}(U,\Delta^{+},\mathscr{D}).$ Therefore,  $\mathscr{R}(U,\Delta^{+},\mathscr{D})=\blacktriangle\mathscr{R}(U,\Delta,\mathscr{D})$. $\Box$

Theorem 3.6 demonstrates how to construct $\mathscr{R}(U,\Delta^{+},\mathscr{D})$ of $(U,\Delta^{+},\mathscr{D})$ based on $\mathscr{R}(U,\Delta,\mathscr{D})$ of $(U,\Delta,\mathscr{D})$ when $POS_{\cup\Delta^{+}}(\mathscr{D}) \neq POS_{\cup\Delta}(\mathscr{D})$. Especially, we can get all reducts of $(U,\Delta^{+},\mathscr{D})$ by Theorem 3.6.

\begin{algorithm}Let $(U,\Delta, \mathscr{D})$ and $(U,\Delta^{+}, \mathscr{D})$ be covering decision information systems, where $U=\{x_{1},x_{2},$ $...,x_{n}\}$, $\Delta=\{\mathscr{C}_{1},\mathscr{C}_{2},...,\mathscr{C}_{m-1},\mathscr{C}_{m}\}$, $\Delta^{+}=\{\mathscr{C}_{1},\mathscr{C}_{2},...,\mathscr{C}_{m-1},\mathscr{C}^{+}_{m}\}$,
and $\mathscr{D}=\{D_{1},D_{2},...,D_{k}\}$. Then

Step 1: Input $(U,\Delta^{+}, \mathscr{D})$;

Step 2: Construct $POS_{\cup\Delta^{+}}(\mathscr{D})$;

Step 3: Compute $R(U,\Delta^{+},\mathscr{D})=\{r^{+}(x)\mid x\in POS_{\cup\Delta^{+}}(\mathscr{D})\}$, where
\begin{eqnarray*}
r^{+}(x)=\left\{
\begin{array}{ccc}
(r(x)\backslash\{\mathscr{C}_{m}\})\cup \{\mathscr{C}^{+}_{m}\},&{\rm if}& x\in \cup\mathscr{A}_{\mathscr{C}^{+}_{m}};\\\\
r(x),&{\rm }& otherwise.
\end{array}
\right.
\end{eqnarray*}

Step 4: Construct $\lozenge{\mathscr{R}}(U,\Delta,\mathscr{D})=\{\Delta_{i}\mid \mathscr{C}_{m}\notin\Delta_{i}\in \mathscr{R}(U,\Delta,\mathscr{D})\}$;

Step 5: Construct $\blacktriangle f(U,\Delta,\mathscr{D})=(\{\mathscr{C}^{+}_{m}\})\bigwedge (\bigwedge_{x\in POS_{\cup\Delta}(\mathscr{D})\wedge x\notin \cup\mathscr{A}_{\mathscr{C}^{+}_{m}}}$ $\bigvee r(x))=\bigvee^{l}_{i=1}\{\bigwedge \Delta'_{i}\mid \Delta'_{i}\subseteq\Delta\}$;

Step 6: Compute $\blacktriangle\mathscr{R}(U,\Delta,\mathscr{D})=\{\Delta'_{j}\mid \overline{\exists}\Delta_{i}\in \mathscr{R}(U,\Delta,\mathscr{D}), \text{ s.t. } \Delta_{i}\subset\Delta'_{j},1\leq j\leq l\}$;

Step 7: Output $\mathscr{R}(U,\Delta^{+},\mathscr{D})=\lozenge{\mathscr{R}}(U,\Delta,\mathscr{D})\cup (\blacktriangle\mathscr{R}(U,\Delta,\mathscr{D}))$.
\end{algorithm}

\begin{example}(Continuation from Example 2.10) $(1)$ Firstly, by Definition 2.8, we have
$r^{+}(x_{1})=\{\mathscr{C}_{1},\mathscr{C}_{4},\mathscr{C}^{+}_{5}\},
$ $r^{+}(x_{2})=\{\mathscr{C}_{1},\mathscr{C}_{4},\mathscr{C}^{+}_{5}\},
r^{+}(x_{3})=\{\mathscr{C}_{1},\mathscr{C}_{2},\mathscr{C}_{3},\mathscr{C}_{4}\},
r^{+}(x_{4})=\{\mathscr{C}_{1},\mathscr{C}_{2},\mathscr{C}_{3},\mathscr{C}_{4}\},
r^{+}(x_{5})=\{\mathscr{C}_{1},\mathscr{C}_{2},\mathscr{C}_{3},\mathscr{C}_{4}\},
r^{+}(x_{6})=\{\mathscr{C}_{1},\mathscr{C}_{2},\mathscr{C}_{3},\mathscr{C}_{4},\mathscr{C}^{+}_{5}\},
r^{+}(x_{7})=\{\mathscr{C}_{2},\mathscr{C}_{3}\}$ and $
r^{+}(x_{8})=\{\mathscr{C}_{2},\mathscr{C}_{3},\mathscr{C}^{+}_{5}\}.$ It follows that $R(U,\Delta^{+},\mathscr{D})=\{\{\mathscr{C}_{1},\mathscr{C}_{4},\mathscr{C}^{+}_{5}\},
$ $\{\mathscr{C}_{1}$, $\mathscr{C}_{2},\mathscr{C}_{3},\mathscr{C}_{4}\},\{\mathscr{C}_{1},\mathscr{C}_{2},\mathscr{C}_{3},$ $\mathscr{C}_{4},\mathscr{C}^{+}_{5}\},\{\mathscr{C}_{2},\mathscr{C}_{3}\},
\{\mathscr{C}_{2},\mathscr{C}_{3},\mathscr{C}^{+}_{5}\}\}.$
By Definition 2.8, we get
\begin{eqnarray*}
f(U,\Delta^{+},\mathscr{D})
&=&(\mathscr{C}_{1}\vee\mathscr{C}_{4}\vee\mathscr{C}^{+}_{5})\wedge(\mathscr{C}_{1}\vee\mathscr{C}_{2}
\vee\mathscr{C}_{3} \vee\mathscr{C}_{4})\wedge
(\mathscr{C}_{1}\vee\mathscr{C}_{2}\vee\mathscr{C}_{3}\vee\mathscr{C}_{4}
\vee\mathscr{C}^{+}_{5})\wedge
(\mathscr{C}_{2}\vee\mathscr{C}_{3})\\&&
\wedge(\mathscr{C}_{2}\vee\mathscr{C}_{3}\vee\mathscr{C}^{+}_{5})\\
&=&(\mathscr{C}_{1}\vee\mathscr{C}_{4}\vee\mathscr{C}^{+}_{5})\wedge(\mathscr{C}_{2}\vee\mathscr{C}_{3})\\
&=&(\mathscr{C}_{1}\wedge\mathscr{C}_{2})\vee (\mathscr{C}_{1}\wedge\mathscr{C}_{3})\vee (\mathscr{C}_{2}\wedge\mathscr{C}_{4})\vee(\mathscr{C}_{2}\wedge\mathscr{C}^{+}_{5})\vee
(\mathscr{C}_{3}\wedge\mathscr{C}_{4})\vee(\mathscr{C}_{3}\wedge\mathscr{C}^{+}_{5}).
\end{eqnarray*}
Therefore, we have $\mathscr{R}(U,\Delta^{+},\mathscr{D})=\{\{\mathscr{C}_{1},\mathscr{C}_{2}\}, \{\mathscr{C}_{1},\mathscr{C}_{3}\}, \{\mathscr{C}_{2},\mathscr{C}_{4}\}, \{\mathscr{C}_{2},\mathscr{C}^{+}_{5}\},
\{\mathscr{C}_{3},\mathscr{C}_{4}\},
\{\mathscr{C}_{3},\mathscr{C}^{+}_{5}\}\}.$

Secondly, according to Theorem 3.5, we get
\begin{eqnarray*}
\lozenge{\mathscr{R}}(U,\Delta,\mathscr{D})&=&\{\{\mathscr{C}_{1},\mathscr{C}_{2}\}, \{\mathscr{C}_{1},\mathscr{C}_{3}\}, \{\mathscr{C}_{2},\mathscr{C}_{4}\},
\{\mathscr{C}_{3},\mathscr{C}_{4}\}\}.\\
\blacktriangle
f(U,\Delta,\mathscr{D})&=&\mathscr{C}^{+}_{5}\wedge(\mathscr{C}_{1}\vee\mathscr{C}_{2}\vee\mathscr{C}_{3}\vee\mathscr{C}_{4})
\wedge(\mathscr{C}_{2}\vee\mathscr{C}_{3})\\
&=&\mathscr{C}^{+}_{5}\wedge(\mathscr{C}_{2}\vee\mathscr{C}_{3})\\
&=&(\mathscr{C}_{2}\wedge\mathscr{C}^{+}_{5})\vee (\mathscr{C}_{3}\wedge\mathscr{C}^{+}_{5}).
\end{eqnarray*}
It implies that $\blacktriangle\mathscr{R}(U,\Delta,\mathscr{D})=\{\{\mathscr{C}_{2},\mathscr{C}^{+}_{5})\}, \{\mathscr{C}_{3},\mathscr{C}^{+}_{5}\}\}.$
Therefore, we obtain $\mathscr{R}(U,\Delta^{+},\mathscr{D})=\{\{\mathscr{C}_{1},
\mathscr{C}_{2}\},\{\mathscr{C}_{1},\\\mathscr{C}_{3}\}, \{\mathscr{C}_{2},
\mathscr{C}_{4}\}, \{\mathscr{C}_{2},\mathscr{C}^{+}_{5}\},
\{\mathscr{C}_{3},\mathscr{C}_{4}\},
\{\mathscr{C}_{3},\mathscr{C}^{+}_{5}\}\}.$

$(2)$ Firstly, by Definition 2.8, we get
$r^{+}(x_{1})=\{\mathscr{C}_{2},\mathscr{C}_{4},\mathscr{C}^{+}_{5}\},
r^{+}(x_{2})$ $=\emptyset,
r^{+}(x_{3})=\emptyset,
r^{+}(x_{4})=\{\mathscr{C}_{2},\mathscr{C}_{3},\mathscr{C}_{4}\},$ $
r^{+}(x_{5})=\{\mathscr{C}_{2},\mathscr{C}_{3},\mathscr{C}_{4}\},
r^{+}(x_{6})=\{\mathscr{C}_{1},\mathscr{C}^{+}_{5}\},
r^{+}(x_{7})=\{\mathscr{C}_{1},\mathscr{C}_{3},\mathscr{C}^{+}_{5}\}$ and $
r^{+}(x_{8})=\{\mathscr{C}_{1},\mathscr{C}_{3},\mathscr{C}^{+}_{5}\}.$
It implies that $R(U,\Delta^{+},\mathscr{D})=\{\{\mathscr{C}_{2},\mathscr{C}_{4},\mathscr{C}^{+}_{5}\},
\{\mathscr{C}_{2},\mathscr{C}_{3},\mathscr{C}_{4}\},\{\mathscr{C}_{1},\mathscr{C}^{+}_{5}\},\{\mathscr{C}_{1},\mathscr{C}_{3},\mathscr{C}^{+}_{5}\}\}$.
By Definition 2.8, we have
\begin{eqnarray*}
f(U,\Delta^{+},\mathscr{D})
&=&(\mathscr{C}_{2}\vee\mathscr{C}_{4}\vee\mathscr{C}^{+}_{5})\wedge(\mathscr{C}_{2}\vee\mathscr{C}_{3}\vee\mathscr{C}_{4})
\wedge(\mathscr{C}_{1}\vee\mathscr{C}^{+}_{5})\wedge(\mathscr{C}_{1}\vee\mathscr{C}_{3}\vee\mathscr{C}^{\vee}_{5})\\
&=&(\mathscr{C}_{2}\vee\mathscr{C}_{4}\vee\mathscr{C}^{+}_{5})\wedge(\mathscr{C}_{2}\vee\mathscr{C}_{3}\vee\mathscr{C}_{4})
\wedge(\mathscr{C}_{1}\vee\mathscr{C}^{+}_{5})\\
&=&(\mathscr{C}_{1}\wedge\mathscr{C}_{2})\vee (\mathscr{C}_{1}\wedge\mathscr{C}_{4})\vee (\mathscr{C}_{2}\wedge\mathscr{C}^{+}_{5})\vee (\mathscr{C}_{3}\wedge\mathscr{C}^{+}_{5})\vee (\mathscr{C}_{4}\wedge\mathscr{C}^{+}_{5}).
\end{eqnarray*}
Therefore, we get $\mathscr{R}(U,\Delta^{+},\mathscr{D})=\{\{\mathscr{C}_{1},\mathscr{C}_{2}\}, \{\mathscr{C}_{1},\mathscr{C}_{4}\}, \{\mathscr{C}_{2},\mathscr{C}^{+}_{5}\}, \{\mathscr{C}_{3},\mathscr{C}^{+}_{5}\}, \{\mathscr{C}_{4},\mathscr{C}^{+}_{5}\}\}.$

Secondly, according to Theorem 3.6, we get
\begin{eqnarray*}
\lozenge{\mathscr{R}}(U,\Delta,\mathscr{D})&=&\{\{\mathscr{C}_{1},\mathscr{C}_{2}\}, \{\mathscr{C}_{1},\mathscr{C}_{4}\}\}.\\
\blacktriangle f(U,\Delta,\mathscr{D})&=&\mathscr{C}^{+}_{5}\wedge(\mathscr{C}_{2}\vee\mathscr{C}_{3}\vee\mathscr{C}_{4}).\\
&=& (\mathscr{C}_{2}\wedge\mathscr{C}^{+}_{5})\vee(\mathscr{C}_{3}\wedge\mathscr{C}^{+}_{5})\vee(\mathscr{C}_{4}\wedge\mathscr{C}^{+}_{5}).
\end{eqnarray*}
It follows that $\blacktriangle\mathscr{R}(U,\Delta,\mathscr{D})=\{\{\mathscr{C}_{2},\mathscr{C}^{+}_{5}\}, \{\mathscr{C}_{3},\mathscr{C}^{+}_{5}\}, \{\mathscr{C}_{4},\mathscr{C}^{+}_{5}\}\}.$ Therefore, we obtain $\mathscr{R}(U,\Delta^{+},\mathscr{D})=\{\{\mathscr{C}_{1},\mathscr{C}_{2}\},\\ \{\mathscr{C}_{1},\mathscr{C}_{4}\},\{\mathscr{C}_{2},\mathscr{C}^{+}_{5}\}, \{\mathscr{C}_{3},\mathscr{C}^{+}_{5}\}, \{\mathscr{C}_{4},\mathscr{C}^{+}_{5}\}\}$.
\end{example}

Example 3.8 shows how to compute attribute reducts in dynamic covering decision information systems with refining coverings by Algorithms 2.9 and 3.7, respectively. We see that Algorithm 3.7 is more effective than Algorithm 2.9 to compute attribute reducts in dynamic covering decision information systems.

Suppose $(U,\Delta^{+}, \mathscr{D})$ and $(U,\Delta, \mathscr{D})$ are covering decision information systems, we denote $SR(U,\Delta^{+},\mathscr{D})$ $=\{r^{+}(x)\in R(U,\Delta^{+},\mathscr{D})\mid x\in POS_{\cup\Delta^{+}}(\mathscr{D}), (\forall y\in POS_{\cup\Delta^{+}}(\mathscr{D}), r^{+}(y)\nsubseteq r^{+}(x)\text{ and } r^{+}(y)\in R(U,\Delta^{+},\mathscr{D}))\}$, and $\|\mathscr{C}\|$ denotes the number of times for a covering $\mathscr{C}$ appeared in $SR(U,\Delta^{+},\mathscr{D})$.

\begin{algorithm}(Heuristic Algorithm of Computing a Reduct of $(U,\Delta^{+},\mathscr{D})$)(IHVR)

Step 1: Input $(U,\Delta^{+}, \mathscr{D})$;

Step 2: Construct $POS_{\cup\Delta^{+}}(\mathscr{D})$;

Step 3: Compute $R(U,\Delta^{+},\mathscr{D})=\{r^{+}(x)\mid x\in POS_{\cup\Delta^{+}}(\mathscr{D})\}$, where
\begin{eqnarray*}
r^{+}(x)=\left\{
\begin{array}{ccc}
(r(x)\backslash\{\mathscr{C}_{m}\})\cup \{\mathscr{C}^{+}_{m}\},&{\rm if}& x\in \cup\mathscr{A}_{\mathscr{C}^{+}_{m}};\\\\
r(x),&{\rm }& otherwise.
\end{array}
\right.
\end{eqnarray*}

Step 4: Construct a reduct $\bigtriangleup^{\ast+}=\{\mathscr{C}_{i_{1}},\mathscr{C}_{i_{2}},...,\mathscr{C}_{i_{j}}\}$, where
$SR_{1}(U,\Delta^{+},\mathscr{D})=SR(U,\Delta^{+},\mathscr{D})$, $\|\mathscr{C}_{i_{1}}\|=max\{\|\mathscr{C}_{i}\|\mid \mathscr{C}_{i}\in r^{+}(x)\in SR_{1}(U,\Delta^{+},\mathscr{D})\}$;
$SR_{2}(U,\Delta^{+},\mathscr{D})=\{r^{+}(x)\in SR(U,\Delta^{+},\mathscr{D})\mid \mathscr{C}_{i_{1}}\notin r^{+}(x)\}$, $\|\mathscr{C}_{i_{2}}\|=max\{\|\mathscr{C}_{i}\|\mid \mathscr{C}_{i}\in r^{+}(x)\in SR_{2}(U,\Delta^{+},\mathscr{D})\}$;
$SR_{3}(U,\Delta^{+},\mathscr{D})=\{r^{+}(x)\in SR(U,\Delta^{+},\mathscr{D})\mid \mathscr{C}_{i_{1}}\notin r^{+}(x)\text{ or } \mathscr{C}_{i_{2}}\notin r^{+}(x)\}$, $\|\mathscr{C}_{i_{3}}\|=max\{\|\mathscr{C}_{i}\|\mid \mathscr{C}_{i}\in r^{+}(x)\in SR_{3}(U,\Delta^{+},\mathscr{D})\}$;
...;
$SR_{j}(U,\Delta^{+},\mathscr{D})=\{r^{+}(x)\in SR(U,\Delta^{+},\mathscr{D})\mid \mathscr{C}_{i_{1}}\notin r^{+}(x)\text{ or } \mathscr{C}_{i_{2}}\notin r^{+}(x)\text{ or }... \text{ or } \mathscr{C}_{i_{j-1}}\notin r^{+}(x)\}$, $\|\mathscr{C}_{i_{j}}\|=max\{\|\mathscr{C}_{i}\|\mid \mathscr{C}_{i}\in r^{+}(x)\in SR_{j}(U,\Delta^{+},\mathscr{D})\}$, and $SR(U,\Delta^{+},\mathscr{D})=\{r^{+}(x)\mid \exists\mathscr{C}_{i_{k}}\in r^{+}(x),1\leq k\leq j\}$;

Step 5: Output the reduct $\bigtriangleup^{\ast+}$.
\end{algorithm}

If there are two coverings $\mathscr{C}_{i}$  and $\mathscr{C}_{j}$ such that $\|\mathscr{C}_{i}\|=\|\mathscr{C}_{j}\|=max\{\|\mathscr{C}_{i}\|\mid \mathscr{C}_{i}\in r^{+}(x)\in SR_{k}(U,\Delta^{+},\mathscr{D})\}$, then we select $\|\mathscr{C}_{i}\|=max\{\|\mathscr{C}_{i}\|\mid \mathscr{C}_{i}\in r^{+}(x)\in SR_{k}(U,\Delta^{+},\mathscr{D})\}$ or $\|\mathscr{C}_{j}\|=max\{\|\mathscr{C}_{i}\|\mid \mathscr{C}_{i}\in r^{+}(x)\in SR_{k}(U,\Delta^{+},\mathscr{D})\}$.

\begin{example}(Continuation from Example 3.8)
$(1)$ In Example 3.8(1), we have $SR(U,\Delta^{+},\mathscr{D})=\{\{\mathscr{C}_{1},\mathscr{C}_{4},$ $\mathscr{C}^{+}_{5}\},\{\mathscr{C}_{2},\mathscr{C}_{3}\}\}.$
By Algorithm 3.9, firstly, we obtain $SR_{1}(U,\Delta^{+},\mathscr{D})=\{\{\mathscr{C}_{1},\mathscr{C}_{4},$ $\mathscr{C}_{5}^{+}\},\{\mathscr{C}_{2},\mathscr{C}_{3}\}\}$ and $\|\mathscr{C}_{1}\|=max\{\|\mathscr{C}_{i}\|\mid \mathscr{C}_{i}\in r^{+}(x)\in SR_{1}(U,\Delta^{+},\mathscr{D})\}$. Secondly, we obtain $SR_{2}(U,\Delta^{+},\mathscr{D})=\{\{\mathscr{C}_{2},\mathscr{C}_{3}\}\},$
$\|\mathscr{C}_{2}\|=max\{\|\mathscr{C}_{i}\|\mid \mathscr{C}_{i}\in r^{+}(x)\in SR_{2}(U,\Delta^{+},\mathscr{D})\}$. Finally, we get a reduct $\bigtriangleup^{\ast+}=\{\mathscr{C}_{1},\mathscr{C}_{2}\}$ of  $(U,\Delta^{+},\mathscr{D})$.

$(2)$ In Example 3.8(2), we get $SR(U,\Delta^{+},\mathscr{D})=\{\{\mathscr{C}_{2},\mathscr{C}_{4},$ $\mathscr{C}^{+}_{5}\},\{\mathscr{C}_{2},\mathscr{C}_{3},\mathscr{C}_{4}\},
\{\mathscr{C}_{1},\mathscr{C}_{5}^{+}\}\}.$
By Algorithm 3.9, firstly, we obtain $SR_{1}(U,\Delta^{+},\mathscr{D})=\{\{\mathscr{C}_{2},\mathscr{C}_{4},$ $\mathscr{C}^{+}_{5}\},\{\mathscr{C}_{2},\mathscr{C}_{3},\mathscr{C}_{4}\},
\{\mathscr{C}_{1},\mathscr{C}^{+}_{5}\}\}$ and $\|\mathscr{C}_{2}\|=max\{\|\mathscr{C}_{i}\|\mid \mathscr{C}_{i}\in r^{+}(x)\in SR_{1}(U,\Delta^{+},\mathscr{D})\}$. Secondly, we get $SR_{2}(U,\Delta^{+},\mathscr{D})=\{\{\mathscr{C}_{1},\mathscr{C}^{+}_{5}\}\},$
$\|\mathscr{C}_{1}\|=max\{\|\mathscr{C}_{i}\|\mid \mathscr{C}_{i}\in r^{+}(x)\in SR_{2}(U,\Delta^{+},\mathscr{D})\}$. Finally, we have a reduct $\bigtriangleup^{\ast+}=\{\mathscr{C}_{1},\mathscr{C}_{2}\}$ of  $(U,\Delta^{+},\mathscr{D})$.
\end{example}

\section{Updating attribute reducts of dynamic covering decision information systems with coarsening coverings}

In this section, we study how to update attribute reducts of dynamic covering decision information systems with coarsening coverings.

\begin{definition}
Let $(U,\Delta, \mathscr{D})$ and $(U,\Delta^{-}, \mathscr{D})$ be covering decision information systems, where $U=\{x_{1},x_{2},$ $...,x_{n}\}$, $\Delta=\{\mathscr{C}_{1},\mathscr{C}_{2},...,\mathscr{C}_{m-1},\mathscr{C}_{m}\}$, $\Delta^{-}=\{\mathscr{C}_{1},\mathscr{C}_{2},...,\mathscr{C}_{m-1},\mathscr{C}^{-}_{m}\}$, and $\mathscr{D}=\{D_{1},D_{2},...,D_{k}\}$.
Then $(U,\Delta^{-}, \mathscr{D})$ is called a dynamic covering decision information system of $(U,\Delta, \mathscr{D})$.
\end{definition}

According to Definition 4.1, if $(U,\Delta, \mathscr{D})$  is a consistent covering decision information system, then $(U,\Delta^{-}, \mathscr{D})$ is a consistent covering decision information system or an inconsistent covering decision information system. Moreover, if $(U,\Delta, \mathscr{D})$ is an inconsistent covering decision information system, then $(U,\Delta^{-}, \mathscr{D})$ is an inconsistent covering decision information system.

In practical situations, there are many types of dynamic covering decision information systems with coarsening coverings. For simplicity, we only consider the dynamic covering decision information system with a coarsening covering in this section.

\begin{example}(Continuation from Example 2.10)
$(1)$ Let $(U,\Delta, \mathscr{D})$ and $(U,\Delta^{-}, \mathscr{D})$ be covering decision information systems, where $U=\{x_{1},x_{2},$ $...,x_{8}\}$, $\Delta=\{\mathscr{C}_{1},\mathscr{C}_{2},\mathscr{C}_{3},\mathscr{C}_{4},\mathscr{C}_{5}\}$,
$\Delta^{-}=\{\mathscr{C}_{1},\mathscr{C}_{2},\mathscr{C}_{3},\mathscr{C}_{4},\mathscr{C}^{-}_{5}\}$, $\mathscr{D}=\{\{x_{1},x_{2}\},\{x_{3},x_{4},x_{5}\},\{x_{6},x_{7},x_{8}\}\}$,
$\mathscr{C}_{5}=\{\{x_{1},x_{2}\},\{x_{2},x_{3},x_{5}\},\{x_{4},x_{5},x_{7}\},\{x_{6}\},\{x_{3},x_{7},x_{8}\}\},$
and $\mathscr{C}^{-}_{5}=\{\{x_{1},x_{2},x_{3},x_{5}\},\{x_{4},x_{5},x_{7}\},\{x_{6}\},\{x_{3},x_{7},x_{8}\}\}.$
Therefore, we see that $(U,\Delta^{-}, \mathscr{D})$ is a consistent covering decision information system.

$(2)$ Let $(U,\Delta,\mathscr{D})$ and $(U,\Delta^{-},\mathscr{D})$ be covering decision information systems, where $U=\{x_{1},x_{2},$ $...,x_{8}\}$, $\Delta=\{\mathscr{C}_{1},\mathscr{C}_{2},\mathscr{C}_{3},\mathscr{C}_{4},\mathscr{C}_{5}\}$,
$\Delta^{-}=\{\mathscr{C}_{1},\mathscr{C}_{2},\mathscr{C}_{3},\mathscr{C}_{4},\mathscr{C}^{-}_{5}\}$,
$\mathscr{D}=\{\{x_{1},x_{2}\},\{x_{3},x_{4},x_{5}\},\{x_{6},x_{7},x_{8}\}\}$,
$\mathscr{C}_{5}=\{\{x_{1},x_{2},\\x_{3},x_{5}\},\{x_{2},x_{4},x_{5},x_{6}\},\{x_{6}\},\{x_{7},x_{8}\}\},$
and $\mathscr{C}^{-}_{5}=\{\{x_{1},x_{2},x_{3},x_{5}\},\{x_{2},x_{4},x_{5},x_{6}\},\{x_{6},x_{7},x_{8}\}\}.$ Therefore, we observe that $(U,\Delta^{-}, \mathscr{D})$ is an inconsistent covering decision information system.
\end{example}

Suppose $(U,\Delta, \mathscr{D})$ and $(U,\Delta^{-},\mathscr{D})$ are covering decision information systems, where $U=\{x_{1},x_{2},...,x_{n}\}$, $\Delta=\{\mathscr{C}_{1},\mathscr{C}_{2},...,\mathscr{C}_{m-1},\mathscr{C}_{m}\}$, and $\Delta^{-}=\{\mathscr{C}_{1},\mathscr{C}_{2},...,\mathscr{C}_{m-1},\mathscr{C}^{-}_{m}\}$,
$\mathscr{A}_{\Delta}=\{C\in \cup \Delta\mid \exists D_{j}\in \mathscr{D}, \text{ s.t. } C\subseteq D_{j}\}$,
$\mathscr{A}_{\Delta^{-}}=\{C\in \cup \Delta^{-}\mid \exists D_{j}\in \mathscr{D}, \text{ s.t. } C\subseteq D_{j}\}$,
$\mathscr{A}_{\mathscr{C}_{m}}=\{C\in \mathscr{C}_{m}\mid \exists D_{j}\in \mathscr{D}, \text{ s.t. } C\subseteq D_{j}\}$,
$\mathscr{A}_{\mathscr{C}^{-}_{m}}=\{C\in \mathscr{C}^{-}_{m}\mid \exists D_{j}\in \mathscr{D}, \text{ s.t. } C\subseteq D_{j}\}$,
$r(x)=\{\mathscr{C}\in \Delta\mid \exists C\in \mathscr{A}_{\Delta}, \text{ s.t. } x\in C\in \mathscr{C}\},$ and  $r^{-}(x)=\{\mathscr{C}\in \Delta^{-}\mid \exists C\in \mathscr{A}_{\Delta^{-}}, \text{ s.t. } x\in C\in \mathscr{C}\}.$

\begin{theorem}
Let $(U,\Delta, \mathscr{D})$ and $(U,\Delta^{-},\mathscr{D})$ be covering decision information systems, where $U=\{x_{1},x_{2},$ $...,x_{n}\}$, $\Delta=\{\mathscr{C}_{1},\mathscr{C}_{2},...,\mathscr{C}_{m-1},\mathscr{C}_{m}\}$,  $\Delta^{-}=\{\mathscr{C}_{1},\mathscr{C}_{2},...,\mathscr{C}_{m-1},\mathscr{C}^{-}_{m}\}$, and $\mathscr{D}=\{D_{1},D_{2},...,D_{k}\}$. Then
we have
\makeatother $$r^{-}(x)=\left\{
\begin{array}{ccc}
(r(x)\backslash\{\mathscr{C}_{m}\})\cup \{\mathscr{C}^{-}_{m}\},&{\rm if}& x\in \cup\mathscr{A}_{\mathscr{C}^{-}_{m}};\\\\
r(x)\backslash\{\mathscr{C}_{m}\},&{\rm }& otherwise.
\end{array}
\right. $$
\end{theorem}

\noindent\textbf{Proof:} By Definitions 2.8, we have $r(x)=\{\mathscr{C}\in \Delta\mid \exists C\in \mathscr{A}_{\Delta}, \text{ s.t. } x\in C\in \mathscr{C}\},$ and  $r^{-}(x)=\{\mathscr{C}\in \Delta^{-}\mid \exists C\in \mathscr{A}_{\Delta^{-}}, \text{ s.t. } x\in C\in \mathscr{C}\}$. Since $\Delta =\{\mathscr{C}_{1},\mathscr{C}_{2},...,\mathscr{C}_{m-1},\mathscr{C}_{m}\},$ and $\Delta^{-}=\{\mathscr{C}_{1},\mathscr{C}_{2},...,\mathscr{C}_{m-1},\mathscr{C}^{-}_{m}\}$,
it follows that $r^{-}(x)=(\{\mathscr{C}\in \Delta\mid \exists C\in \mathscr{A}_{\Delta}, \text{ s.t. } x\in C\in \mathscr{C}\}\backslash\{\mathscr{C}_{m}\})\cup \{\mathscr{C}^{-}_{m}\mid \exists C\in \mathscr{A}_{\mathscr{C}^{-}_{m}}, \text{ s.t. } x\in C\in \mathscr{C}^{-}_{m}\}$ for $x\in U$.
For convenience, we denote $\cup\mathscr{A}_{\mathscr{C}_{m}}=\cup\{C\in \mathscr{C}_{m}\mid  \exists D_{i}\in \mathscr{D}, \text{ s.t. } C \subseteq D_{i}\}$ and
$\cup\mathscr{A}_{\mathscr{C}^{-}_{m}}=\cup\{ C\in \mathscr{C}^{-}_{m}\mid \exists D_{i}\in \mathscr{D}, \text{ s.t. } C \subseteq D_{i}\}$.
Obviously, we get $\cup\mathscr{A}_{\mathscr{C}^{-}_{m}}$ $\subseteq$ $\cup\mathscr{A}_{\mathscr{C}_{m}}$,
so we get $r^{-}(x)=(r(x)\backslash\{\mathscr{C}_{m}\})\cup \{\mathscr{C}^{-}_{m}\}$
and $r^{-}(y)=r(y)\backslash\{\mathscr{C}_{m}\}$ for $x\in \cup\mathscr{A}_{\mathscr{C}^{-}_{m}}$ and $y\notin \cup\mathscr{A}_{\mathscr{C}^{-}_{m}}$, respectively. Therefore, we have
\makeatother $$r^{-}(x)=\left\{
\begin{array}{ccc}
(r(x)\backslash\{\mathscr{C}_{m}\})\cup \{\mathscr{C}^{-}_{m}\},&{\rm if}& x\in \cup\mathscr{A}_{\mathscr{C}^{-}_{m}};\\\\
r(x)\backslash\{\mathscr{C}_{m}\},&{\rm }& otherwise.
\end{array}
\right. \Box$$

\begin{theorem}
Let $(U,\Delta, \mathscr{D})$ and $(U,\Delta^{-},\mathscr{D})$ be covering decision information systems, where $U=\{x_{1},x_{2},$ $...,x_{n}\}$, $\Delta=\{\mathscr{C}_{1},\mathscr{C}_{2},...,\mathscr{C}_{m-1},\mathscr{C}_{m}\}$, $\Delta^{-}=\{\mathscr{C}_{1},\mathscr{C}_{2},...,\mathscr{C}_{m-1},\mathscr{C}^{-}_{m}\}$,
$\mathscr{D}=\{D_{1},D_{2},...,D_{k}\}$,
$\lozenge{\mathscr{R}}(U,\Delta,\mathscr{D})=\{\Delta_{i}\mid \mathscr{C}_{m}\notin\Delta_{i}\in \mathscr{R}(U,\Delta,\mathscr{D})\},$
$\blacktriangle f(U,\Delta,\mathscr{D})=\{\mathscr{C}^{-}_{m}\}\bigwedge (\bigwedge_{x\in POS_{\cup\Delta}(\mathscr{D})\wedge x\notin \cup\mathscr{A}_{\mathscr{C}^{-}_{m}}}$ $\bigvee r^{-}(x))=\bigvee^{l}_{i=1}\{\bigwedge \Delta'_{i}\mid \Delta'_{i}\subseteq\Delta\}$, and $\blacktriangle\mathscr{R}(U,\Delta,\mathscr{D})=\{\Delta'_{j}\mid \overline{\exists}\Delta_{i}\in \mathscr{R}(U,\Delta,\mathscr{D}), \text{ s.t. } \Delta_{i}\subset\Delta'_{j},1\leq j\leq l\}$. If $POS_{\cup\Delta^{-}}(\mathscr{D})=POS_{\cup\Delta}(\mathscr{D})$, then  $\mathscr{R}(U,\Delta^{-},\mathscr{D})=\lozenge{\mathscr{R}}(U,\Delta,\mathscr{D})\cup (\blacktriangle\mathscr{R}(U,\Delta,\mathscr{D}))$.
\end{theorem}

\noindent\textbf{Proof:} The proof is similar to Theorem 3.5. $\Box$

Theorem 4.4 shows how to construct $\mathscr{R}(U,\Delta^{-},\mathscr{D})$ of $(U,\Delta^{-},\mathscr{D})$ based on $\mathscr{R}(U,\Delta,\mathscr{D})$ of $(U,\Delta,\mathscr{D})$ when $POS_{\cup\Delta}(\mathscr{D})=POS_{\cup\Delta^{-}}(\mathscr{D})$.

\begin{theorem}
Let $(U,\Delta,\mathscr{D})$ and $(U,\Delta^{-},\mathscr{D})$ be covering decision information systems, where $U=\{x_{1},x_{2},$ $...,x_{n}\}$, $\Delta=\{\mathscr{C}_{1},\mathscr{C}_{2},...,\mathscr{C}_{m-1},\mathscr{C}_{m}\}$,  $\Delta^{-}=\{\mathscr{C}_{1},\mathscr{C}_{2},...,\mathscr{C}_{m-1},\mathscr{C}^{-}_{m}\}$,
and $\mathscr{D}=\{D_{1},D_{2},...,D_{k}\}$. If $POS_{\cup\Delta}(\mathscr{D}) \neq POS_{\cup\Delta^{-}}(\mathscr{D})$, then we have
$r(x)=\{\mathscr{C}_{m}\}$ and $r^{-}(x)=\emptyset$ for $x\in POS_{\cup\Delta}(\mathscr{D})\backslash POS_{\cup\Delta^{-}}(\mathscr{D})$.
\end{theorem}

\noindent\textbf{Proof:} The proof is straightforward by Definition 2.8. $\Box$

\begin{theorem}
Let $(U,\Delta, \mathscr{D})$ and $(U,\Delta^{-}, \mathscr{D})$ be covering decision information systems, where $U=\{x_{1},x_{2},$ $...,x_{n}\}$, $\Delta=\{\mathscr{C}_{1},\mathscr{C}_{2},...,\mathscr{C}_{m-1},\mathscr{C}_{m}\}$, $\Delta^{-}=\{\mathscr{C}_{1},\mathscr{C}_{2},...,\mathscr{C}_{m-1},\mathscr{C}^{-}_{m}\}$, $\mathscr{D}=\{D_{1},D_{2},...,D_{k}\}$, $\blacktriangle f(U,\Delta,\mathscr{D})=\bigwedge (\bigwedge_{x\in POS_{\cup\Delta^{-}}(\mathscr{D})}$ $\bigvee r^{-}(x))=\bigvee^{l}_{i=1}\{\bigwedge \Delta'_{i}\mid \Delta'_{i}\subseteq\Delta\}$, and $\blacktriangle\mathscr{R}(U,\Delta,\mathscr{D})=\{\Delta'_{j}\mid 1\leq j\leq l\}$. If $POS_{\cup\Delta^{-}}(\mathscr{D})\neq POS_{\cup\Delta}(\mathscr{D})$, then $\mathscr{R}(U,\Delta^{-},\mathscr{D})=\blacktriangle\mathscr{R}(U,\Delta,\mathscr{D})$.
\end{theorem}

\noindent\textbf{Proof:} The proof is straightforward by Definition 2.8. $\Box$

\begin{algorithm}Let $(U,\Delta, \mathscr{D})$ and $(U,\Delta^{-}, \mathscr{D})$ be covering decision information systems, where $U=\{x_{1},x_{2},$ $...,x_{n}\}$, $\Delta=\{\mathscr{C}_{1},\mathscr{C}_{2},...,\mathscr{C}_{m-1},\mathscr{C}_{m}\}$, $\Delta^{-}=\{\mathscr{C}_{1},\mathscr{C}_{2},...,\mathscr{C}_{m-1},\mathscr{C}^{-}_{m}\}$,
$\mathscr{D}=\{D_{1},D_{2},...,D_{k}\}$. Then

Step 1: Input $(U,\Delta^{-},\mathscr{D})$;

Step 2: Construct $POS_{\cup\Delta^{-}}(\mathscr{D})$;

Step 3: Compute $R(U,\Delta^{-},\mathscr{D})=\{r^{-}(x)\mid x\in POS_{\cup\Delta^{-}}(\mathscr{D})\}$, where
\begin{eqnarray*}
r^{-}(x)=\left\{
\begin{array}{ccc}
(r(x)\backslash\{\mathscr{C}_{m}\})\cup \{\mathscr{C}^{-}_{m}\},&{\rm if}& x\in \cup\mathscr{A}_{\mathscr{C}^{-}_{m}};\\\\
r(x)\backslash\{\mathscr{C}_{m}\},&{\rm }& otherwise.
\end{array}
\right.
\end{eqnarray*}

Step 4: Construct $\lozenge{\mathscr{R}}(U,\Delta,\mathscr{D})=\{\Delta_{i}\mid \mathscr{C}_{m}\notin\Delta_{i}\in \mathscr{R}(U,\Delta,\mathscr{D})\}$;

Step 5: Construct
\begin{eqnarray*}
\blacktriangle f(U,\Delta,\mathscr{D})=\left\{
\begin{array}{ccc}
(\{\mathscr{C}^{-}_{m}\})\bigwedge (\bigwedge_{x\notin \cup\mathscr{A}_{\mathscr{C}^{-}_{m}}}\bigvee r^{-}(x)),&{\rm if}& POS_{\cup\Delta^{-}}(\mathscr{D})= POS_{\cup\Delta}(\mathscr{D});\\\\
\bigwedge_{x\in POS_{\cup\Delta^{-}}(\mathscr{D})}\bigvee r^{-}(x),&{\rm }& POS_{\cup\Delta^{-}}(\mathscr{D})\neq POS_{\cup\Delta}(\mathscr{D}).
\end{array}
\right.
\end{eqnarray*}

Step 6: Compute $\blacktriangle\mathscr{R}(U,\Delta,\mathscr{D})=\{\Delta'_{j}\mid \overline{\exists}\Delta_{i}\in \mathscr{R}(U,\Delta,\mathscr{D}), \text{ s.t. } \Delta_{i}\subset\Delta'_{j},1\leq j\leq l\}$;

Step 7: Output $\mathscr{R}(U,\Delta^{-},\mathscr{D})=\lozenge{\mathscr{R}}(U,\Delta,\mathscr{D})\cup (\blacktriangle\mathscr{R}(U,\Delta,\mathscr{D}))$.
\end{algorithm}

\begin{example}(Continuation from Example 2.10)
$(1)$ Firstly, by Definition 2.8, we have
$r^{-}(x_{1})=\{\mathscr{C}_{1},\mathscr{C}_{4}\},
$ $r^{-}(x_{2})$ $=\{\mathscr{C}_{1},\mathscr{C}_{4}\},
r^{-}(x_{3})=\{\mathscr{C}_{1},\mathscr{C}_{2},\mathscr{C}_{3},\mathscr{C}_{4}\},
r^{-}(x_{4})=\{\mathscr{C}_{1},\mathscr{C}_{2},\mathscr{C}_{3},\mathscr{C}_{4}\},
r^{-}(x_{5})=\{\mathscr{C}_{1},\mathscr{C}_{2},\mathscr{C}_{3},\mathscr{C}_{4}\},
r^{-}(x_{6})=\{\mathscr{C}_{1},$ $\mathscr{C}_{2},\mathscr{C}_{3},\mathscr{C}_{4},\mathscr{C}^{-}_{5}\},$ $
r^{-}(x_{7})=\{\mathscr{C}_{2},\mathscr{C}_{3}\},$ and $
r^{-}(x_{8})=\{\mathscr{C}_{2},\mathscr{C}_{3}\}$. It implies that
$R(U,\Delta^{-},\mathscr{D})=\{\{\mathscr{C}_{1},\mathscr{C}_{4}\},
\{\mathscr{C}_{1},$ $\mathscr{C}_{2},\mathscr{C}_{3},\mathscr{C}_{4}\},\{\mathscr{C}_{1},\mathscr{C}_{2},\mathscr{C}_{3},$ $\mathscr{C}_{4},\mathscr{C}^{-}_{5}\},\{\mathscr{C}_{2},\mathscr{C}_{3}\}\}.$
By Definition 2.8, we get
\begin{eqnarray*}
f(U,\Delta^{-},\mathscr{D})
&=&(\mathscr{C}_{1}\vee\mathscr{C}_{4})\wedge(\mathscr{C}_{1}\vee\mathscr{C}_{2}\vee\mathscr{C}_{3}\vee\mathscr{C}_{4})\wedge
(\mathscr{C}_{1}\vee\mathscr{C}_{2}\vee\mathscr{C}_{3}\vee\mathscr{C}_{4}\vee\mathscr{C}^{-}_{5})\wedge(\mathscr{C}_{2}\vee\mathscr{C}_{3})\\
&=&(\mathscr{C}_{1}\vee\mathscr{C}_{4})\wedge(\mathscr{C}_{2}\vee\mathscr{C}_{3})\\
&=&(\mathscr{C}_{1}\wedge\mathscr{C}_{2})\vee(\mathscr{C}_{1}\wedge\mathscr{C}_{3})
\vee(\mathscr{C}_{2}\wedge\mathscr{C}_{4})\vee(\mathscr{C}_{3}\wedge\mathscr{C}_{4}).
\end{eqnarray*}
Therefore, we have $\mathscr{R}(U,\Delta^{-},\mathscr{D})=\{\{\mathscr{C}_{1},\mathscr{C}_{2}\}, \{\mathscr{C}_{1},\mathscr{C}_{3}\}, \{\mathscr{C}_{2},\mathscr{C}_{4}\},\{\mathscr{C}_{3},\mathscr{C}_{4}\}\}.$

Secondly, according to Theorem 4.4, we get
\begin{eqnarray*}
\lozenge{\mathscr{R}}(U,\Delta,\mathscr{D})&=&\{\{\mathscr{C}_{1},\mathscr{C}_{2}\}, \{\mathscr{C}_{1},\mathscr{C}_{3}\}, \{\mathscr{C}_{2},\mathscr{C}_{4}\},
\{\mathscr{C}_{3},\mathscr{C}_{4}\}\}.\\
\blacktriangle
f(U,\Delta,\mathscr{D})&=&\mathscr{C}^{-}_{5}\wedge(\mathscr{C}_{1}\vee\mathscr{C}_{4})\wedge(\mathscr{C}_{1}\vee\mathscr{C}_{2}\vee\mathscr{C}_{3}\vee\mathscr{C}_{4})
\wedge(\mathscr{C}_{2}\vee\mathscr{C}_{3})\\
&=&\mathscr{C}^{-}_{5}\wedge(\mathscr{C}_{1}\vee\mathscr{C}_{4})\wedge(\mathscr{C}_{2}\vee\mathscr{C}_{3})\\
&=&(\mathscr{C}_{1}\wedge\mathscr{C}_{2}\wedge\mathscr{C}_{5}^{-})\vee(\mathscr{C}_{1}\wedge\mathscr{C}_{3}\wedge\mathscr{C}_{5}^{-})
\vee(\mathscr{C}_{2}\wedge\mathscr{C}_{4}\wedge\mathscr{C}_{5}^{-})\vee(\mathscr{C}_{3}\wedge\mathscr{C}_{4}\wedge\mathscr{C}_{5}^{-}).
\end{eqnarray*}
It follows that $\blacktriangle\mathscr{R}(U,\Delta,\mathscr{D})= \emptyset.$
Therefore, $\mathscr{R}(U,\Delta^{-},\mathscr{D})=\{\{\mathscr{C}_{1},\mathscr{C}_{2}\}, \{\mathscr{C}_{1},\mathscr{C}_{3}\}, \{\mathscr{C}_{2},\mathscr{C}_{4}\}, \{\mathscr{C}_{3},\mathscr{C}_{4}\}\}.$

$(2)$ Firstly, by Definition 2.8, we get
$r^{-}(x_{1})=\{\mathscr{C}_{2},\mathscr{C}_{4}\},
r^{-}(x_{2})$ $=\emptyset,
r^{-}(x_{3})=\emptyset,
r^{-}(x_{4})=\{\mathscr{C}_{2},\mathscr{C}_{3},\mathscr{C}_{4}\},$ $
r^{-}(x_{5})=\{\mathscr{C}_{2},\mathscr{C}_{3},\mathscr{C}_{4}\},
r^{-}(x_{6})=\{\mathscr{C}_{1},\mathscr{C}^{-}_{5}\},
r^{-}(x_{7})=\{\mathscr{C}_{1},\mathscr{C}_{3},\mathscr{C}^{-}_{5}\},$ and $
r^{-}(x_{8})=\{\mathscr{C}_{1},\mathscr{C}_{3},\mathscr{C}^{-}_{5}\}.$
It follows that $R(U,\Delta^{-},\mathscr{D})=\{\{\mathscr{C}_{2},\mathscr{C}_{4}\},
\{\mathscr{C}_{2},\mathscr{C}_{3},\mathscr{C}_{4}\},\{\mathscr{C}_{1},\mathscr{C}^{-}_{5}\},\{\mathscr{C}_{1},\mathscr{C}_{3},\mathscr{C}^{-}_{5}\}\}$.
By Definition 2.8, we obtain
\begin{eqnarray*}
f(U,\Delta^{-},\mathscr{D})
&=&(\mathscr{C}_{2}\vee\mathscr{C}_{4})\wedge(\mathscr{C}_{2}\vee\mathscr{C}_{3}\vee\mathscr{C}_{4})
\wedge(\mathscr{C}_{1}\vee\mathscr{C}^{-}_{5})\wedge(\mathscr{C}_{1}\vee\mathscr{C}_{3}\vee\mathscr{C}^{-}_{5})\\
&=&(\mathscr{C}_{2}\vee\mathscr{C}_{4})\wedge(\mathscr{C}_{1}\vee\mathscr{C}^{-}_{5})\\
&=&(\mathscr{C}_{1}\wedge\mathscr{C}_{2})\vee (\mathscr{C}_{1}\wedge\mathscr{C}_{4})\vee (\mathscr{C}_{2}\wedge\mathscr{C}^{-}_{5})\vee (\mathscr{C}_{4}\wedge\mathscr{C}^{-}_{5}).
\end{eqnarray*}
Therefore, we have $\mathscr{R}(U,\Delta^{-},\mathscr{D})=\{\{\mathscr{C}_{1},\mathscr{C}_{2}\}, \{\mathscr{C}_{1},\mathscr{C}_{4}\}, \{\mathscr{C}_{2},\mathscr{C}^{-}_{5}\}, \{\mathscr{C}_{4},\mathscr{C}^{-}_{5}\}\}.$

Secondly, according to Theorem 4.5, we have
\begin{eqnarray*}
\lozenge{\mathscr{R}}(U,\Delta,\mathscr{D})&=&\{\{\mathscr{C}_{1},\mathscr{C}_{2}\}, \{\mathscr{C}_{1},\mathscr{C}_{4}\}\}.\\
\blacktriangle f(U,\Delta,\mathscr{D})&=&\mathscr{C}^{-}_{5}\wedge(\mathscr{C}_{2}\vee\mathscr{C}_{4})\wedge(\mathscr{C}_{2}\vee\mathscr{C}_{3}\vee\mathscr{C}_{4})\\
&=& (\mathscr{C}_{2}\wedge\mathscr{C}^{-}_{5})\vee(\mathscr{C}_{4}\wedge\mathscr{C}^{-}_{5}).
\end{eqnarray*}
It implies that $\blacktriangle
\mathscr{R}(U,\Delta,\mathscr{D})=\{\{\mathscr{C}_{2},\mathscr{C}^{-}_{5}\},\{\mathscr{C}_{4},\mathscr{C}^{-}_{5}\}\}.$ Therefore, $\mathscr{R}(U,\Delta^{-},\mathscr{D})=\{\{\mathscr{C}_{1},\mathscr{C}_{2}\}, \{\mathscr{C}_{1}, \mathscr{C}_{4}\}, \{\mathscr{C}_{2},\mathscr{C}^{-}_{5}\},$ $\{\mathscr{C}_{4},\mathscr{C}^{-}_{5}\}\}$.
\end{example}

Example 4.8 shows how to compute attribute reducts in dynamic covering decision information systems with coarsening coverings by Algorithms 2.9 and 4.7, respectively. We see that Algorithm 4.7 is more effective than Algorithm 2.9 to compute attribute reducts in dynamic covering decision information systems with coarsening coverings.

Suppose $(U,\Delta^{-}, \mathscr{D})$ and $(U,\Delta, \mathscr{D})$ are covering decision information systems, we denote $SR(U,\Delta^{-},\mathscr{D})=\{r^{-}(x)\in R(U,\Delta^{-},\mathscr{D})\mid x\in POS_{\cup\Delta^{-}}(\mathscr{D})\wedge (\forall y\in POS_{\cup\Delta^{+}}(\mathscr{D}), r^{-}(y)\nsubseteq r^{-}(x), r^{-}(y)\in R(U,\Delta^{-},\mathscr{D}))\}$, and $\|\mathscr{C}\|$ denotes the number of times for a covering $\mathscr{C}$ appeared in $SR(U,\Delta^{-},\mathscr{D})$.

\begin{algorithm}(Heuristic Algorithm of Computing a Reduct of $(U,\Delta^{-},\mathscr{D})$)(IHVC)

Step 1: Input $(U,\Delta^{-},\mathscr{D})$;

Step 2: Construct $POS_{\cup\Delta^{-}}(\mathscr{D})$;

Step 3:  Compute $R(U,\Delta^{-},\mathscr{D})=\{r^{-}(x)\mid x\in POS_{\cup\Delta^{-}}(\mathscr{D})\}$, where
\begin{eqnarray*}
r^{-}(x)=\left\{
\begin{array}{ccc}
(r(x)\backslash\{\mathscr{C}_{m}\})\cup \{\mathscr{C}^{-}_{m}\},&{\rm if}& x\in \cup\mathscr{A}_{\mathscr{C}^{-}_{m}};\\\\
r(x)\backslash\{\mathscr{C}_{m}\},&{\rm }& otherwise.
\end{array}
\right.
\end{eqnarray*}

Step 4: Construct a reduct $\bigtriangleup^{\ast-}=\{\mathscr{C}_{i_{1}},\mathscr{C}_{i_{2}},...,\mathscr{C}_{i_{j}}\}$, where
$SR_{1}(U,\Delta^{-},\mathscr{D})=SR(U,\Delta^{-},\mathscr{D})$, $\|\mathscr{C}_{i_{1}}\|=max\{\|\mathscr{C}_{i}\|\mid \mathscr{C}_{i}\in r^{-}(x)\in R_{1}(U,\Delta^{-},\mathscr{D})\}$;
$SR_{2}(U,\Delta^{-},\mathscr{D})=\{r^{-}(x)\in SR(U,\Delta^{-},\mathscr{D})\mid \mathscr{C}_{i_{1}}\notin r^{-}(x)\}$, $\|\mathscr{C}_{i_{2}}\|=max\{\|\mathscr{C}_{i}\|\mid \mathscr{C}_{i}\in r^{-}(x)\in SR_{2}(U,\Delta^{-},\mathscr{D})\}$;
$SR_{3}(U,\Delta^{-},\mathscr{D})=\{r^{-}(x)\in SR(U,\Delta^{-},\mathscr{D})\mid \mathscr{C}_{i_{1}}\notin r^{-}(x)\text{ or } \mathscr{C}_{i_{2}}\notin r^{-}(x)\}$, $\|\mathscr{C}_{i_{3}}\|=max\{\|\mathscr{C}_{i}\|\mid \mathscr{C}_{i}\in r^{-}(x)\in SR_{3}(U,\Delta^{-},\mathscr{D})\}$;
...;
$SR_{j}(U,\Delta^{-},\mathscr{D})=\{r^{-}(x)\in SR(U,\Delta^{-},\mathscr{D})\mid \mathscr{C}_{i_{1}}\notin r^{-}(x)\text{ or } \mathscr{C}_{i_{2}}\notin r^{-}(x)\text{ or }... \text{ or } \mathscr{C}_{i_{j-1}}\notin r^{-}(x)\}$, $\|\mathscr{C}_{i_{j}}\|=max\{\|\mathscr{C}_{i}\|\mid \mathscr{C}_{i}\in r^{-}(x)\in SR_{j}(U,\Delta^{-},\mathscr{D})\}$, and $SR(U,\Delta^{-},\mathscr{D})=\{r^{-}(x)\mid \exists\mathscr{C}_{i_{k}}\in r^{-}(x),1 \leq k\leq j\}$;

Step 5: Output the reduct $\bigtriangleup^{\ast-}$.
\end{algorithm}

If there exist two coverings $\mathscr{C}_{i}$  and $\mathscr{C}_{j}$ such that $\|\mathscr{C}_{i}\|=\|\mathscr{C}_{j}\|=max\{\|\mathscr{C}_{i}\|\mid \mathscr{C}_{i}\in r^{-}(x)\in SR_{k}(U,\Delta^{-},\mathscr{D})\}$, then we select $\|\mathscr{C}_{i}\|=max\{\|\mathscr{C}_{i}\|\mid \mathscr{C}_{i}\in r^{-}(x)\in SR_{k}(U,\Delta^{-},\mathscr{D})\}$ or $\|\mathscr{C}_{j}\|=max\{\|\mathscr{C}_{i}\|\mid \mathscr{C}_{i}\in r^{-}(x)\in SR_{k}(U,\Delta^{-},\mathscr{D})\}$.

\begin{example}(Continuation from Example 4.8)
$(1)$ In Example 4.8(1), we derive $SR(U,\Delta^{-},\mathscr{D})=\{\{\mathscr{C}_{1},\mathscr{C}_{4}\},$ $
\{\mathscr{C}_{2},\mathscr{C}_{3}\}\}.$
By Algorithm 4.9, firstly, we obtain $SR_{1}(U,\Delta^{-},\mathscr{D})=\{\{\mathscr{C}_{1},\mathscr{C}_{4}\},
\{\mathscr{C}_{2},\mathscr{C}_{3}\}\}$ and $\|\mathscr{C}_{1}\|=max\{\|\mathscr{C}_{i}\|\mid \mathscr{C}_{i}\in r^{-}(x)\in SR_{1}(U,\Delta^{-},\mathscr{D})\}$. Secondly, we have $SR_{2}(U,\Delta^{-},\mathscr{D})=\{\{\mathscr{C}_{2},\mathscr{C}_{3}\}\},$
$\|\mathscr{C}_{2}\|=max\{\|\mathscr{C}_{i}\|\mid \mathscr{C}_{i}\in r^{-}(x)\in SR_{2}(U,\Delta^{-},\mathscr{D})\}$. Finally, we get a reduct $\bigtriangleup^{\ast-}=\{\mathscr{C}_{1},\mathscr{C}_{2}\}$ of  $(U,\Delta^{-},\mathscr{D})$.

$(2)$ In Example 4.8(2), we have $SR(U,\Delta^{-},\mathscr{D})=\{\{ \mathscr{C}_{1},\mathscr{C}^{-}_{5}\},\{\mathscr{C}_{2},\mathscr{C}_{4}\}\}.$
Firstly, by Algorithm 4.9, we obtain $SR_{1}(U,\Delta^{-},\mathscr{D})=\{\{ \mathscr{C}_{1},\mathscr{C}^{-}_{5}\},\{\mathscr{C}_{2},\mathscr{C}_{4}\}\}$ and $\|\mathscr{C}_{1}\|=max\{\|\mathscr{C}_{i}\|\mid \mathscr{C}_{i}\in r^{-}(x)\in SR_{1}(U,\Delta^{-},\mathscr{D})\}$. Secondly, we get $SR_{2}(U,\Delta^{-},\mathscr{D})=\{\{\mathscr{C}_{2},\mathscr{C}_{4}\}\},$
$\|\mathscr{C}_{2}\|=max\{\|\mathscr{C}_{i}\|\mid \mathscr{C}_{i}\in r^{-}(x)\in SR_{2}(U,\Delta^{-},\mathscr{D})\}$. Finally, we obtain a reduct $\bigtriangleup^{\ast-}=\{\mathscr{C}_{1},\mathscr{C}_{2}\}$ of  $(U,\Delta^{-},\mathscr{D})$.
\end{example}

\section{Experimental results }

In this section, we employ the experimental results to demonstrate that IHVR and IHVC are feasible and efficient to perform attribute reduction of dynamic covering decision information systems with refining and coarsening coverings.

To test NIHV, IHVR and IHVC, we transform eight data sets depicted by Table 1, which are downloaded from UCI Machine Learning Repository\cite{Frank1}, into covering decision information systems. Concretely, we normalize all attribute values of these data sets into the interval $[0,1]$ and employ the neighborhood operator $N(x)=\{y|d(x,y)\leq 0.05,y\in U\}$ for $x\in U$ to derive covering decision information systems, where $U$ is the object set, $A$ is the conditional attribute set and $d(x,y)=[\sum_{c\in A} |c(x)-c(y)|^{2} ]^{\frac{1}{2}}$.
Moreover, we perform all computations
on a PC with a Intel(R) Dual-Core(TM) i7-7700K CPU $@$ 4.20 GHZ, 32
GB memory, 64-bit Windows 10 and 64-bit
Matlab R2016a.

\begin{table}[H]\footnotesize
\caption{Data sets downloaded from UCI Machine Learning Repository.
} \tabcolsep0.16in
\begin{tabular}{ccccc}
\hline
No.&Name&Samples&Conditional Attributes &Decision Attribute\\\hline
1  &Wine  & 178 &13& 1\\
2  &Breast Cancer Wisconsin(wdbc)  & 569 &30& 1\\
3  &Seismic-Bumps  & 2584 &18& 1\\
4  &Abalone  & 4177 &8& 1\\
5  &Car Evaluation  & 1728 &6& 1\\
6  &Chess (King-Rook vs. King-Pawn)  & 3196 &36& 1\\
7  &Optical Recognition of Handwritten Digits  & 5620 &64& 1\\
8  &Letter Recognition  & 20000 &16& 1\\
\hline
\end{tabular}
\end{table}

\subsection{Stability of NIHV, IHVR and IHVC}

In this section, we employ the experimental results to demonstrate the stability of NIHV, IHVR and IHVC for attribute reduction of dynamic covering decision information systems.

Firstly, to test NIHV and IHVR, we derive the covering decision information systems  $\{(U_{i},\Delta_{i},$ $\mathscr{D}_{i})\mid 1\leq i\leq 8\}$ by transforming data sets in Table 1 and obtain the dynamic covering decision information system  $(U_{i},\Delta^{+}_{i},\mathscr{D}_{i})$ by refining the last covering of $(U_{i},\Delta_{i},\mathscr{D}_{i})$, where $1\leq i\leq 8$. Concretely, we part some blocks of the last covering into smaller blocks randomly.
Especially, we get ten dynamic covering decision information systems $(U_{i1},\Delta^{+}_{i1},\mathscr{D}_{i1})$,$(U_{i2},\Delta^{+}_{i2},\mathscr{D}_{i2}),...,$ and
$(U_{i10},\Delta^{+}_{i10},\mathscr{D}_{i10})$, which contain $10\%, 20\%,...,100\%$ of objects of $U_{i}$, respectively.
Subsequently, we run NIHV and IHVR ten times on $(U_{ij},\Delta^{+}_{ij},\mathscr{D}_{ij})$, where $1\leq i\leq 8$ and $1\leq j\leq 10$.
Especially, we compute the average time and standard deviation of ten computational times for the dynamic covering decision information system $(U_{ij},\Delta^{+}_{ij},\mathscr{D}_{ij})$ and depict the results by Tables 2 and 3.

Secondly, to test NIHV and IHVC, we derive the covering decision information systems  $\{(U_{i},\Delta_{i},$ $\mathscr{D}_{i})\mid 1\leq i\leq 8\}$ by transforming data sets in Table 1 and obtain the dynamic covering decision information system  $(U_{i},\Delta^{-}_{i},\mathscr{D}_{i})$ by coarsening the last covering of $(U_{i},\Delta_{i},\mathscr{D}_{i})$, where $1\leq i\leq 8$. Concretely, we combine some blocks of the last covering into large blocks randomly.
Especially, we get ten dynamic covering decision information systems $(U_{i1},\Delta^{-}_{i1},\mathscr{D}_{i1})$,$(U_{i2},\Delta^{-}_{i2},\mathscr{D}_{i2}),...,$ and
$(U_{i10},\Delta^{-}_{i10},\mathscr{D}_{i10})$, which contain $10\%, 20\%,...,100\%$ of objects of $U_{i}$, respectively.
Subsequently, we run NIHV and IHVC ten times on $(U_{ij},\Delta^{-}_{ij},\mathscr{D}_{ij})$, where $1\leq i\leq 8$ and $1\leq j\leq 10$.
Especially, we compute the average time and standard deviation of ten computational times for the dynamic covering decision information systems $(U_{ij},\Delta^{-}_{ij},\mathscr{D}_{ij})$ and depict the results by Tables 4 and 5.

\begin{table}[H]\footnotesize
\caption{Computational times using NIHV and IHVR} \tabcolsep0.06in
\label{bigtable0}
\begin{center}
\begin{tabular}{ c c c c c c c c c c c c}
\hline
No$\setminus$ t(s) & Algo. & 10\% & 20\% & 30\% & 40\% & 50\% & 60\% & 70\% & 80\% & 90\% & 100\%  \\ \hline
\multirow{2}*{$(U_{1},\Delta^{+}_{1},\mathscr{D}_{1})$} & NIHV & 0.0142 & 0.0320 & 0.0561 & 0.1050 & 0.1663 & 0.2155 & 0.2713 & 0.3572 & 0.4461 & 0.5357 \\
& IHVR & 0.0068 & 0.0135 & 0.0199 & 0.0355 & 0.0597 & 0.0662 & 0.0735 & 0.0942 & 0.1219 & 0.1419 \\
\hline
\multirow{2}*{$(U_{2},\Delta^{+}_{2},\mathscr{D}_{2})$} & NIHV & 0.1846 & 0.6007 & 1.2797 & 2.1838 & 3.3089 & 4.7283 & 6.2382 & 8.0147 & 9.9181 & 12.1967 \\
& IHVR & 0.0545 & 0.1669 & 0.3160 & 0.4909 & 0.7090 & 0.9946 & 1.2518 & 1.5665 & 1.8482 & 2.2205 \\
\hline
\multirow{2}*{$(U_{3},\Delta^{+}_{3},\mathscr{D}_{3})$} & NIHV & 0.3692 & 1.4264 & 2.7412 & 4.0880 & 5.5206 & 7.3213 & 9.3704 & 11.3816 & 13.5673 & 15.7106 \\
& IHVR & 0.0361 & 0.0851 & 0.1414 & 0.1872 & 0.2346 & 0.2650 & 0.3153 & 0.3717 & 0.4353 & 0.5101 \\
\hline
\multirow{2}*{$(U_{4},\Delta^{+}_{4},\mathscr{D}_{4})$} & NIHV & 1.3725 & 4.4995 & 9.3057 & 15.3776 & 22.8759 & 30.8318 & 40.4298 & 50.3114 & 60.7472 & 71.7886 \\
& IHVR & 0.3068 & 0.7897 & 1.6550 & 2.7504 & 4.2078 & 5.3572 & 7.1887 & 8.8296 & 10.7476 & 12.6319 \\
\hline
\multirow{2}*{$(U_{5},\Delta^{+}_{5},\mathscr{D}_{5})$} & NIHV & 0.0245 & 0.0651 & 0.1087 & 0.1549 & 0.2065 & 0.2622 & 0.3227 & 0.3698 & 0.4335 & 0.4999 \\
& IHVR & 0.0050 & 0.0151 & 0.0246 & 0.0337 & 0.0442 & 0.0523 & 0.0639 & 0.0732 & 0.0867 & 0.1007 \\
\hline
\multirow{2}*{$(U_{6},\Delta^{+}_{6},\mathscr{D}_{6})$} & NIHV & 0.3250 & 0.9588 & 1.7954 & 2.2586 & 2.6766 & 3.4308 & 4.3235 & 5.3157 & 6.0635 & 7.1550 \\
& IHVR & 0.0125 & 0.0312 & 0.0646 & 0.0685 & 0.0889 & 0.0984 & 0.1259 & 0.1524 & 0.1744 & 0.2047 \\
\hline
\multirow{2}*{$(U_{7},\Delta^{+}_{7},\mathscr{D}_{7})$} & NIHV & 2.5748 & 5.6880 & 9.2705 & 13.0273 & 16.7961 & 20.8991 & 25.5264 & 30.2107 & 35.1496 & 40.3688 \\
& IHVR & 0.0992 & 0.1879 & 0.2799 & 0.3753 & 0.4692 & 0.6077 & 0.7225 & 0.8666 & 0.9831 & 1.1410 \\
\hline
\multirow{2}*{$(U_{8},\Delta^{+}_{8},\mathscr{D}_{8})$} & NIHV & 2.9520 & 6.7087 & 11.1105 & 15.8891 & 20.8680 & 26.5046 & 32.0929 & 38.5874 & 45.4555 & 52.9705 \\
& IHVR & 0.2441 & 0.5405 & 0.8566 & 1.2156 & 1.5683 & 1.9666 & 2.3681 & 2.8123 & 3.3406 & 3.8437 \\
\hline
\end{tabular}
\end{center}
\end{table}



\begin{table}[H]\footnotesize
\caption{Standard deviations using NIHV and IHVR} \tabcolsep0.08in
\label{bigtable2}
\begin{center}
\begin{tabular}{ c c c c c c c c c c c c}
\hline
No $\setminus$ SD& Algo. & 10\% & 20\% & 30\% & 40\% & 50\% & 60\% & 70\% & 80\% & 90\% & 100\%  \\ \hline
\multirow{2}*{$(U_{1},\Delta^{+}_{1},\mathscr{D}_{1})$} & NIHV & 0.0035 & 0.0002 & 0.0008 & 0.0016 & 0.0009 & 0.0016 & 0.0010 & 0.0013 & 0.0010 & 0.0013 \\
& IHVR & 0.0007 & 0.0000 & 0.0004 & 0.0007 & 0.0036 & 0.0004 & 0.0004 & 0.0004 & 0.0007 & 0.0003 \\
\hline
\multirow{2}*{$(U_{2},\Delta^{+}_{2},\mathscr{D}_{2})$} & NIHV & 0.0521 & 0.0022 & 0.0043 & 0.0064 & 0.0039 & 0.0093 & 0.0340 & 0.0318 & 0.0294 & 0.0711 \\
& IHVR & 0.0001 & 0.0007 & 0.0020 & 0.0018 & 0.0034 & 0.0036 & 0.0061 & 0.0049 & 0.0058 & 0.0075 \\
\hline
\multirow{2}*{$(U_{3},\Delta^{+}_{3},\mathscr{D}_{3})$} & NIHV & 0.0018 & 0.0034 & 0.0052 & 0.0127 & 0.0108 & 0.0179 & 0.0490 & 0.0193 & 0.0173 & 0.0284 \\
& IHVR & 0.0002 & 0.0003 & 0.0006 & 0.0008 & 0.0004 & 0.0042 & 0.0039 & 0.0072 & 0.0055 & 0.0103 \\
\hline
\multirow{2}*{$(U_{4},\Delta^{+}_{4},\mathscr{D}_{4})$} & NIHV & 0.0074 & 0.0266 & 0.0227 & 0.0169 & 0.0707 & 0.1061 & 0.0605 & 0.1953 & 0.2152 & 0.2163 \\
& IHVR & 0.0011 & 0.0013 & 0.0031 & 0.0056 & 0.0103 & 0.0220 & 0.0074 & 0.0173 & 0.0242 & 0.0156 \\
\hline
\multirow{2}*{$(U_{5},\Delta^{+}_{5},\mathscr{D}_{5})$} & NIHV & 0.0001 & 0.0005 & 0.0008 & 0.0007 & 0.0008 & 0.0008 & 0.0028 & 0.0017 & 0.0017 & 0.0013 \\
& IHVR & 0.0001 & 0.0001 & 0.0001 & 0.0003 & 0.0003 & 0.0005 & 0.0002 & 0.0002 & 0.0004 & 0.0004 \\
\hline
\multirow{2}*{$(U_{6},\Delta^{+}_{6},\mathscr{D}_{6})$} & NIHV & 0.0012 & 0.0017 & 0.0024 & 0.0051 & 0.0116 & 0.0265 & 0.0320 & 0.0397 & 0.0356 & 0.0355 \\
& IHVR & 0.0001 & 0.0001 & 0.0004 & 0.0003 & 0.0003 & 0.0027 & 0.0016 & 0.0017 & 0.0018 & 0.0034 \\
\hline
\multirow{2}*{$(U_{7},\Delta^{+}_{7},\mathscr{D}_{7})$} & NIHV & 0.0060 & 0.0073 & 0.0138 & 0.0420 & 0.0388 & 0.0292 & 0.1189 & 0.0382 & 0.0365 & 0.1400 \\
& IHVR & 0.0005 & 0.0011 & 0.0016 & 0.0074 & 0.0104 & 0.0115 & 0.0138 & 0.0181 & 0.0207 & 0.0187 \\
\hline
\multirow{2}*{$(U_{8},\Delta^{+}_{8},\mathscr{D}_{8})$} & NIHV & 0.0047 & 0.0125 & 0.0220 & 0.0287 & 0.1268 & 0.0298 & 0.0295 & 0.1429 & 0.0576 & 0.1275 \\
& IHVR & 0.0014 & 0.0011 & 0.0044 & 0.0044 & 0.0049 & 0.0080 & 0.0064 & 0.0070 & 0.0098 & 0.0101 \\
\hline
\end{tabular}
\end{center}
\end{table}


From Tables 3 and 5, we see that the standard deviations of ten computational times by NIHV, IHVR and IHVR are very small, which illustrates that these algorithms are stable for computing attribute reducts of dynamic covering decision information systems. For example, from Row 2 in Table 3, we see that
$\{0.0035, 0.0002, 0.0008, 0.0016, 0.0009, 0.0016, 0.0010, 0.0013, 0.0010, 0.0013\}$ and
$\{0.0007, 0.0000, 0.0004,$ $ 0.0007, 0.0036, 0.0004, 0.0004, 0.0004, 0.0007, 0.0003\}$ are the standard deviations of computational times with NIHV and IHVR, respectively, in $\{(U_{1j},\Delta^{+}_{1j},\mathscr{D}_{1j})\mid 1\leq j\leq 10\}$; from Row 2 in Table 5, we also observe that $\{0.0001, 0.0003, 0.0003, 0.0004, 0.0208, 0.0014, 0.0016, 0.0013, 0.0012, 0.0016\}$ and
$\{0.0007, $ $0.0001, 0.0001, 0.0006, 0.0016, 0.0004, 0.0004, 0.0002, 0.0004, 0.0034\}$ are the standard deviations of computational times by NIHV and IHVC, respectively, in $\{(U_{1j},\Delta^{-}_{1j},$ $\mathscr{D}_{1j})\mid 1\leq j\leq 10\}$.
Furthermore, the standard deviations of ten computational times by IHVR and IHVC are almost smaller than those by NIHV, which demonstrates that IHVR and IHVC are more stable than NIHV for computing attribute reducts of dynamic covering decision information systems.

\noindent\textbf{Remark:} In Tables 2 and 4, $t(s)$ denotes the measure of time is
in seconds; in Tables 3 and 5, SD means the standard deviation.



\begin{table}[H]\footnotesize
\caption{Computational times using NIHV and IHVC} \tabcolsep0.06in
\label{bigtable1}
\begin{center}
\begin{tabular}{ c c c c c c c c c c c c}
\hline
No$\setminus$ t(s) & Algo. & 10\% & 20\% & 30\% & 40\% & 50\% & 60\% & 70\% & 80\% & 90\% & 100\%  \\ \hline
\multirow{2}*{$(U_{1},\Delta^{-}_{1},\mathscr{D}_{1})$} & NIHV & 0.0116 & 0.0314 & 0.0553 & 0.1029 & 0.1778 & 0.2166 & 0.2701 & 0.3510 & 0.4456 & 0.5342 \\
& IHVC & 0.0067 & 0.0131 & 0.0190 & 0.0353 & 0.0567 & 0.0687 & 0.0728 & 0.0874 & 0.1208 & 0.1432 \\
\hline
\multirow{2}*{$(U_{2},\Delta^{-}_{2},\mathscr{D}_{2})$} & NIHV & 0.1672 & 0.6017 & 1.2799 & 2.1817 & 3.2991 & 4.7194 & 6.2004 & 7.9794 & 9.9513 & 12.2012 \\
& IHVC & 0.0542 & 0.1675 & 0.3171 & 0.4927 & 0.7006 & 0.9882 & 1.2476 & 1.5693 & 1.8611 & 2.2195 \\
\hline
\multirow{2}*{$(U_{3},\Delta^{-}_{3},\mathscr{D}_{3})$} & NIHV & 0.3674 & 1.4266 & 2.7282 & 4.0706 & 5.4995 & 7.3041 & 9.3304 & 11.3418 & 13.5462 & 15.6447 \\
& IHVC & 0.0356 & 0.0855 & 0.1306 & 0.1801 & 0.2282 & 0.2522 & 0.3011 & 0.3564 & 0.4173 & 0.4860 \\
\hline
\multirow{2}*{$(U_{4},\Delta^{-}_{4},\mathscr{D}_{4})$} & NIHV & 1.3508 & 4.4415 & 9.1954 & 15.2757 & 22.5871 & 30.4979 & 39.6721 & 49.3271 & 59.9633 & 70.3547 \\
& IHVC & 0.2904 & 0.7461 & 1.5506 & 2.6326 & 3.9416 & 4.9916 & 6.5337 & 8.0904 & 9.7584 & 11.3507 \\
\hline
\multirow{2}*{$(U_{5},\Delta^{-}_{5},\mathscr{D}_{5})$} & NIHV & 0.0241 & 0.0612 & 0.1026 & 0.1492 & 0.2023 & 0.2663 & 0.3217 & 0.3730 & 0.4385 & 0.5081 \\
& IHVC & 0.0043 & 0.0114 & 0.0202 & 0.0299 & 0.0418 & 0.0583 & 0.0681 & 0.0814 & 0.0966 & 0.1151 \\
\hline
\multirow{2}*{$(U_{6},\Delta^{-}_{6},\mathscr{D}_{6})$} & NIHV & 0.3252 & 0.9643 & 1.7980 & 2.2592 & 2.6785 & 3.4523 & 4.3257 & 5.3294 & 6.1436 & 7.1852 \\
& IHVC & 0.0128 & 0.0333 & 0.0720 & 0.0771 & 0.0854 & 0.1087 & 0.1451 & 0.1749 & 0.2001 & 0.2465 \\
\hline
\multirow{2}*{$(U_{7},\Delta^{-}_{7},\mathscr{D}_{7})$} & NIHV & 2.5722 & 5.6679 & 9.2415 & 13.0057 & 16.8011 & 20.8637 & 25.3666 & 30.1462 & 35.1028 & 40.1981 \\
& IHVC & 0.0990 & 0.1834 & 0.2720 & 0.3669 & 0.4552 & 0.5776 & 0.6857 & 0.8078 & 0.9356 & 1.0832 \\
\hline
\multirow{2}*{$(U_{8},\Delta^{-}_{8},\mathscr{D}_{8})$} & NIHV & 2.9316 & 6.6747 & 11.0356 & 15.8119 & 20.6779 & 26.3517 & 31.9347 & 38.3283 & 45.4949 & 52.8093 \\
& IHVC & 0.2288 & 0.5231 & 0.8105 & 1.1971 & 1.4945 & 1.8806 & 2.2693 & 2.7013 & 3.5746 & 3.7357 \\
\hline
\end{tabular}
\end{center}
\end{table}



\begin{table}[H]\footnotesize
\caption{Standard deviations using NIHV and IHVC} \tabcolsep0.08in
\label{bigtable3}
\begin{center}
\begin{tabular}{ c c c c c c c c c c c c}
\hline
No $\setminus$ SD& Algo. & 10\% & 20\% & 30\% & 40\% & 50\% & 60\% & 70\% & 80\% & 90\% & 100\%  \\ \hline
\multirow{2}*{$(U_{1},\Delta^{-}_{1},\mathscr{D}_{1})$} & NIHV & 0.0001 & 0.0003 & 0.0003 & 0.0004 & 0.0208 & 0.0014 & 0.0016 & 0.0013 & 0.0012 & 0.0016 \\
& IHVC & 0.0007 & 0.0001 & 0.0001 & 0.0006 & 0.0016 & 0.0004 & 0.0004 & 0.0002 & 0.0004 & 0.0034 \\
\hline
\multirow{2}*{$(U_{2},\Delta^{-}_{2},\mathscr{D}_{2})$} & NIHV & 0.0003 & 0.0026 & 0.0032 & 0.0045 & 0.0060 & 0.0123 & 0.0263 & 0.0091 & 0.0120 & 0.0617 \\
& IHVC & 0.0002 & 0.0006 & 0.0009 & 0.0025 & 0.0032 & 0.0034 & 0.0056 & 0.0057 & 0.0071 & 0.0028 \\
\hline
\multirow{2}*{$(U_{3},\Delta^{-}_{3},\mathscr{D}_{3})$} & NIHV & 0.0012 & 0.0039 & 0.0040 & 0.0094 & 0.0057 & 0.0198 & 0.0151 & 0.0278 & 0.0191 & 0.0310 \\
& IHVC & 0.0007 & 0.0004 & 0.0007 & 0.0006 & 0.0007 & 0.0033 & 0.0040 & 0.0047 & 0.0080 & 0.0056 \\
\hline
\multirow{2}*{$(U_{4},\Delta^{-}_{4},\mathscr{D}_{4})$} & NIHV & 0.0027 & 0.0046 & 0.0315 & 0.0319 & 0.0258 & 0.0916 & 0.1081 & 0.0393 & 0.1319 & 0.1020 \\
& IHVC & 0.0004 & 0.0021 & 0.0074 & 0.0040 & 0.0063 & 0.0059 & 0.0105 & 0.0219 & 0.0108 & 0.0261 \\
\hline
\multirow{2}*{$(U_{5},\Delta^{-}_{5},\mathscr{D}_{5})$} & NIHV & 0.0000 & 0.0015 & 0.0003 & 0.0007 & 0.0011 & 0.0011 & 0.0015 & 0.0022 & 0.0017 & 0.0021 \\
& IHVC & 0.0000 & 0.0001 & 0.0003 & 0.0001 & 0.0004 & 0.0002 & 0.0004 & 0.0003 & 0.0003 & 0.0005 \\
\hline
\multirow{2}*{$(U_{6},\Delta^{-}_{6},\mathscr{D}_{6})$} & NIHV & 0.0010 & 0.0037 & 0.0048 & 0.0043 & 0.0166 & 0.0193 & 0.0278 & 0.0338 & 0.0422 & 0.0443 \\
& IHVC & 0.0004 & 0.0001 & 0.0008 & 0.0003 & 0.0027 & 0.0044 & 0.0030 & 0.0021 & 0.0050 & 0.0063 \\
\hline
\multirow{2}*{$(U_{7},\Delta^{-}_{7},\mathscr{D}_{7})$} & NIHV & 0.0063 & 0.0090 & 0.0091 & 0.0462 & 0.0405 & 0.0295 & 0.0728 & 0.0327 & 0.1167 & 0.0657 \\
& IHVC & 0.0004 & 0.0010 & 0.0009 & 0.0080 & 0.0048 & 0.0114 & 0.0114 & 0.0134 & 0.0106 & 0.0175 \\
\hline
\multirow{2}*{$(U_{8},\Delta^{-}_{8},\mathscr{D}_{8})$} & NIHV & 0.0065 & 0.0065 & 0.0090 & 0.0170 & 0.0167 & 0.0190 & 0.1008 & 0.0477 & 0.0326 & 0.1652 \\
& IHVC & 0.0017 & 0.0008 & 0.0020 & 0.0031 & 0.0031 & 0.0039 & 0.0070 & 0.0089 & 0.0206 & 0.0188 \\
\hline
\end{tabular}
\end{center}
\end{table}

\subsection{Comparison of NIHV and IHVR}

In this section, we employ the experimental results to illustrate that IHVR is more effective than NIHV for computing attribute reducts of dynamic
covering decision information systems when refining coverings.

Firstly, we compare the running
times of IHVR with those of NIHV in dynamic
covering decision information systems when refining coverings.
From Table 2, we find the times of computing attribute reducts with NIHV and IHVR in dynamic covering decision information systems $\{(U_{ij},\Delta^{+}_{ij},\mathscr{D}_{ij})\mid 1\leq i\leq 8,1\leq j\leq 10\}$. Especially, we observe that IHVR runs faster than NIHV for computing attribute reducts of dynamic covering decision information systems $\{(U_{ij},\Delta^{+}_{ij},\mathscr{D}_{ij})\mid 1\leq i\leq 8,1\leq j\leq 10\}$. For example, from Row 2 of Table 2,
we have the computational times $\{0.0142, 0.0320, 0.0561, 0.1050, 0.1663,$ $ 0.2155, 0.2713, 0.3572, 0.4461, 0.5357\}$ and
$\{0.0068, 0.0135, 0.0199, $ $0.0355, 0.0597, 0.0662, 0.0735, 0.0942,$ $ 0.1219, 0.1419\}$ with NIHV and IHVR, respectively, in dynamic covering decision information systems  $\{(U_{1j},\Delta^{+}_{1j},\mathscr{D}_{1j})\mid 1\leq j\leq 10\}$. It is obvious that the computational times of NIHV are larger than those of IHVR in dynamic covering decision information systems $\{(U_{ij},\Delta^{+}_{ij},\mathscr{D}_{ij})\mid 1\leq i\leq 8,1\leq j\leq 10\}$.

Secondly, we employ Figure 1 to illustrate the experimental results with NIHV and IHVR in dynamic covering decision information systems $\{(U_{ij},\Delta^{+}_{ij},\mathscr{D}_{ij})\mid 1\leq i\leq 8,1\leq j\leq 10\}$, where Figure 1(i) illustrates the computational times with NIHV and IHVR in $\{(U_{ij},\Delta^{+}_{ij},\mathscr{D}_{ij})\mid 1\leq j\leq 10\}$, $-\ast-$ and $-\circ-$ denote NIHV and IHVR, respectively. From Figure 1, we see that IHVR performs faster than NIHV in $\{(U_{ij},\Delta^{+}_{ij},\mathscr{D}_{ij})\mid 1\leq i\leq 8, 1\leq j\leq 10\}$. Especially, the computational time of NIHV increases faster than IHVR with the increase of the cardinality of object set.

\begin{figure}[H]
\begin{minipage}{0.5\linewidth}
\centerline{\includegraphics[width=4.5cm]{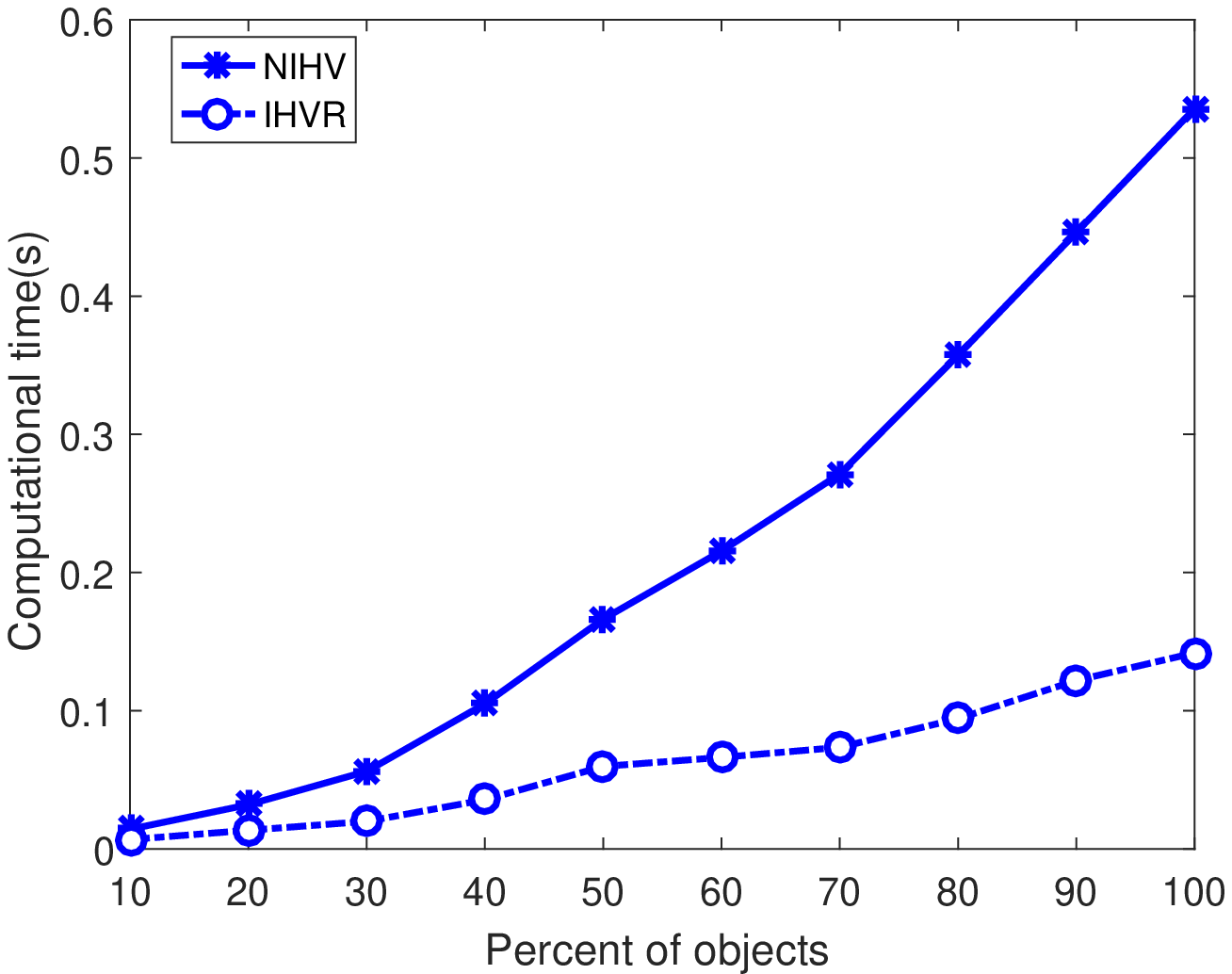}}
\centerline{($1$)}
\end{minipage}
\hfill
\begin{minipage}{0.6\linewidth}
\centerline{\includegraphics[width=4.5cm]{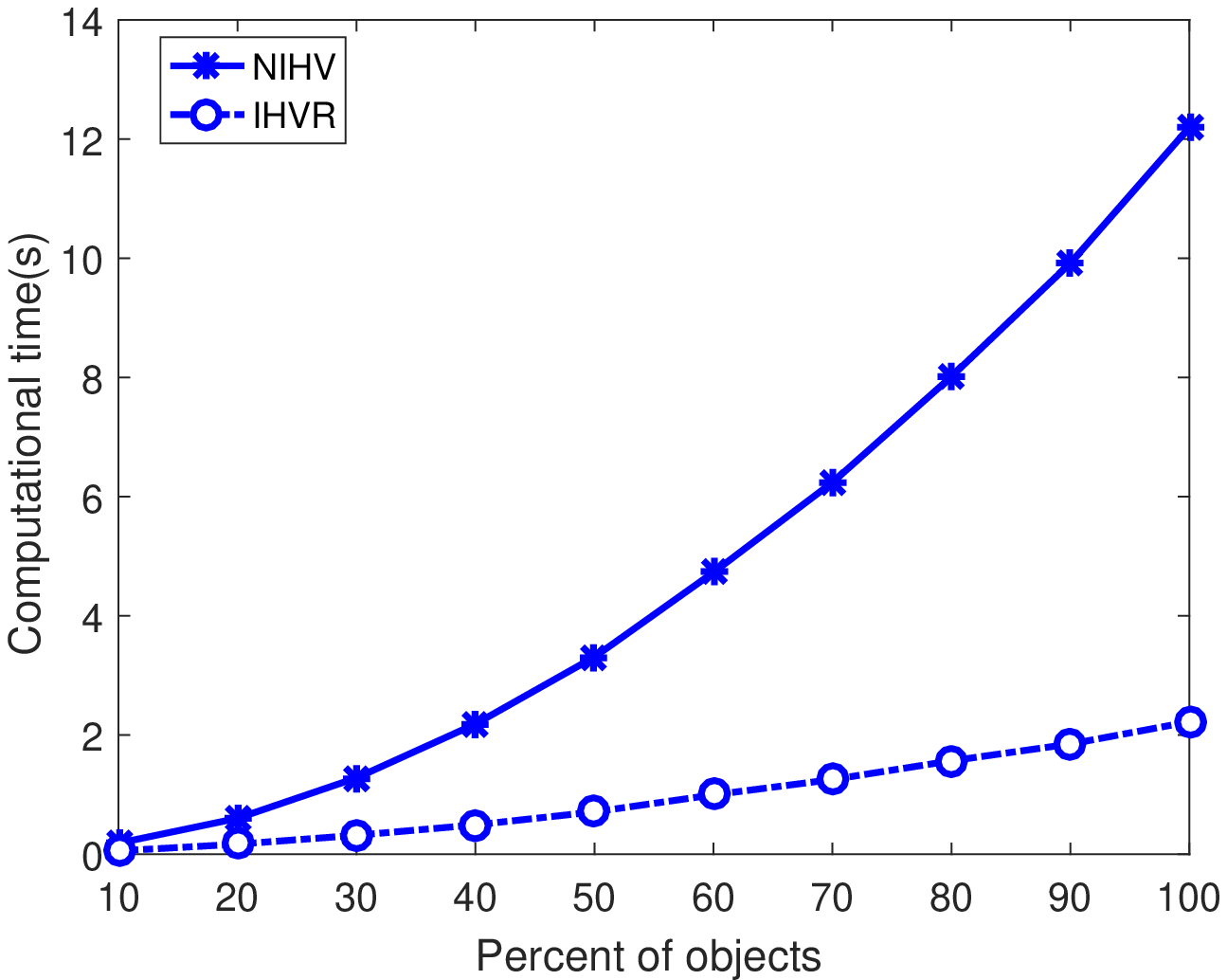}}
\centerline{($2$)}
\end{minipage}
\end{figure}

\begin{figure}[H]
\addtocounter{subfigure}{3} 
\begin{minipage}{0.5\linewidth}
\centerline{\includegraphics[width=4.5cm]{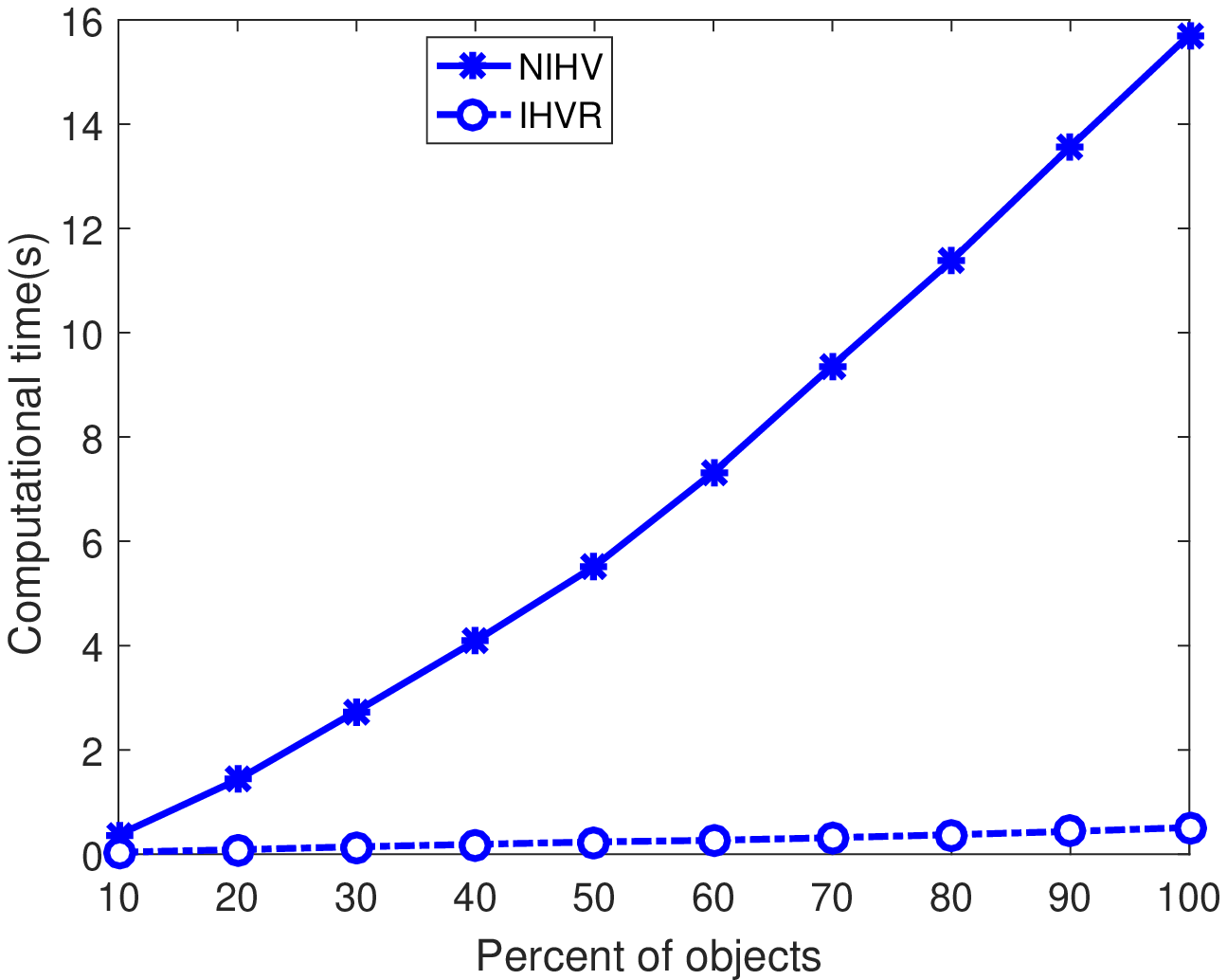}}
\centerline{($3$)}
\end{minipage}
\hfill
\begin{minipage}{0.6\linewidth}
\centerline{\includegraphics[width=4.5cm]{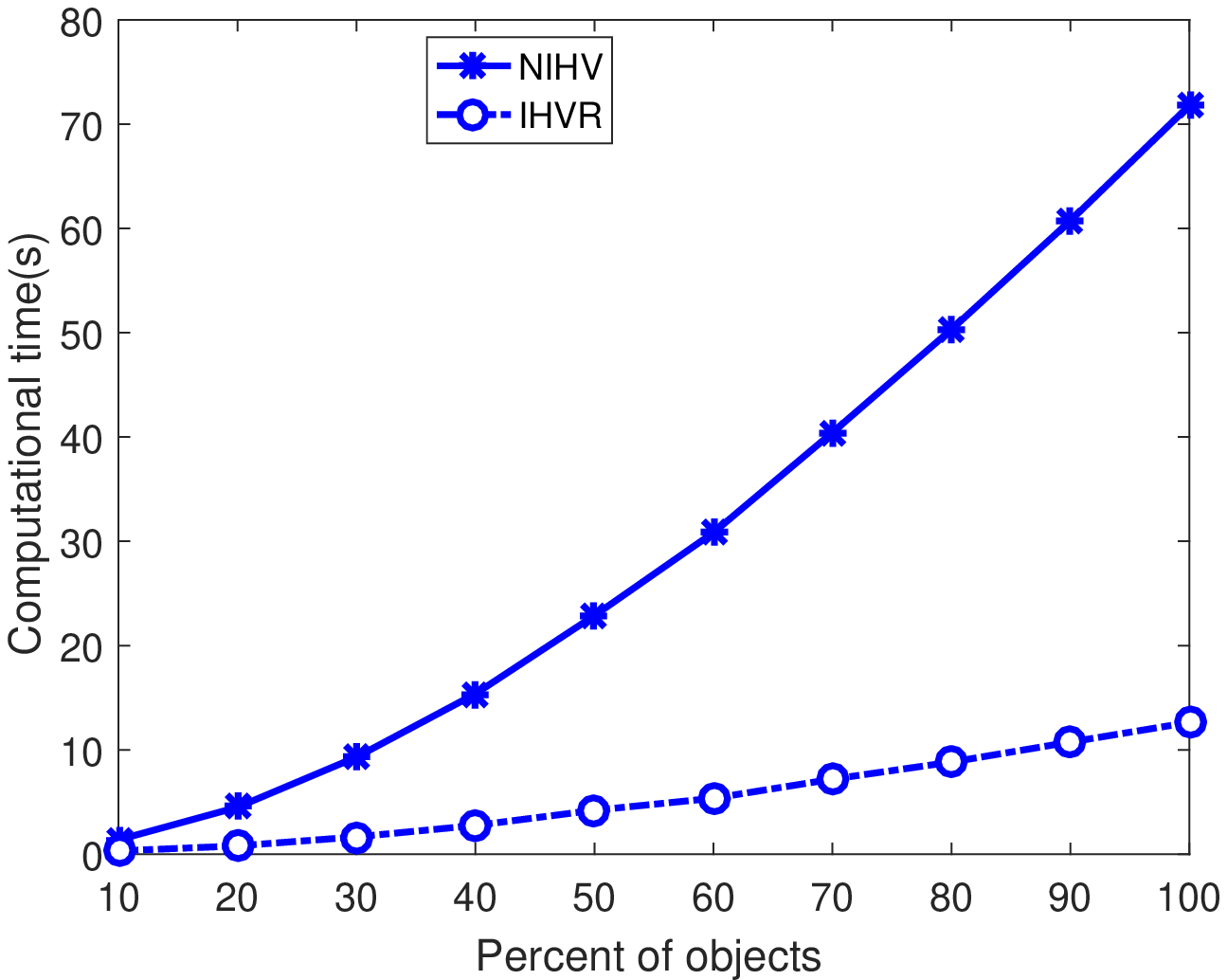}}
\centerline{($4$)}
\end{minipage}
\hfill
\end{figure}

\begin{figure}[H]
\addtocounter{subfigure}{6} 
\begin{minipage}{0.5\linewidth}
\centerline{\includegraphics[width=4.5cm]{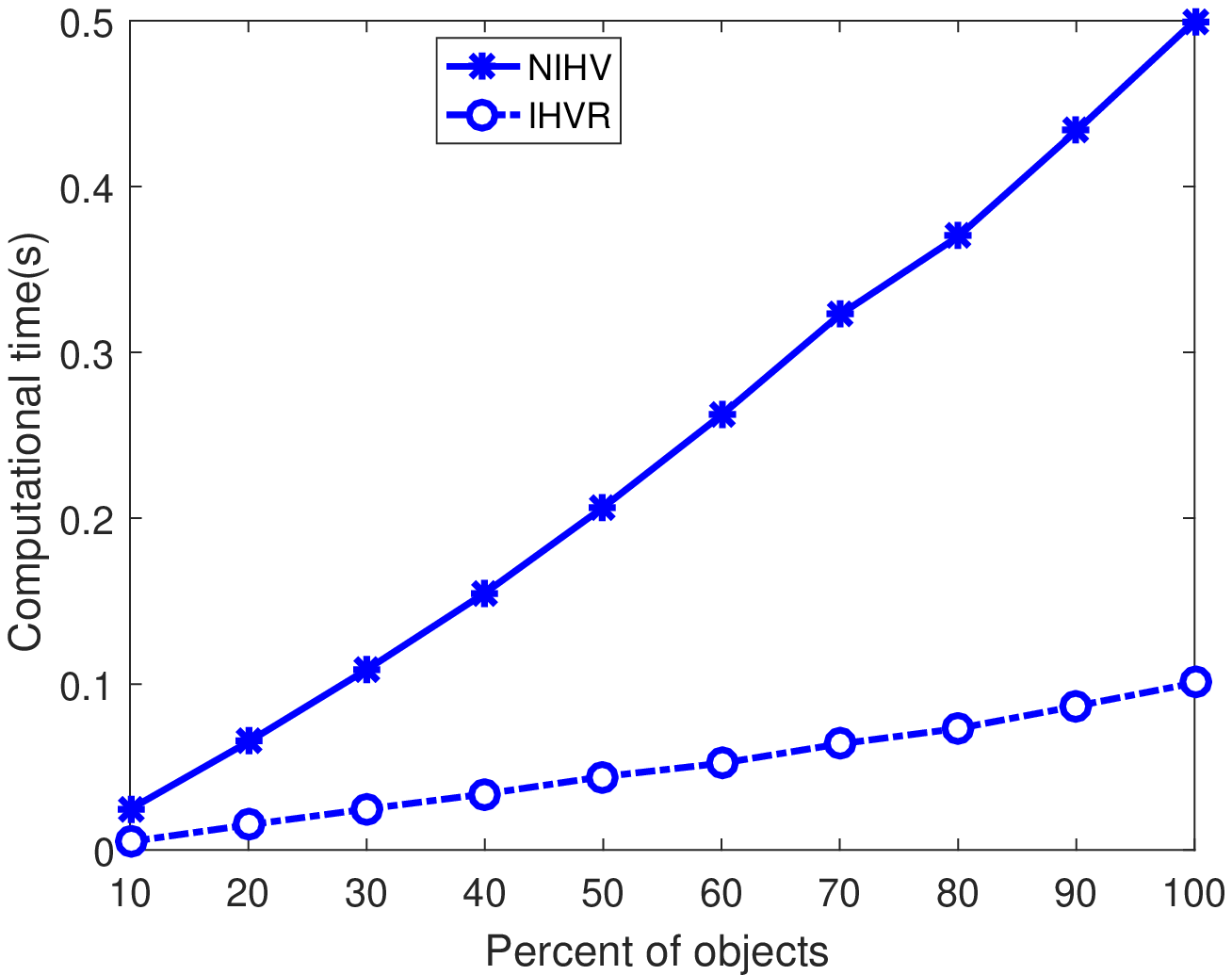}}
\centerline{($5$)}
\end{minipage}
\hfill
\begin{minipage}{0.6\linewidth}
\centerline{\includegraphics[width=4.5cm]{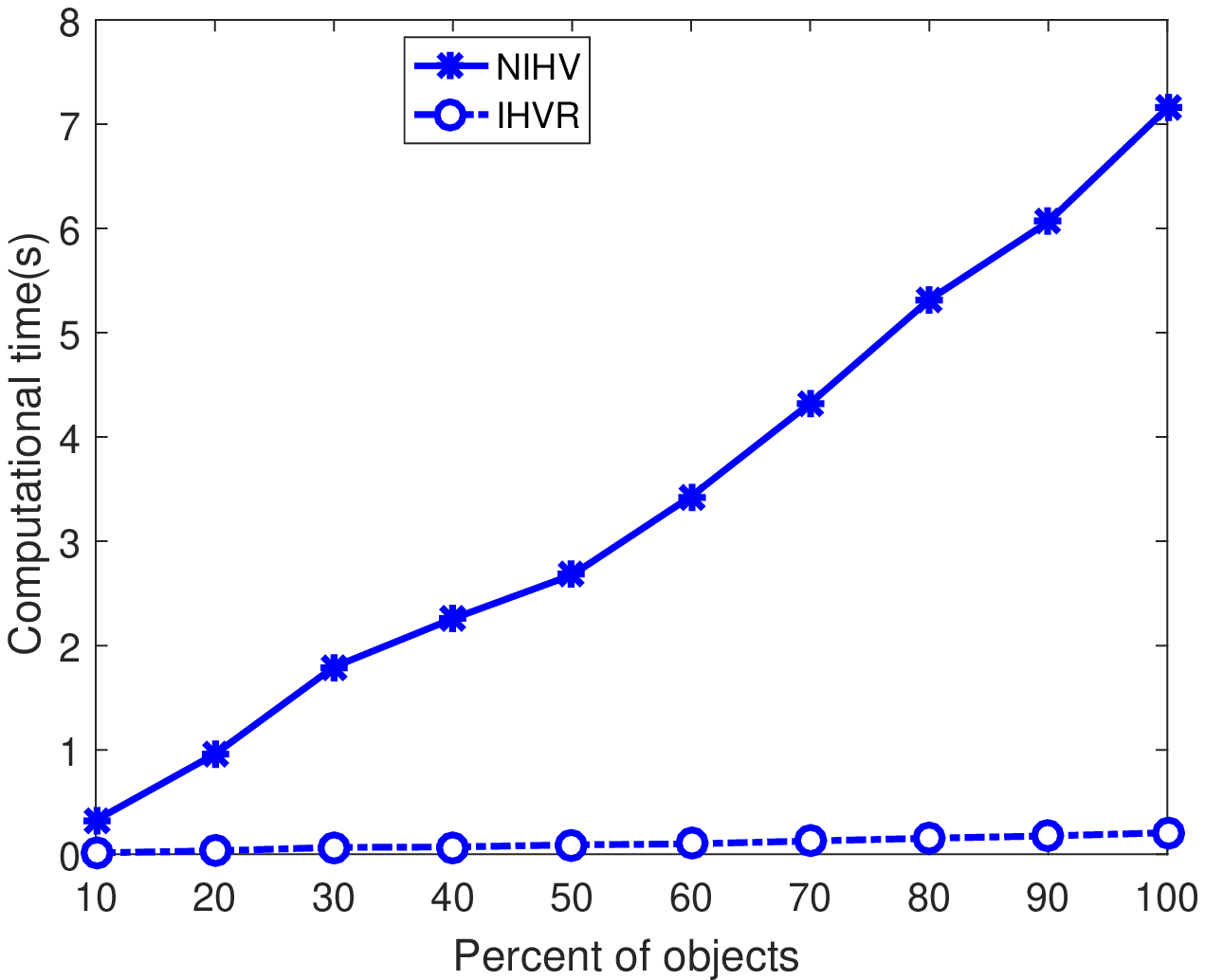}}
\centerline{($6$)}
\end{minipage}
\hfill
\end{figure}

\begin{figure}[!htp]
\begin{minipage}{0.5\linewidth}
\centerline{\includegraphics[width=4.5cm]{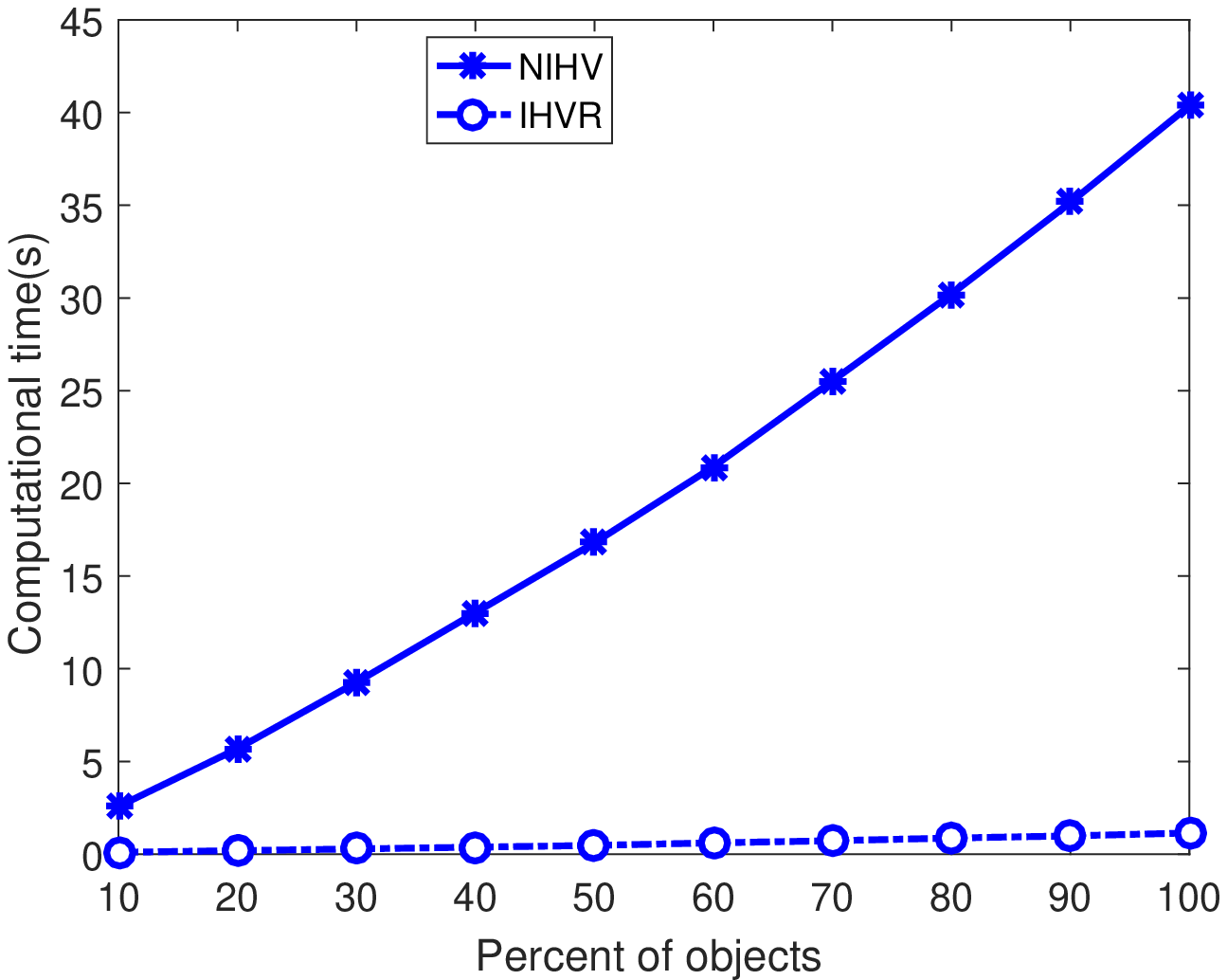}}
\centerline{($7$)}
\end{minipage}
\hfill
\begin{minipage}{0.6\linewidth}
\centerline{\includegraphics[width=4.5cm]{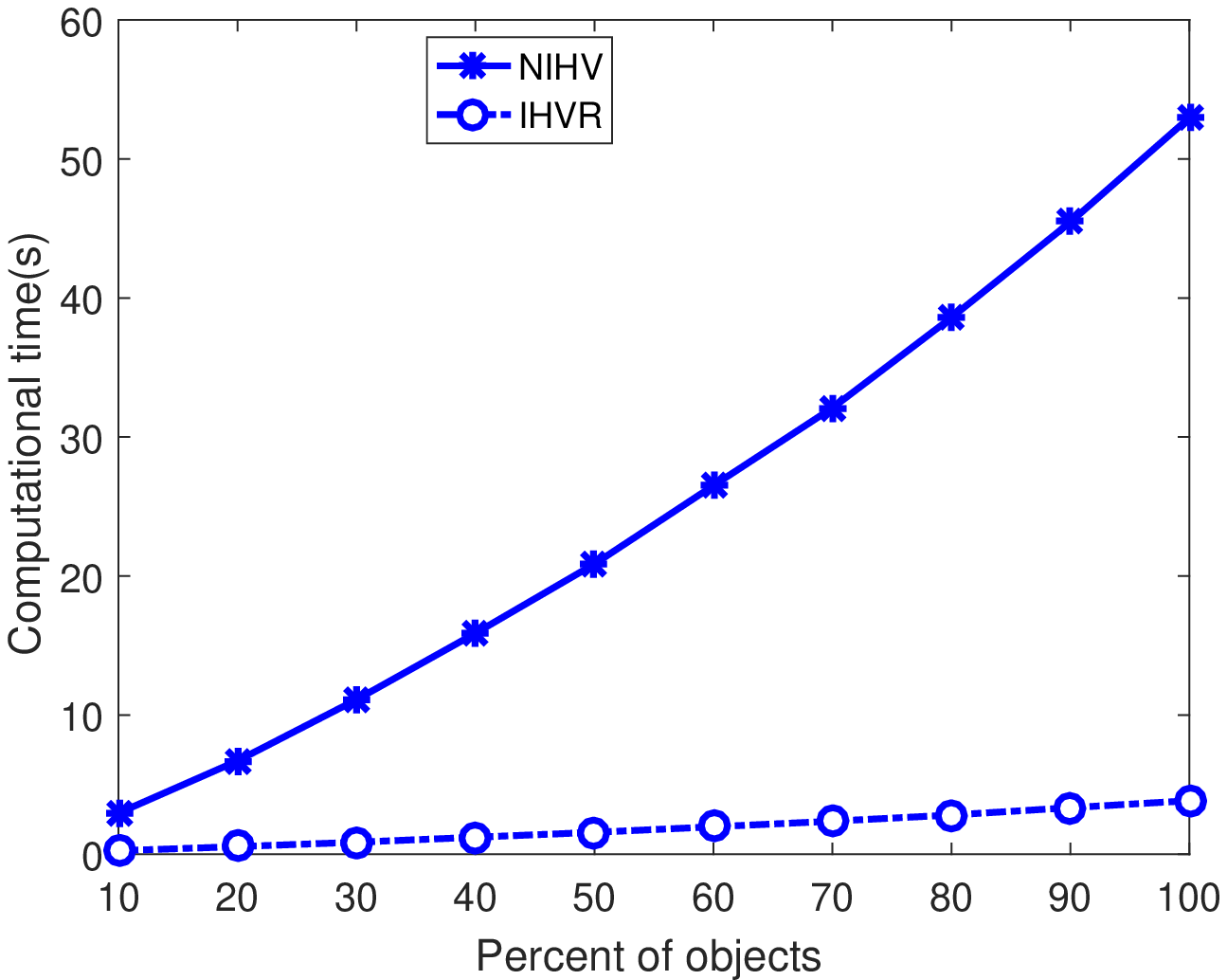}}
\centerline{($8$)}
\end{minipage}
\hfill
\caption{Computational times  using NIHV and IHVR.}
\label{figTZ_objaddall}
\end{figure}

Therefore, the experimental results in Table 2 and Figure 1 illustrate that IHVR is more efficient and feasible than NIHV for computing attribute reducts of dynamic
covering decision information systems with refining coverings.

%
%


\subsection{Comparison of NIHV and IHVC}

In this section, we employ the experimental results to illustrate that IHVC is more effective than NIHV in dynamic
covering decision information systems when coarsening coverings.

Firstly, we compare the computational
times of IHVC with those of NIHV in dynamic
covering decision information systems when coarsening coverings.
From Table 4, we see the times of computing attribute reducts with  NIHV and IHVC in dynamic covering decision information systems $\{(U_{ij},\Delta^{-}_{ij},\mathscr{D}_{ij})\mid 1\leq i\leq 8,1\leq j\leq 10\}$. Especially, we see that IHVC performs faster than NIHV in computing attribute reduct of dynamic covering decision information systems $\{(U_{1j},\Delta^{-}_{1j},\mathscr{D}_{1j})\mid 1\leq i\leq 8, 1\leq j\leq 10\}$.
For example, from Row 2 of Table 4,
we have the computational times
$\{0.0116, 0.0314, 0.0553, 0.1029, 0.1778,$ $ 0.2166, 0.2701, 0.3510, 0.4456, 0.5342\}$ and
$\{0.0067, 0.0131, 0.0190,$ $ 0.0353, 0.0567, 0.0687, 0.0728, 0.0874, $ $0.1208, 0.1432\}$
with NIHV and IHVC, respectively, in dynamic covering decision information system  $\{(U_{1j},\Delta^{-}_{1j},\mathscr{D}_{1j})\mid 1\leq j\leq 10\}$. It is obvious that the computational times of NIHV are larger than those of IHVC in dynamic covering decision information systems $\{(U_{1j},\Delta^{-}_{1j},\mathscr{D}_{1j})\mid 1\leq j\leq 10\}$.



\begin{figure}[H]
\begin{minipage}{0.5\linewidth}
\centerline{\includegraphics[width=4.5cm]{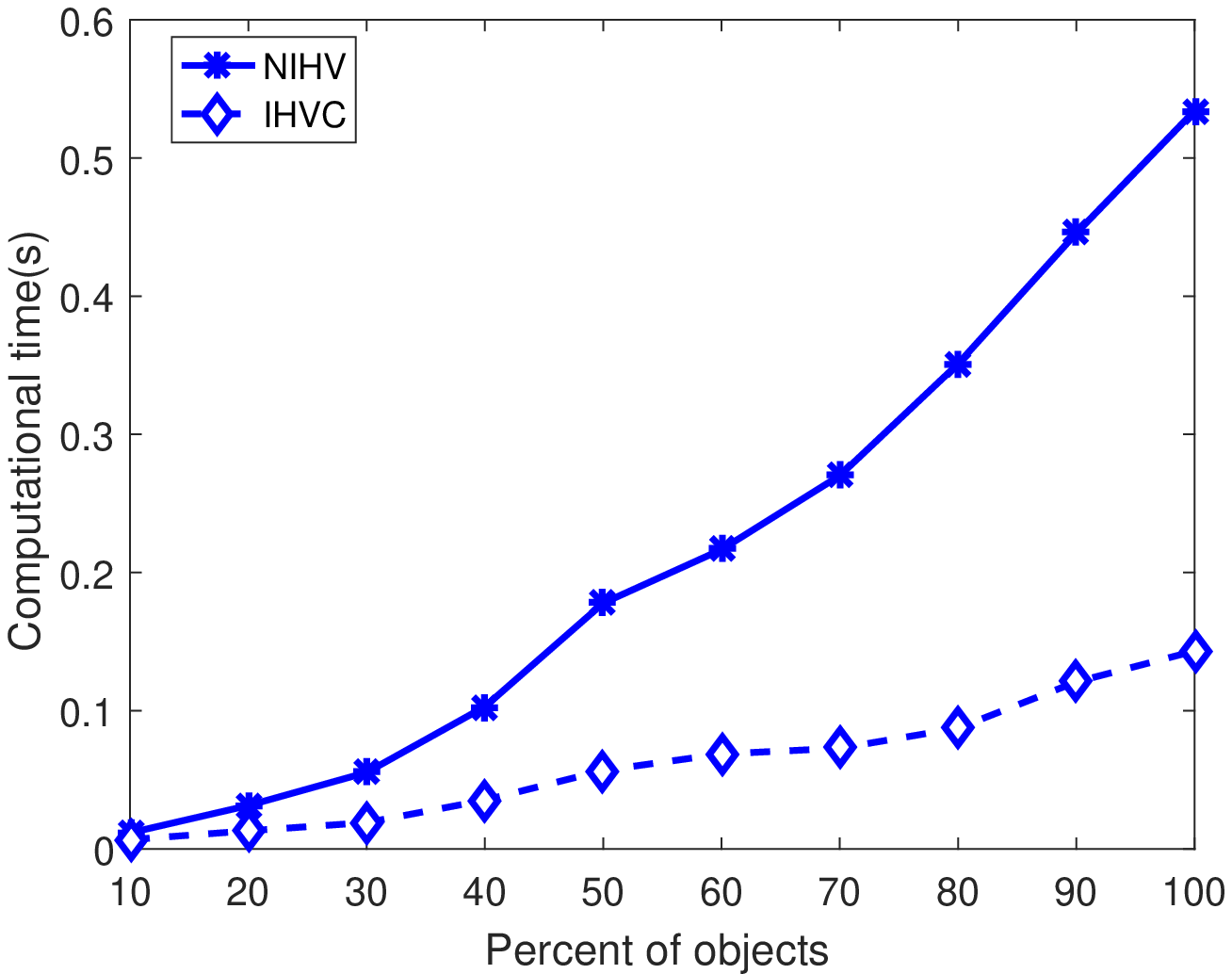}}
\centerline{($1$)}
\end{minipage}
\hfill
\begin{minipage}{0.6\linewidth}
\centerline{\includegraphics[width=4.5cm]{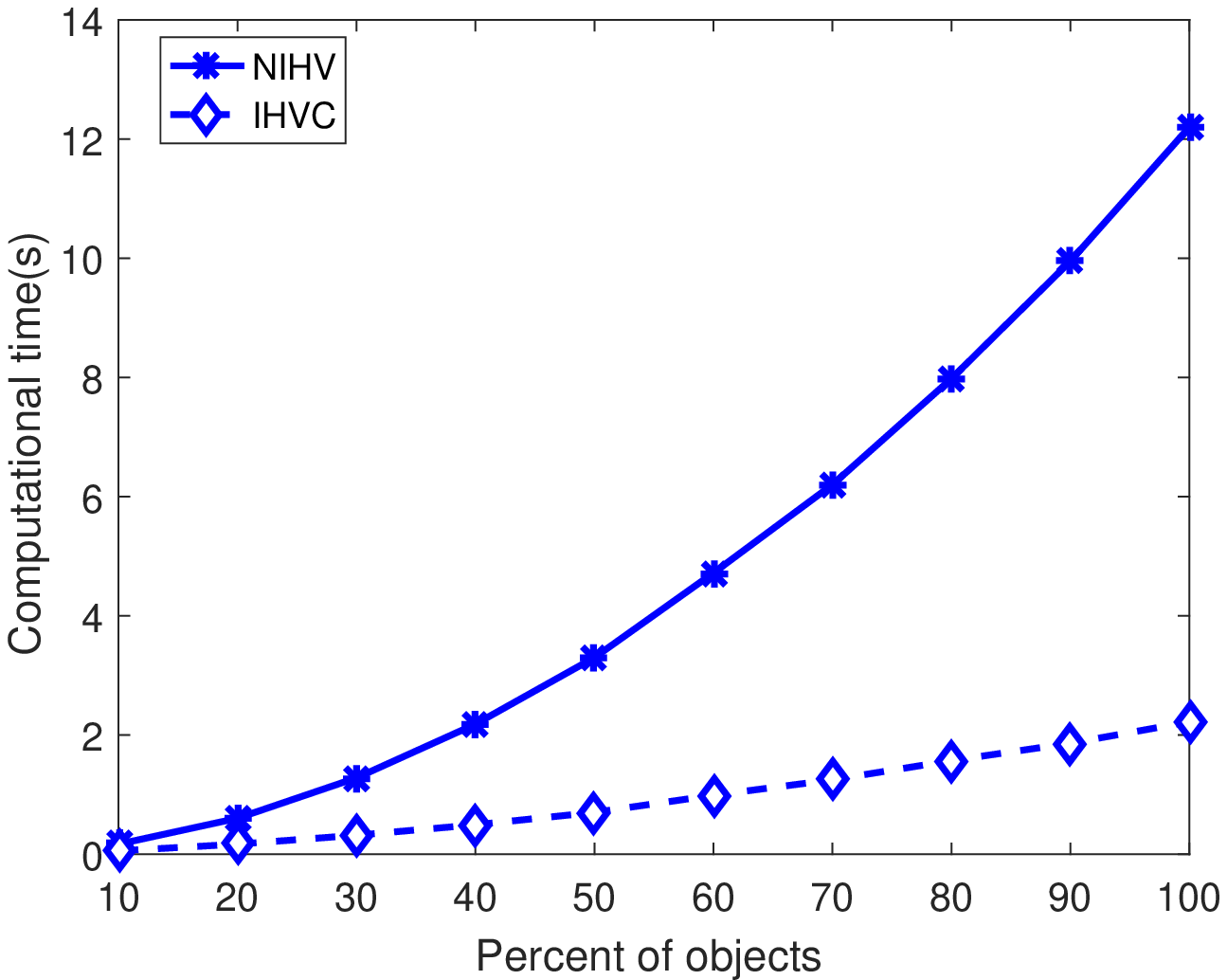}}
\centerline{($2$)}
\end{minipage}
\end{figure}

\begin{figure}[H]
\addtocounter{subfigure}{3} 
\begin{minipage}{0.5\linewidth}
\centerline{\includegraphics[width=4.5cm]{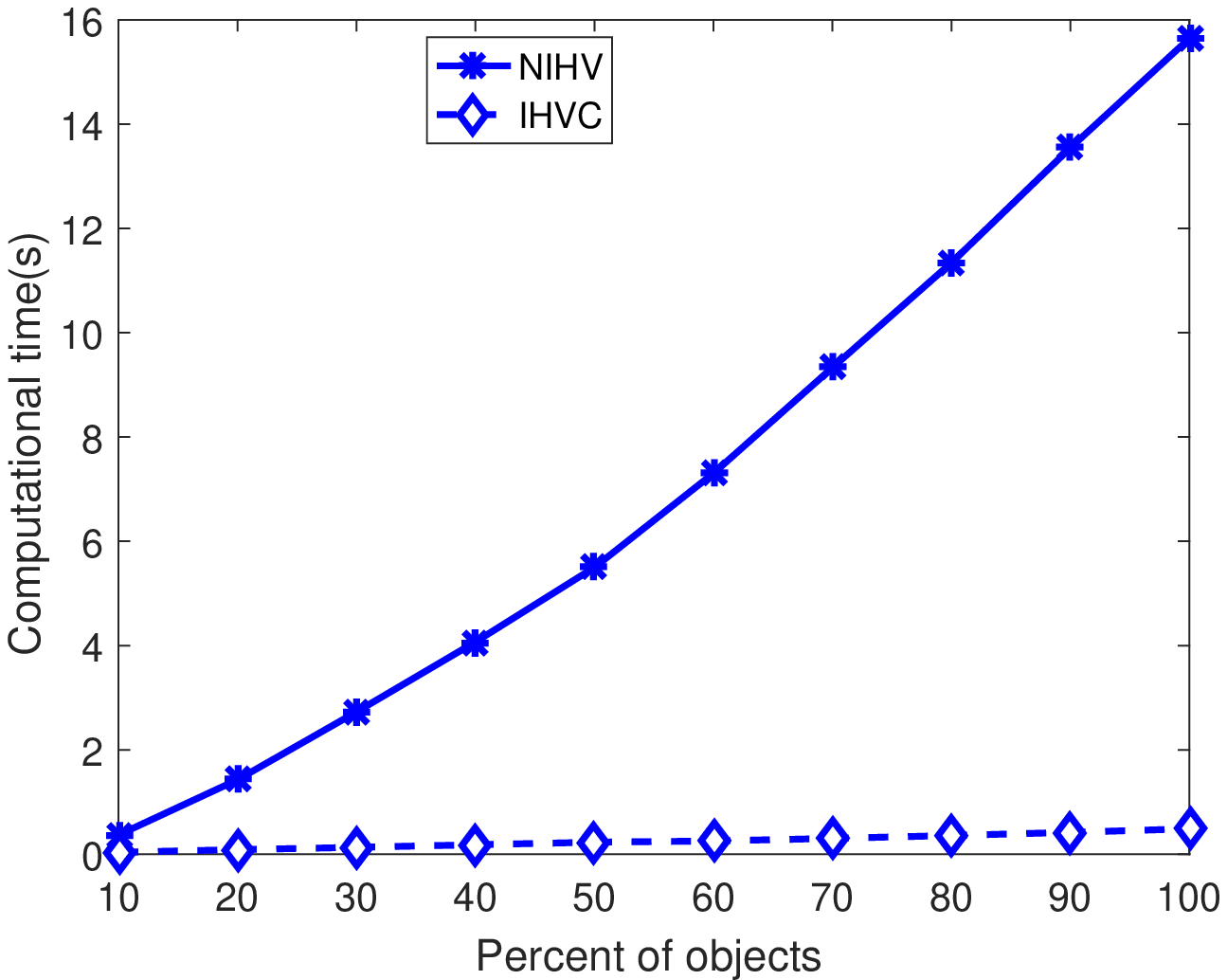}}
\centerline{($3$)}
\end{minipage}
\hfill
\begin{minipage}{0.6\linewidth}
\centerline{\includegraphics[width=4.5cm]{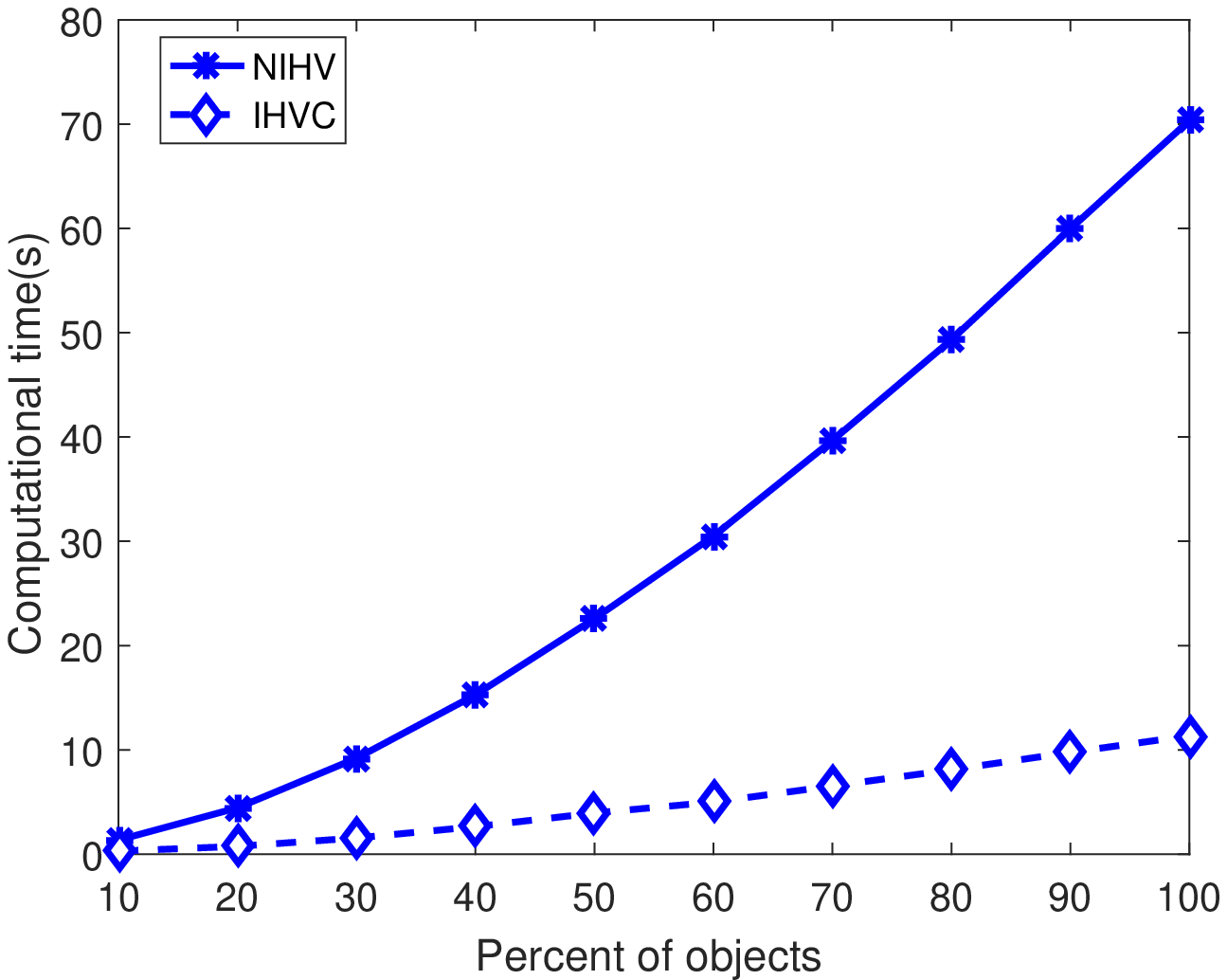}}
\centerline{($4$)}
\end{minipage}
\hfill
\end{figure}

\begin{figure}[H]
\addtocounter{subfigure}{6} 
\begin{minipage}{0.5\linewidth}
\centerline{\includegraphics[width=4.5cm]{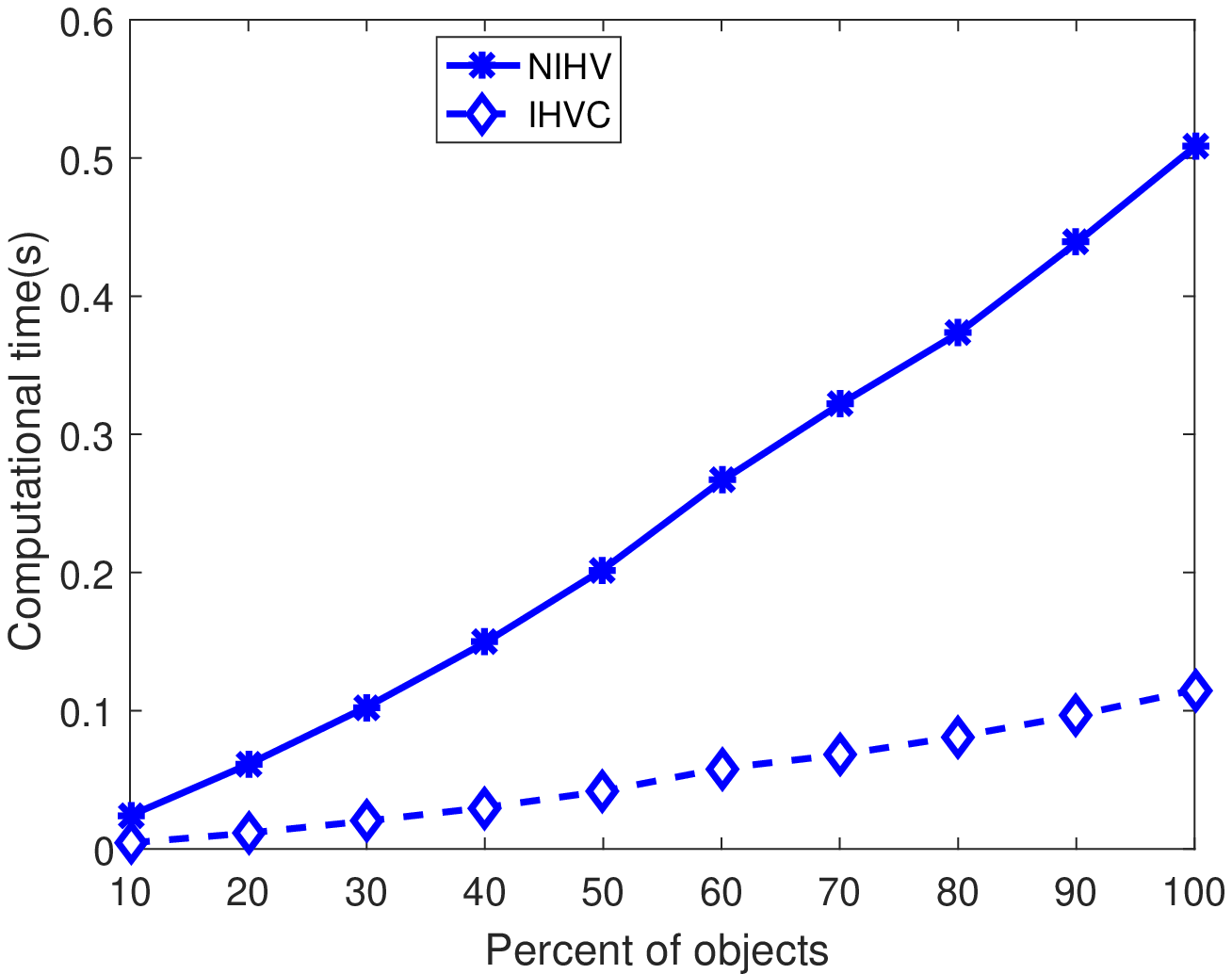}}
\centerline{($5$)}
\end{minipage}
\hfill
\begin{minipage}{0.6\linewidth}
\centerline{\includegraphics[width=4.5cm]{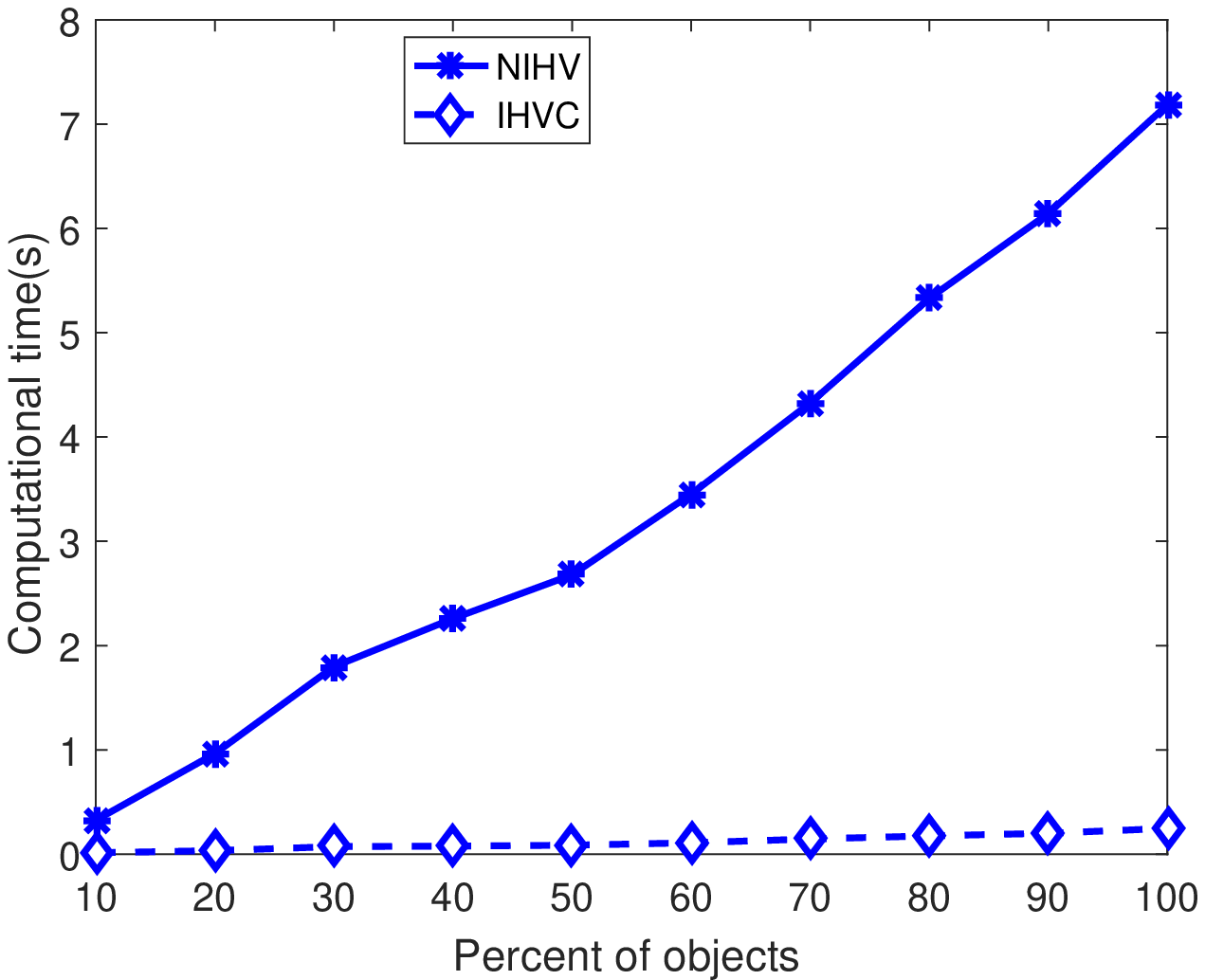}}
\centerline{($6$)}
\end{minipage}
\hfill
\end{figure}

\begin{figure}[!htp]
\begin{minipage}{0.5\linewidth}
\centerline{\includegraphics[width=4.5cm]{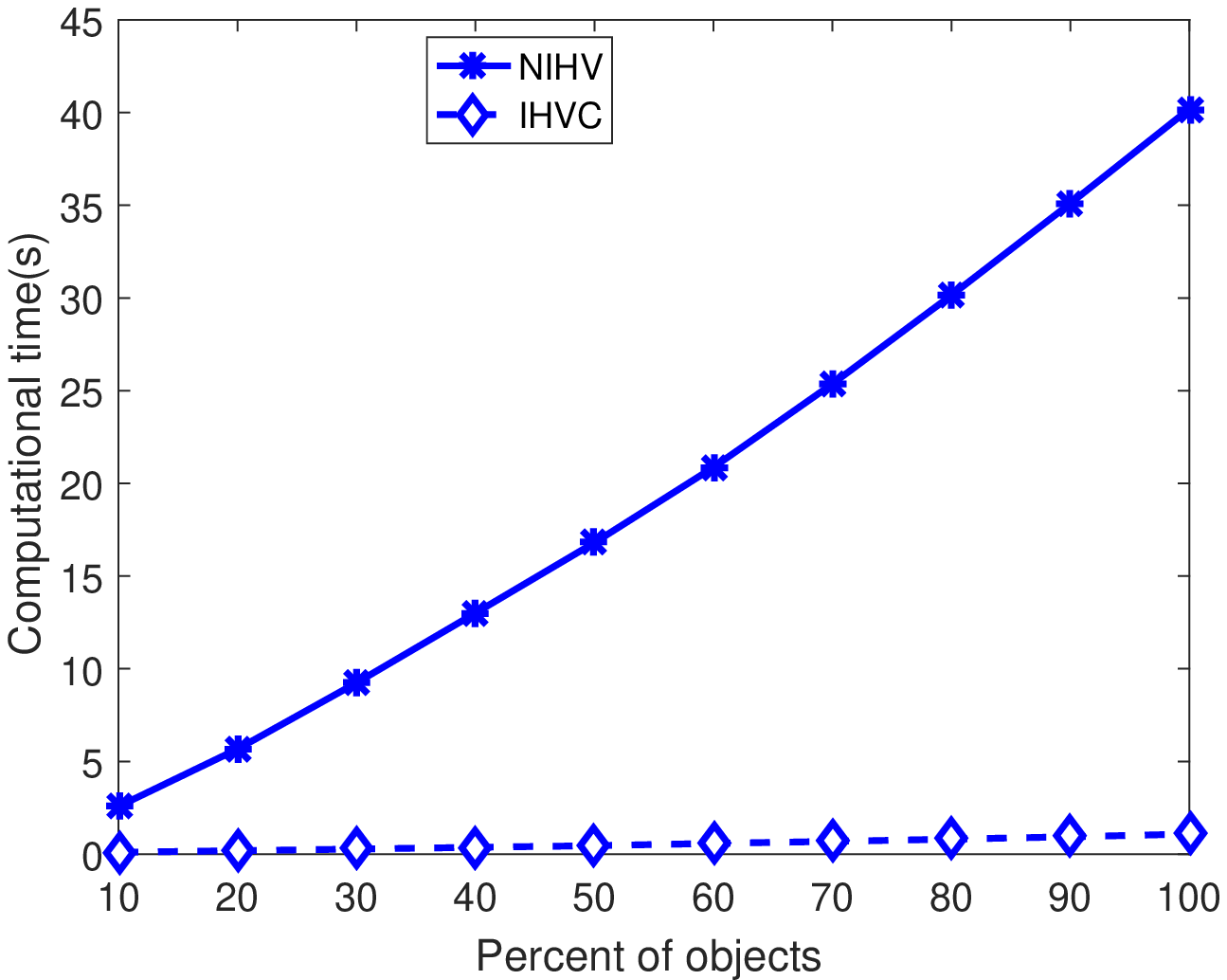}}
\centerline{($7$)}
\end{minipage}
\hfill
\begin{minipage}{0.6\linewidth}
\centerline{\includegraphics[width=4.5cm]{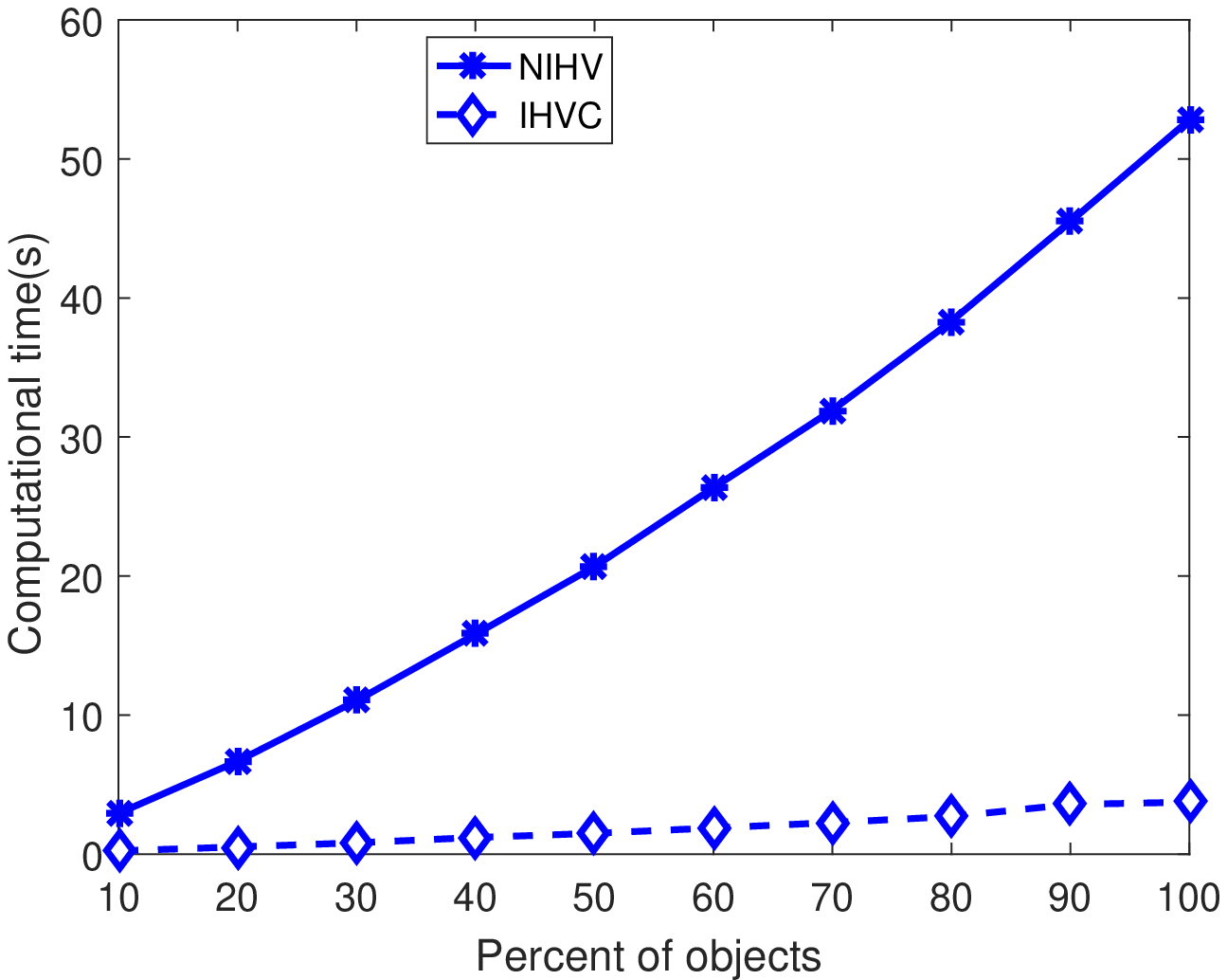}}
\centerline{($8$)}
\end{minipage}
\hfill
\caption{Computational times  using NIHV and IHVC.}
\label{figTZ_objaddall}
\end{figure}

Secondly, we employ Figure 2 to illustrate the experimental results with NIHV and IHVC in $\{(U_{ij},\Delta^{-}_{ij},$ $\mathscr{D}_{ij})\mid 1\leq i\leq 8,1\leq j\leq 10\}$, where Figure 1(i) illustrates the computational times with NIHV and IHVC in $\{(U_{ij},\Delta^{-}_{ij},\mathscr{D}_{ij})\mid 1\leq j\leq 10\}$, $-\ast-$ and $-\diamondsuit-$ denote NIHV and IHVC, respectively. Especially, we observe that IHVC performs faster than NIHV for computing attribute reducts of dynamic covering decision information systems. For example, Figure 1(1) illustrates IHVC runs faster than NIHV in $\{(U_{1j},\Delta^{-}_{1j},\mathscr{D}_{1j})\mid 1\leq j\leq 10\}$. Especially, the computational time of NIHV increases faster than IHVC with the increase of the cardinality of object set.


Therefore, the experimental results in Table 3 and Figure 2 illustrate that IHVC is more effective and feasible than NIHV for attribute reduction of dynamic
covering decision information systems when coarsening coverings.

\section{Conclusions}

Knowledge reduction of dynamic covering decision information systems is an important topic of covering-based rough set theory. In this paper, firstly, we have shown the mechanisms of updating related families of dynamic covering decision information systems when refining and coarsening covering. We have investigated how to construct attribute reducts with the updated related families for dynamic covering decision information systems. Furthermore, we have provided the incremental algorithms for computing
attribute reducts of dynamic covering decision information systems, and employed several examples to illustrate how to perform attribute reduction of dynamic covering decision information systems. Finally, we have performed the experiments to illustrate that the proposed algorithms are effective and feasible to compute attribute reducts of dynamic covering decision information systems.

In the future, we will provide incremental learning methods for dynamic covering decision information systems with variations of object sets. Especially, we will provide incremental algorithms for attribute reduction of dynamic covering decision information systems when object sets are varying with time.

\section*{ Acknowledgments}

We would like to thank the anonymous reviewers very much for their
professional comments and valuable suggestions. This work is
supported by the National Natural Science Foundation of China (Nos.61603063,61673301,11771059,61573255),
Hunan Provincial Natural Science Foundation of China(Nos.2018JJ2027, 2018JJ3518),
the Scientific
Research Fund of Hunan Provincial Education Department(No.15B004), the Scientific
Research Fund of Hunan Provincial Science and Technology Department(No.2015RS4049).

\end{document}